# A Comprehensive Bibliometric Analysis on Social Network Anonymization: Current Approaches and Future Directions

Navid Yazdanjue[1], Hossein Yazdanjouei[†2], Hassan Gharoun[†1], Mohammad Sadegh Khorshidi[†1], Morteza Rakhshaninejad[†3], Amir H. Gandomi[*1,4]

[1] Faculty of Engineering and Information Technology, University of Technology Sydney, Ultimo, NSW, 2007, Australia.

[2]Microelectronics Research Laboratory, Urmia University, Urmia, Iran.

[3]School of Industrial Engineering, Iran University of Science and Technology, Tehran, Iran.

[4]University Research and Innovation Center (EKIK), Óbuda University, 1034, Budapest, Hungary

†: These authors contributed equally to this work
*: The Corresponding Author: Gandomi@uts.edu.au

# A Comprehensive Bibliometric Analysis on Social Network Anonymization: Current Approaches and Future Directions

## Abstract

In recent decades, social network anonymization has become a crucial research field due to its pivotal role in preserving users' privacy. However, the high diversity of approaches introduced in relevant studies poses a challenge to gaining a profound understanding of the field. In response to this, the current study presents an exhaustive and well-structured bibliometric analysis of the social network anonymization field. To begin our research, related studies from the period of 2007-2022 were collected from the Scopus Database then pre-processed. Following this, the VOSviewer was used to visualize the network of authors' keywords. Subsequently, extensive statistical and network analyses were performed to identify the most prominent keywords and trending topics. Additionally, the application of co-word analysis through SciMAT and the Alluvial diagram allowed us to explore the themes of social network anonymization and scrutinize their evolution over time. These analyses culminated in an innovative taxonomy of the existing approaches and anticipation of potential trends in this domain. To the best of our knowledge, this is the first bibliometric analysis in the social network anonymization field, which offers a deeper understanding of the current state and an insightful roadmap for future research in this domain.

# 1- Introduction

Over the past century, social network applications have experienced exponential growth, fundamentally transforming the way we communicate, share information, and connect with others. With millions of users worldwide, these platforms have become an integral part of our daily lives. They are used for a wide variety of purposes, from staying in touch with friends and family, business networking, to accessing news and entertainment, and even for political engagement and social activities. In addition to their vast user bases, social network applications have also evolved in terms of complexity and sophistication and now offer a multitude of features, such as live streaming, eCommerce, virtual reality, and much more. Furthermore, with the advent of technologies like Artificial Intelligence (AI) and Machine Learning (ML), social networks have become more personalized and interactive, enhancing the user experience while expanding the amount of personal data that is collected and processed. However, the rapid growth and increasing complexity of social network applications have also brought forth significant privacy concerns. Given the volume and sensitivity of data shared on these platforms, ensuring user privacy has become paramount (Hay et al., 2007; Külcü & Henkoğlu, 2014; Zheleva & Getoor, 2007).

Privacy threats, including risks associated with identity, relationship, and attribute exposure, emerge when online platforms choose to make social network data publicly accessible or distribute it to third-party entities for research or commercial endeavors. To address this challenge, social network anonymization has emerged as a key solution to preserve user privacy. Anonymization techniques aim to hide the identity of users, their relationships, or specific sensitive attributes while allowing the useful characteristics of the network and its data to be studied and used.

Due to the mentioned exponential growth of social network applications and the corresponding abundance in user data, a surge of scholarly interest in this field has been recently sparked. Academics, researchers, and developers have been focusing on creating and enhancing anonymization techniques for social networks, recognizing the urgency to reconcile the dual demands of data utility and privacy protection. This high interest is evidenced by the increasing number of publications related to this topic in esteemed academic journals and conferences. Researchers are continually proposing novel anonymization techniques to keep up with the evolving landscape of social network applications. A key aspect of these research studies is to find a balance between ensuring user privacy and maintaining data utility. While the primary objective is to protect user privacy, it is equally important to ensure that anonymized data remains useful for research and business purposes. Hence, many proposed techniques focus on generating synthetic data or modifying social network graphs in a way that anonymizes user identities while preserving the overall network characteristics.

Regarding the importance of social network anonymization in preserving users' privacy in today's data-driven world, the main goal of this work was to collect and scrutinize related studies, especially those employing anonymization techniques, in the social network privacy preservation field and present a taxonomy of these approaches. Furthermore, the absence of any existing bibliometric literature review in the realm of social network anonymization, coupled with the advantages of bibliometric analysis, inspired us to contribute thus study in the social network anonymization domain. This research carries significant importance in this growing field, aiding scholars and industry professionals to understand the latest research trends and developments to develop novel and efficient anonymization techniques.

In this direction, we first extracted publications from 2007, the year that this concept emerged, to 2022 from the open-source Elsevier Developer Portal using Scopus API (Elsevier 2022). From the extracted articles, a total of 315 relevant articles were selected for inclusion in this study. Then, the authors' keywords were pre-processed and visualized using network visualization tools. Afterward, we conducted two primary analyses, namely statistical measures and network analysis, along with a co-word analysis. These were performed to detect the themes and topics and study their evolution, interrelations, and trends within the social network anonymization domain. Ultimately, the findings allowed us to categorize the popular approaches from the inception of this field of study into a novel taxonomy.

The rest of this study is organized as follows. Section 2 delivers an overview of the associated surveys and studies that have been conducted in the realm of social network anonymization. In Section 3, we delve into the methodology applied in this review and explain the process of gathering pertinent papers as well as the pre-processing phase. The statistical measures and analysis of the network are also presented. Moving forward, the co-word analysis and trajectory of various themes and topics' evolution are provided. At the end of this section, the prevalent approaches utilized in the field of social network anonymization are categorized based on the conducted analyses. Section 4 discusses the emerging research trends and areas of potential interest in the field of social network anonymization. Consequently, Section 5 wraps up the paper with concluding remarks.

# 2- Related Works

In recent years, the field of social network anonymization has been a focal point of intensive research, leading to a substantial and ever-growing body of literature dedicated to the domain. This extensive research activity reflects the critical importance of privacy in our increasingly interconnected world. Several surveys and literature reviews have been conducted, providing a comprehensive overview of the various anonymization techniques, their evolution, and their effectiveness in different social network types. Such studies have played a significant role in advancing our understanding of the challenges and complexities involved in social network anonymization. They have also helped identify areas that require further exploration. These outstanding surveys and literature review studies are elaborated on in the following paragraphs.

In 2008, Zhou et al. conducted a comprehensive review of existing techniques for social network anonymization (Zhou et al., 2008). They performed an organized examination of the methods employed to maintain user privacy during the sharing or publication of such data. They provided a systematic overview of the field, offering insights into the strengths and weaknesses of various approaches, their applicability to different types of data, and their effectiveness in maintaining a balance between privacy and data utility. Also, the authors recognized the challenges in privacy preservation methods within social network data, especially when compared to the traditional relational data cases that have been extensively studied. The analyses of social network anonymization methods were focused on three key aspects: preserving privacy; understanding the background knowledge available to the adversary; and maintaining data utility, which is the value of the data for further use or analysis. To better structure their review, the authors categorized the existing anonymization methods into two principal types: clustering-based approaches and graph modification approaches. Clustering-based approaches function by grouping nodes (individuals or entities) and their edges (connections) into larger collective units known as super-nodes and super-edges, respectively.

Each of these larger units (super-nodes and super-edges) is then subjected to an anonymization process, effectively hiding individual identities within the clusters. The clustering-based approach is further divided into the following four subcategories.

- Vertex Clustering: This method groups nodes, or vertices, based on some similarity metric. The similarity might be based on the attributes of the nodes or on the structure of the graph like nodes having similar connectivity (Hay et al., 2010; Hay et al., 2008). Once the nodes are grouped into clusters, they can be replaced with a single super-node to anonymize the individual nodes in each cluster. For each super-node, two features are represented, namely the number of nodes and the number of edges within the cluster, to maintain the utility of the anonymized network.
- Edge Clustering: In this method, the edges are the main focus to preserve the sensitive relationships (Zheleva & Getoor, 2007). Edges with similar properties (such as weight or type of relationship) are grouped into clusters. Similar to vertex clustering, these grouped edges can be replaced with a super-edge, anonymizing the original individual relationships. Also, the number of edges within each super-edge will be represented to maintain the utility of the network.
- Vertex and Edge Clustering: This approach, also called generalization, combines both vertex and edge clustering (CampanA, 2008), grouping both nodes and edges. It provides a more comprehensive anonymization, as it anonymizes both individual identities (nodes) and their relationships (edges). Once nodes and edges are grouped into clusters, they can be replaced with super-nodes and super-edges, effectively hiding individual information within the network. More precisely, the process of anonymizing node attributes utilizes a generalization technique, which is extensively researched in the context of relational data. For structural anonymization, this method utilizes edge generalization, which is similar to the method outlined by (Zheleva & Getoor, 2007). However, a significant distinction exists in this method, as it incorporates both the loss of generalization information loss and structural information loss during the clustering process.
- Vertex-Attribute Mapping Clustering: This method was originally used for anonymizing the bipartite graphs (Cormode et al., 2008, 2010). When publishing a bipartite graph, the structure of the graph is preserved. The nodes are organized into clusters, and the relationship or mapping between these clusters in the original graph and those in the published graph is made publicly available. This process enables the anonymization of individual nodes while maintaining the overall structure and relationships of the graph for analysis or research purposes. Consequently, it is necessary to carefully construct the mapping between the clusters. This helps ensure that the anonymization process is effective and maintains the integrity of the structural relationships present in the original graph.

On the other hand, graph modification approaches alter the structure of the network graph to conceal individual identities while preserving the overall characteristics of the network. The authors also considered three subcategories for the graph modification approach, which are subsequently defined.

- Optimization Graph Construction: This method involves constructing a new optimized version of the original graph with a new degree sequence that is K-degree anonymous, maintaining the overall structure while ensuring that individual identities are concealed (Liu & Terzi, 2008). Specifically, the main idea behind K-degree is to modify the original network so that each node in the network has the same degree as at least 'K-1' other nodes.
- Randomized Graph Modification: This method introduces randomness into the graph modification process to ensure anonymity. They may randomly add, delete, or modify nodes and edges within the network while attempting to preserve the overall structure and characteristics. The randomness introduced by these methods makes it more difficult to de-anonymize the data, thus providing an additional layer of privacy protection. Zhou et al. categorized the introduced techniques in this method into three general categories, including Randomized Edge Construction, Randomized Spectrum Preserving, and Randomized Weight Perturbation.

- Greedy Graph Modification: This method iteratively modifies the graph to provide the K-anonymity-based models, which results in protecting individual identities. At each iteration, the method makes the modification that provides the greatest immediate benefit according to a specific objective, such as maximizing privacy or minimizing information loss. While these methods can be simpler and faster than other methods, they may not always provide the optimal solution, as the modifications are done based on immediate gains rather than long-term optimization.

The social network anonymization taxonomy proposed by Zhou et al. is presented in Fig. 1.

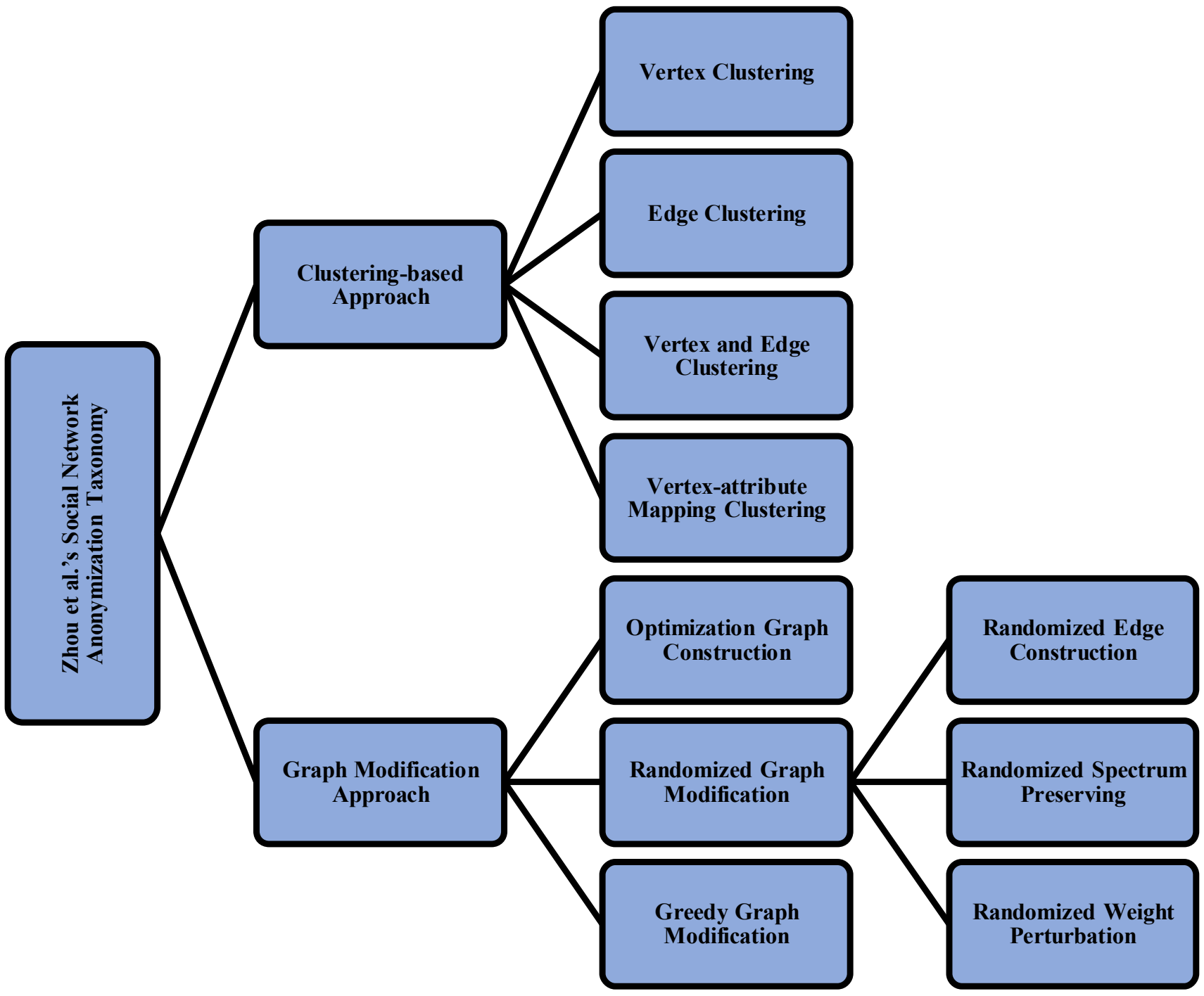

*Figure 1. Social Network Anonymization Taxonomy introduced by Zhou et al. (2008)*

In 2010, Wu et al. investigated a review of the advancements in research concerning the privacy-preserving publishing of graph and social network data (X. Wu et al., 2010). Their analysis is oriented toward understanding how to anonymize and publish such data while maintaining user privacy. Besides, the authors divided the anonymization strategies for simple graphs into three main categories. The first category is K-anonymity-based privacy preservation via edge modification, which involves altering edges within the graph in a way that at least 'K' nodes share similar identifiable characteristics, thereby preserving privacy. In this category, they consider three subcategories, including K-degree generalization, K-neighbourhood anonymity, and K-automorphism anonymity, which are subsequently described.

In the K-degree generalization or K-degree anonymity model, the structure of the graph is modified in such a way that at least 'K' nodes share the same degree. Also, the K-neighbourhood anonymity focuses on the neighborhood of each node. A node's neighborhood in a graph is all the nodes it is directly connected to. K-neighborhood anonymity ensures that each node shares the same neighborhood with at least 'K-1' other nodes. This means that for each node, there are at least 'K' nodes in the graph, including itself, that are connected to the same set of nodes. Besides, an automorphism in graph theory is an isomorphism from a graph to itself. It essentially means a reordering of the nodes of the graph in such a way that the new graph is exactly the same as the original. In the context of the K-anonymity model, K-automorphism anonymity means that each node belongs to a set of at least 'K' nodes that are indistinguishable from each other when considering the entire structure of the graph.

The second category is probabilistic privacy preservation via randomization. In this approach, the connections between nodes are randomized to add a level of uncertainty, effectively anonymizing the data while still preserving the overall structure and characteristics of the graph. The authors explored two edge-based randomization strategies that are frequently used in social networks: random addition/deletion and random switch. The random addition/deletion approach involves adding or deleting edges at random, while the random switch approach involves randomly switching a pair of existing edges.

The third category is privacy preservation via generalization which is identical to the clustering-based approach. Following the categorization of simple graph anonymization methods, the authors turned their attention to rich graphs that are more complex and contain additional information, like edge direction, edge weights, or additional attributes on the nodes or edges. Since anonymizing these graphs requires more sophisticated methods, the authors reviewed current approaches for handling this increased complexity. Their proposed social network anonymization taxonomy is illustrated in Fig. 2.

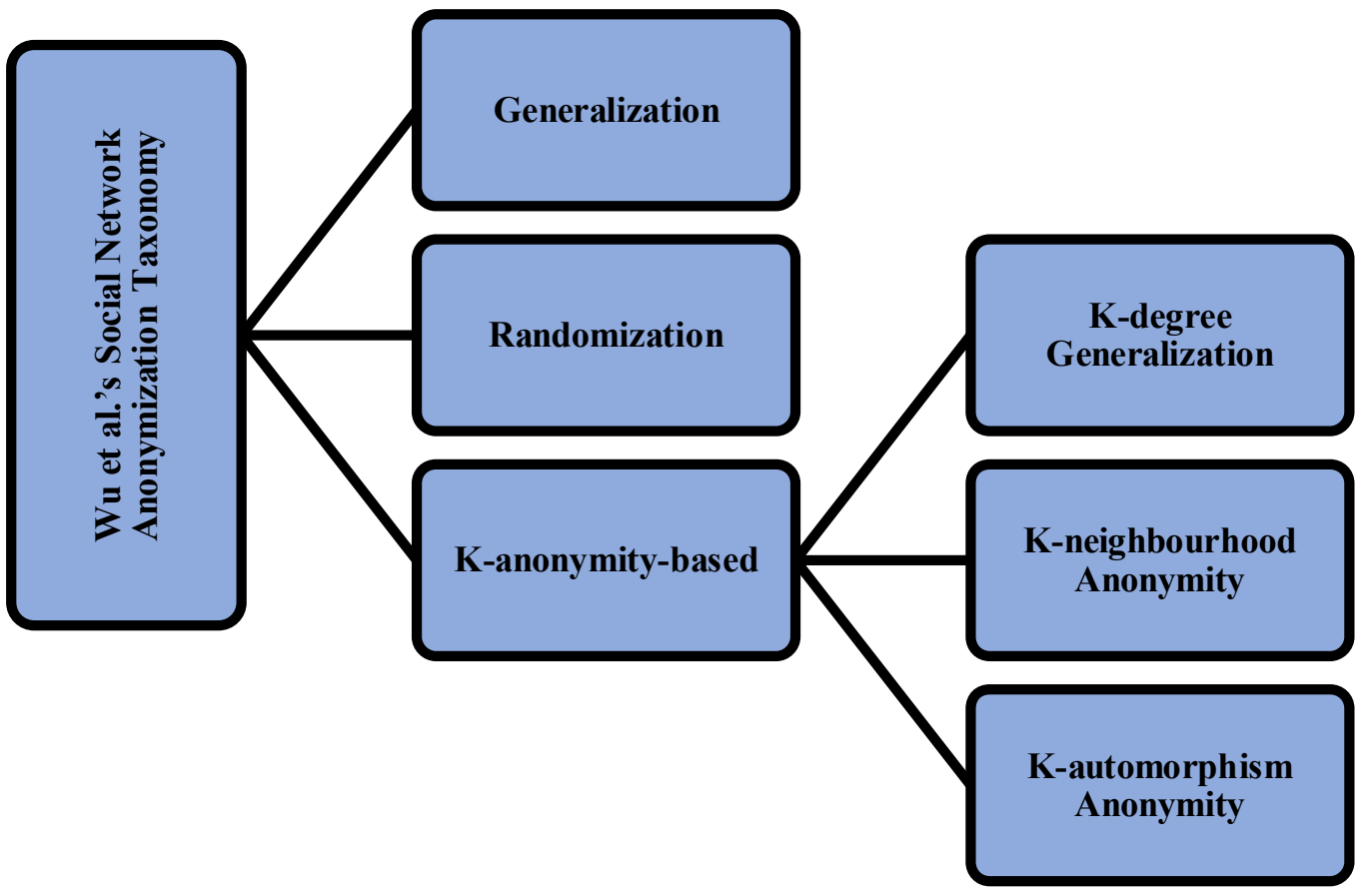

*Figure 2. Social Network Anonymization Taxonomy proposed by Wu et al. (2010)*

In 2016, Abawajy et al. provided an overview of the recent advancements in techniques for social network anonymization when releasing social network data publicly, as well as the challenges and potential research directions (Abawajy et al., 2016). They also covered various privacy threats and attacks that adversaries might use to exploit anonymized data from social networks.

The authors of this work organized the methods for anonymizing social network data into two primary groups: Non-perturbation Privacy Preservation Models and Differential Privacy Models. Four subdivisions are identified within the Non-perturbation Privacy Preservation Models: random graph editing, k-anonymization methods, clustering-based techniques, and probabilistic privacy preservation approaches. Importantly, the Probabilistic Privacy Preservation Approach was introduced as new category for anonymizing social networks. This innovative method, also known as the Uncertain Graph Approach, attributes a specific probability value to each network connection, indicating the likelihood of its existence (Boldi et al., 2012). This means each connection in the network is not certain but has a chance of being there, which is represented by a probability value. In the context of social network anonymization, an uncertain graph can help preserve user privacy. By introducing uncertainty into the connections between individuals, it becomes more challenging to infer specific details about an individual based on their connections within the network. The presence of uncertainty can make it harder to re-identify individuals in the network, thereby protecting their privacy.

Furthermore, differential privacy is a mathematical definition of privacy that was first proposed by Cynthia Dwork in 2006 and has since become a popular method for ensuring privacy in data analysis (Dwork et al., 2006, 2016). In tabular databases, it gives a guarantee that the removal or addition of a single database entry does not significantly affect the results of any statistical queries. Within the Differential Privacy category in the context of social network anonymization, Abawajy et al. (2016) delineated two subgroups based on whether differential privacy is applied at the node or edge level, which are elaborated in the following:

- Node-level Differential Privacy: This technique focuses on protecting the privacy of the individual nodes and their adjacent edges (their direct connections). When node-level differential privacy is applied, it makes it hard to infer the presence of a specific individual in the social network. The noise is introduced in such a way that whether a particular person is part of the network or not cannot be confidently determined. Moreover, node-level differential privacy provides protection to the edges adjacent to a node. This means that even if an individual is known to be part of the network, the presence or absence of a particular edge of that individual in the anonymized data cannot be determined with certainty.
- Edge-level Differential Privacy: This technique focuses on protecting the connections within a social network, represented as edges in the network graph. The principle behind it is that the presence or absence of specific edges should be concealed to maintain privacy, while the overall pattern of edges can be made public. Edge-level differential privacy aims to add enough "noise" to these edges such that it becomes hard to determine with certainty whether any particular edge exists or not.

The taxonomy proposed by Abawajy et al. (2016) is demonstrated in Fig. 3.

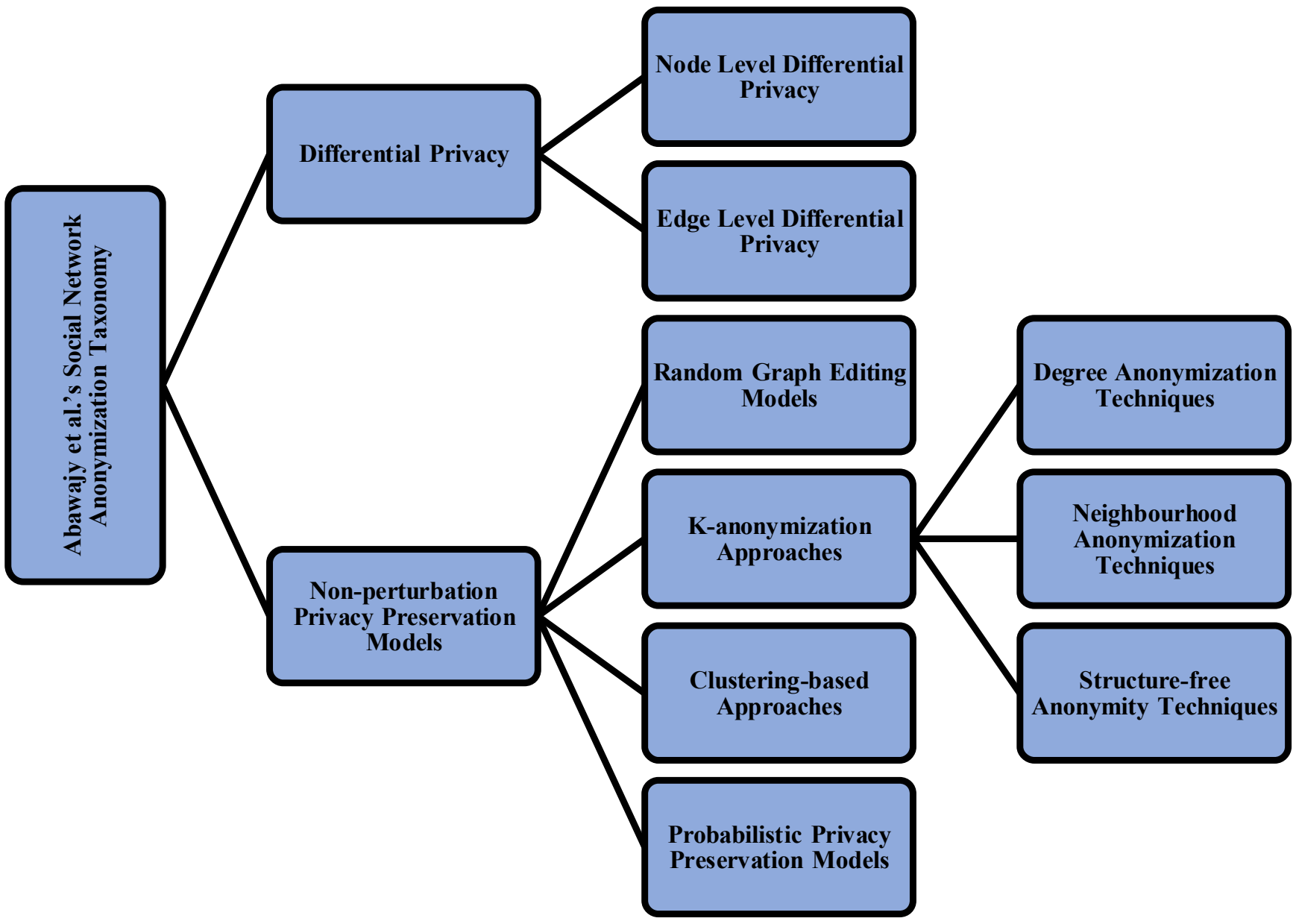

*Figure 3. Social Network Anonymization Taxonomy developed by Abawajy et al. (2016)*

During the same year, Ji et al. conducted a study on the methods of both anonymizing and de-anonymizing graph data (Ji et al., 2016). In their analysis, they reviewed various scholarly articles and classified graph anonymization strategies into six main categories: Naive ID Removal; Edge Editing-based Anonymization that is identical to the edge modification in the graph modification approach; K-anonymity; Aggregation/Class/Cluster based Anonymization; Differential Privacy; and Random Walk-based Anonymization. In their taxonomy, aside from the Naive ID Removal and Random Walk-based Anonymization methods, all other approaches align with those specified in earlier taxonomies. Therefore, Naive ID Removal and Random Walk-based Anonymization are elaborated as follows.

- Naive ID Removal is one of the simplest and earliest approaches for social network anonymization, in which explicit identifiers of the nodes (such as names, usernames, and any unique identifiers) in the network are simply removed or replaced with non-identifying labels.
- Random Walk-based Anonymization is a privacy preservation approach designed to protect the edges between nodes in a social network (Mittal et al., 2012). This approach replaces an existing edge between two nodes with a new edge that's determined through a process known as a random walk. This approach effectively randomizes the relationships in the network, making it difficult to ascertain the true relationships while maintaining the overall structure and characteristics of the network.

The taxonomy proposed by Ji et al. (2016) is represented Fig. 4.

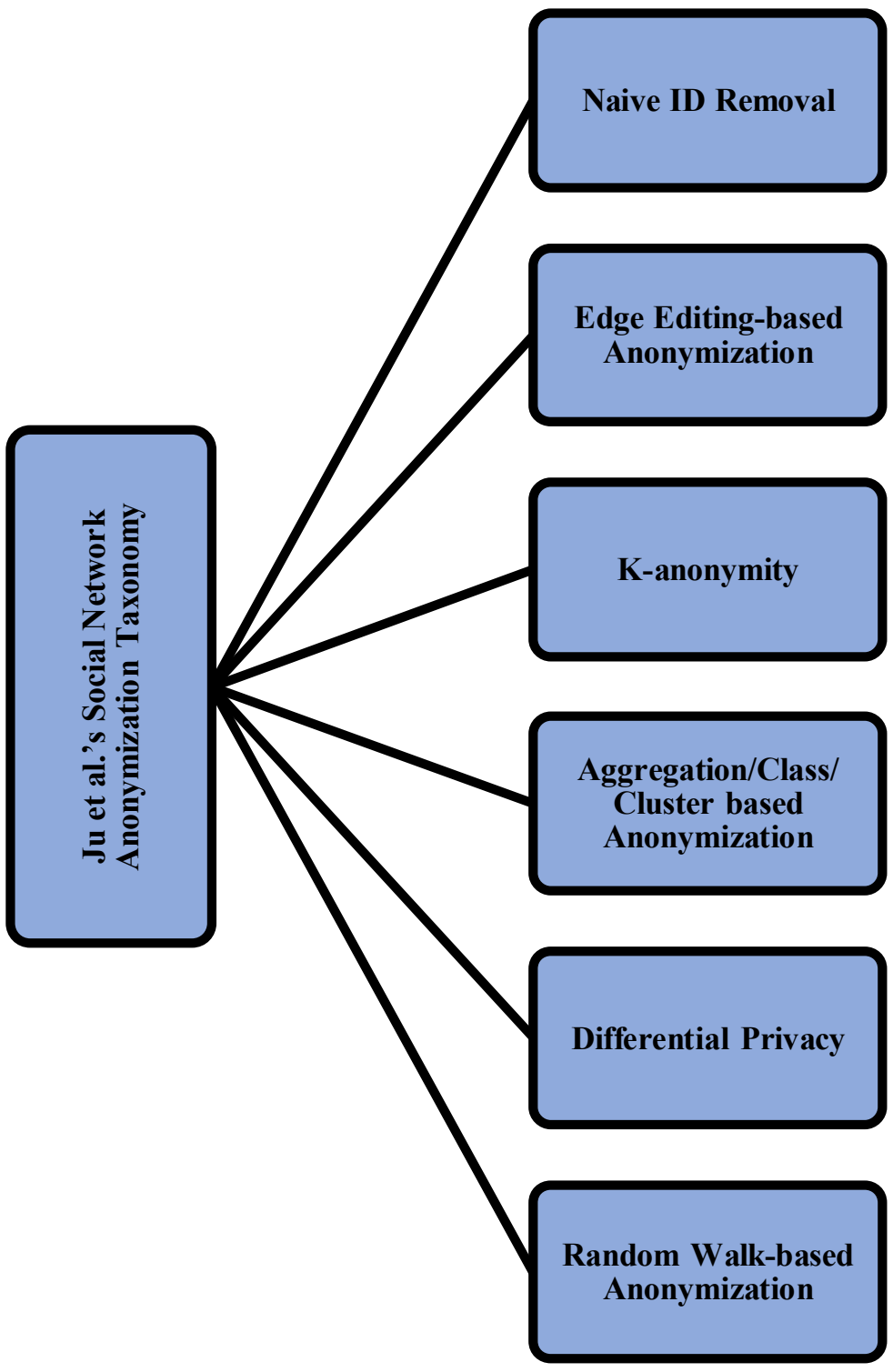

*Figure 4. Social Network Anonymization Taxonomy by Ji et al. (2016*

In 2017, Casas-Roma et al. conducted a review of research papers on social network anonymization, focusing on those that suggested graph modification techniques to ensure network anonymity (Casas-Roma et al., 2017b). They provided a comprehensive discussion about the advantages and drawbacks of each method. The researchers organized the graph modification strategies into three primary categories: Edge and Vertex Modification, Uncertain Graphs, and Generalization and Clustering-based approaches. Further, they subdivided the Edge and Vertex Modification techniques into three additional subcategories, namely K-anonymity, Extending K-anonymity, and Beyond K-anonymity methods. Their taxonomy is illustrated visually in Fig. 5.

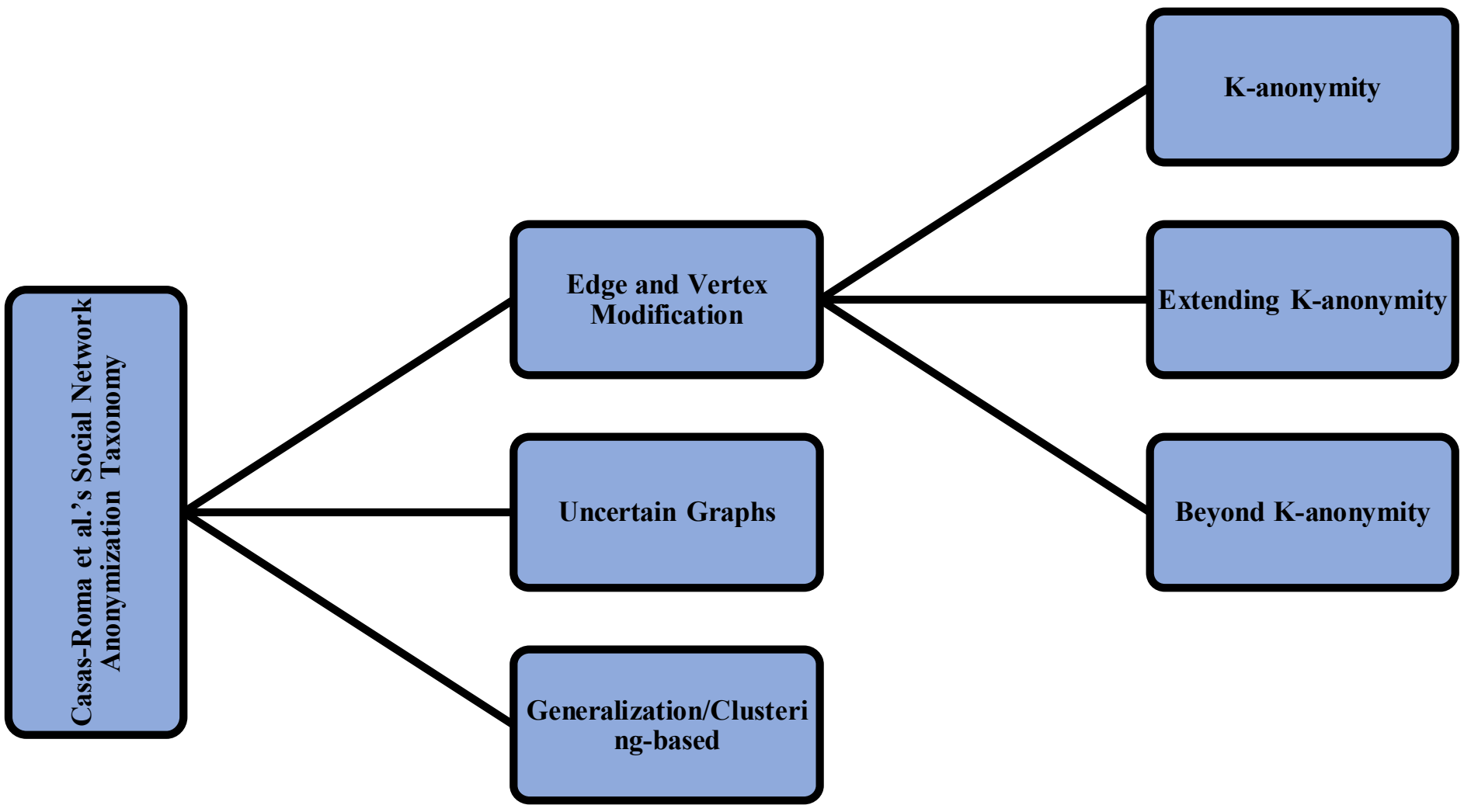

*Figure 5. Social Network Anonymization Taxonomy by Casas-Roma et al. (2017)*

In 2018, Siddula et al. conducted a review of the methods designed to preserve the privacy of users and their relationships in social networks. Based on a previous study (Casas-Roma et al., 2017b), they classified the privacy concerns of social networks into three categories: Node Privacy, Attribute Privacy, and Link Privacy. Nevertheless, their article focused only on studies suggesting anonymization methods to protect node and link privacy. As previously highlighted, the authors' focus was exclusively on node and link privacy issues. For the category of node privacy, they examined studies that implemented anonymization through Naïve Anonymization and Node Perturbation methods. After analyzing the surveyed research, they concluded that there are predominantly two strategies to perturb nodes in social networks, specifically Random Perturbation and Constrained Perturbation. Additionally, in terms of link privacy, they examined research introducing social network anonymization techniques based on Edge Perturbation and Random Walk methodologies, from which they determined that there are generally five distinct approaches to perturb the edges in a social network. These methods include Intact Edges, Partial-Edge Removal, Cluster-Edge Anonymization, Cluster-Edge Anonymization with Constraints, and Removed Edges.

It is worth mentioning that Siddula et al. did not propose a taxonomy for techniques used in social network anonymization. Nevertheless, we have derived a classification scheme based on the studies they reviewed, which is graphically illustrated in Fig. 6.

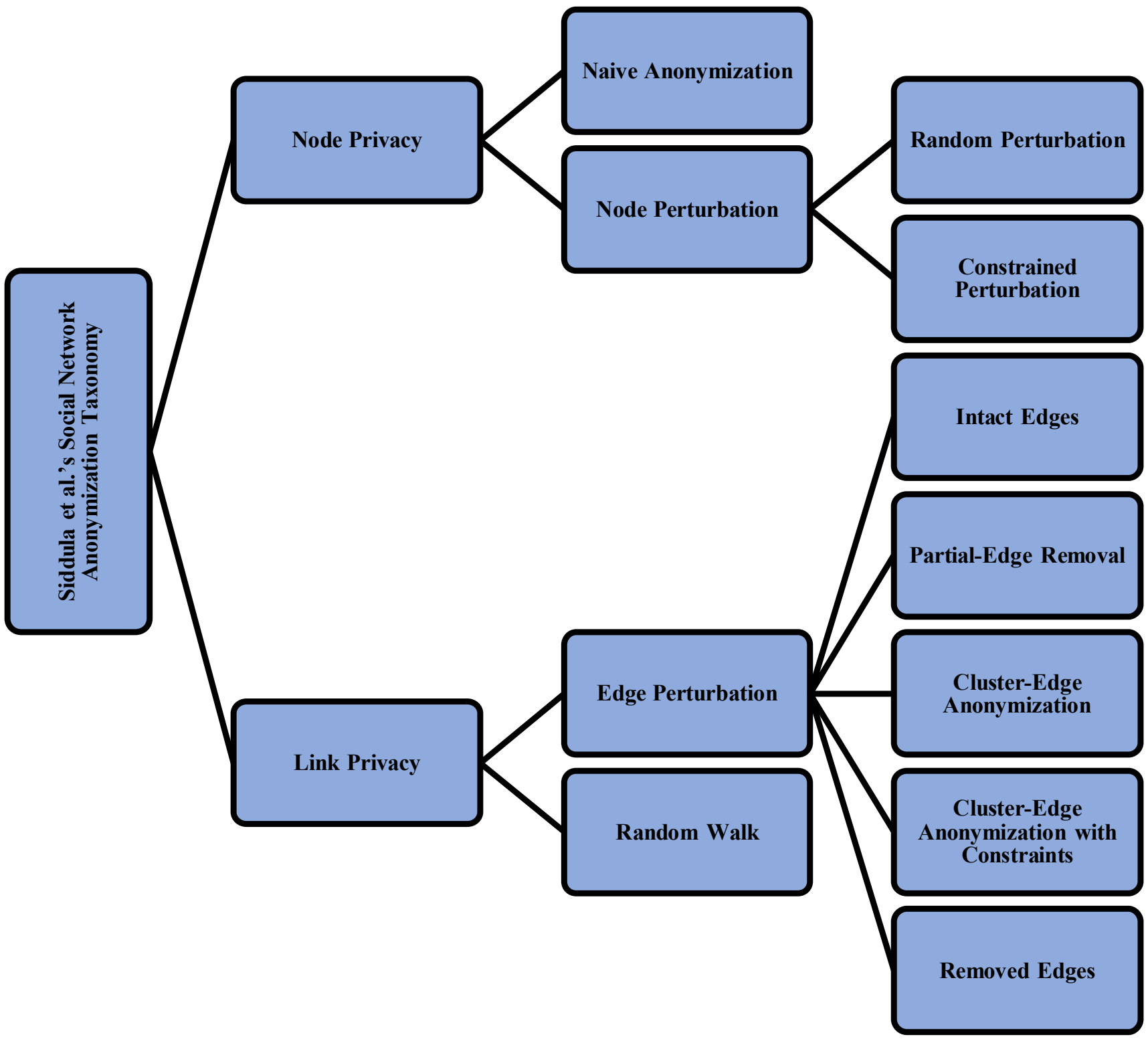

*Figure 6. Social Network Anonymization Taxonomy based on Siddula et al. (2018)*

In 2019, Sathiya Devi et al. conducted a literature survey concerning anonymization methods in social networks (Sathiya Devi & Indhumathi, 2019). Their primary focus was on techniques used to maintain the attribute privacy of social network users. In line with this, they grouped these methods into five distinct categories: K-anonymity, L-diversity, T-closeness, Slicing, and Differential Privacy. In their suggested classification, three concepts, namely L-diversity, T-closeness, and Slicing, are elaborated upon as follows.

- L-diversity: This model is an extension of K-anonymity, introduced to overcome some of its limitations. The principle of L-diversity is that within each group of 'K' indistinguishable individuals, there should be at least 'L' "well-represented" values for each sensitive attribute.
- T-closeness: This model is another extension of K-anonymity and L-diversity, introduced to address their remaining weaknesses. The principle of T-closeness is that the distribution of a sensitive attribute within any group of 'K' indistinguishable individuals must be close to the overall distribution of that attribute in the entire dataset.
- Slicing: This technique divides data both horizontally and vertically. The horizontal division groups together similar records. The vertical division separates the dataset into different "slices," each containing a subset of the attributes. Each slice is then independently anonymized by permuting the order of the records within the slice.

A visual representation of their taxonomy is provided in Fig. 7.

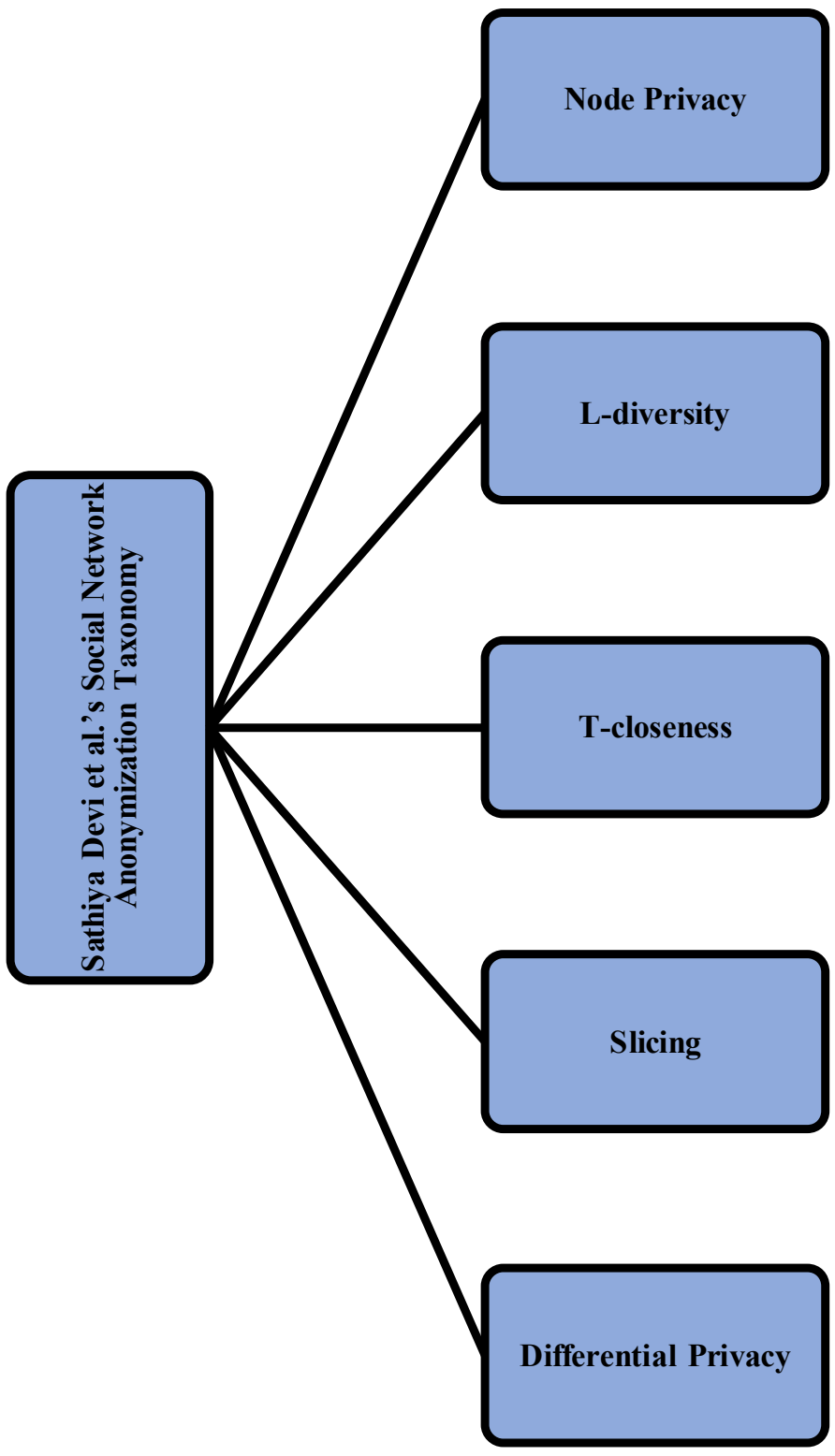

*Figure 7. Social Network Anonymization Taxonomy proposed by Sathiya Devi et al. (2019)*

In 2020, Majeed et al. reviewed the techniques employed in anonymizing data to preserve the privacy of published data (Majeed & Lee, 2020). As part of their study, they offered a taxonomy for the anonymization techniques used to protect the privacy of social network data. In their proposed taxonomy, privacy-aware graph computation refers to a strategy where instead of sharing the entire graph data (which might include sensitive information), only specific aggregate properties or statistics of the graph are computed and released in response to queries from data analysts. Moreover, hybrid graph anonymity methods combine different techniques for anonymizing social networks, aiming to create an appropriately anonymized version of the network. This approach seeks to address the trade-off between maintaining privacy and preserving usefulness. The schematic representation of their proposed taxonomy can be seen in Fig. 8.

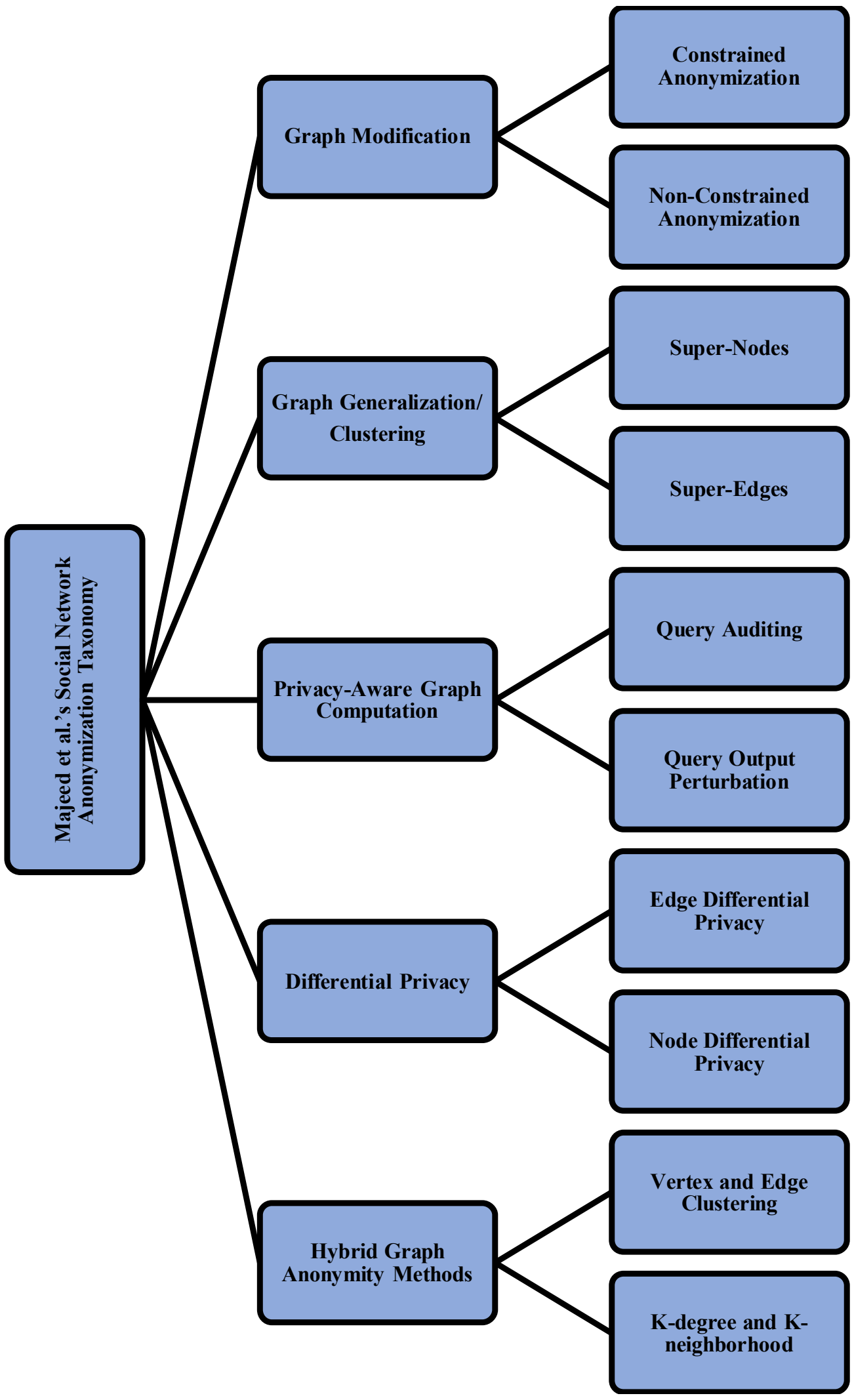

*Figure 8. Social Network Anonymization Taxonomy developed by Majeed et al. (2020)*

In a recent study, Kiranmayi et al. scrutinized research focused on anonymizing social networks in order to preserve user privacy (Kiranmayi & Maheswari, 2021). They proposed a classification of methods used for social network anonymization, which is visually represented in Fig. 9.

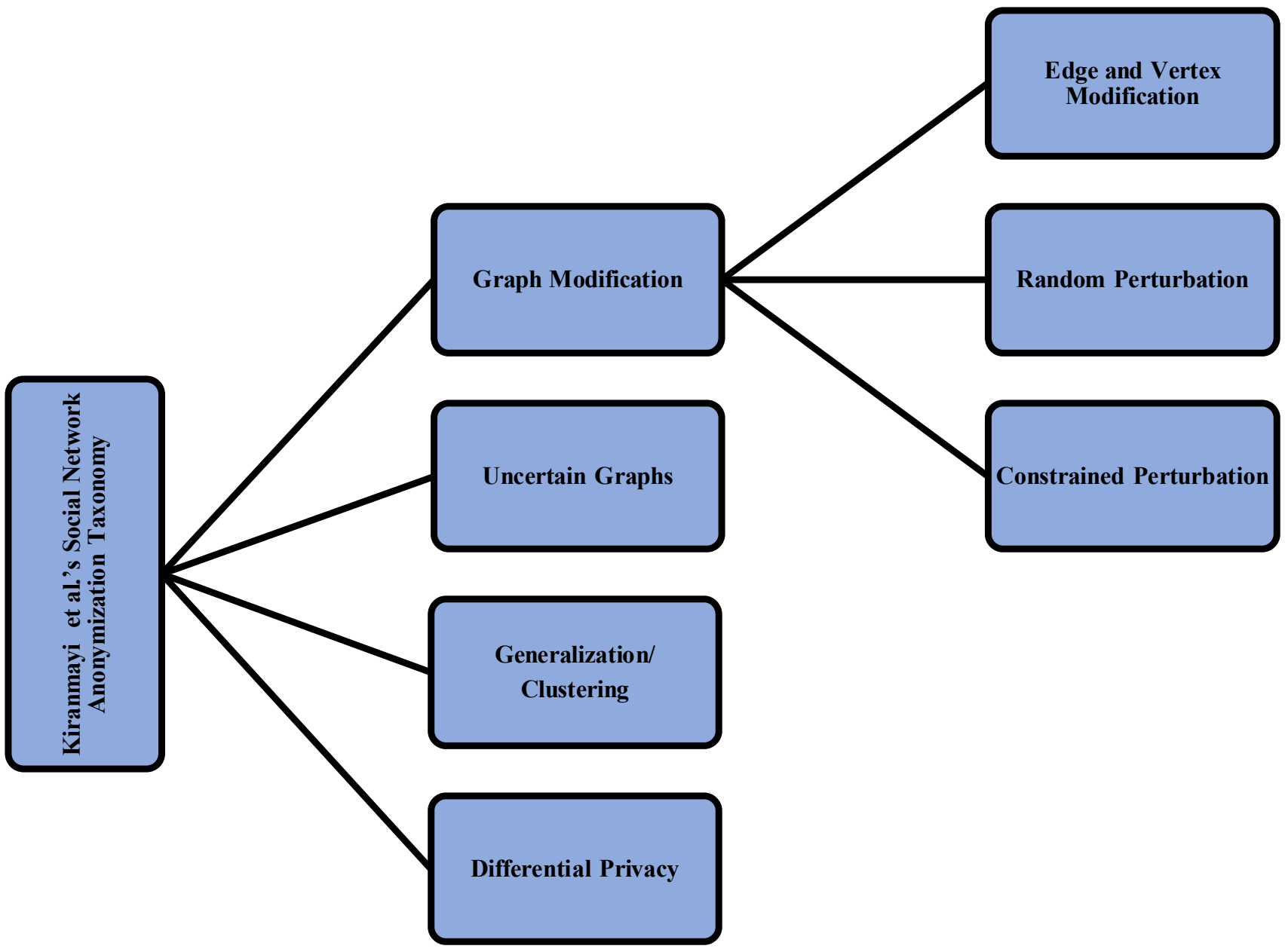

*Figure 9. Social Network Anonymization Taxonomy proposed by Kiranmayi et al. (2021)*

While the previously mentioned studies have indeed provided a valuable understanding of the various social network anonymization approaches and taxonomies, to the best of our knowledge, there has yet to be a comprehensive bibliometric literature review conducted within this area. Such a study could offer insight into emerging trends, key themes, and potential areas of interest for future research.

Consequently, this article presents a bibliometric analysis of social network anonymization studies to identify the current trends in this field, the main approaches of social network anonymization, as well as the recent advances. In this regard, we collected all the papers conducted from 2007 to 2022 from the open-source Elsevier Developer Portal using the Scopus API (Elsevier 2022). After pre-processing these works, we performed statistical, network, and co-word analyses to detect the high-trend topics and themes in this field. Additionally, a new taxonomy of the current social network anonymization approaches is provided. The main contributions of this work can be summarized as follows:

(i) All published studies related to the social network anonymization field were collected and pre-processed.
(ii) A network visualization of the used keywords was created by the authors to understand the most frequent keywords.
(iii) Statistical and network analyses were performed to identify the main keywords, themes, and topics.
(iv) A co-word analysis was conducted to detect the prominent themes and topics.
(v) The evolution of the social network anonymization themes was investigated.
(vi) A novel taxonomy of the social network anonymization approaches was developed based on the conducted analyses.
(vii) Future research trends in the social network anonymization field are presented.

# 3- Research Methodology

The methodology of this study draws upon the bibliometric analysis of the previous research studies conducted in the field of social network anonymization. In the pursuit of this objective, we utilized the keywords provided by the authors of existing scholarly articles to construct our bibliometric analysis methodology, which are illustrated in Fig. 10. This approach allowed us to investigate, assess, and analyze the literature in this field.

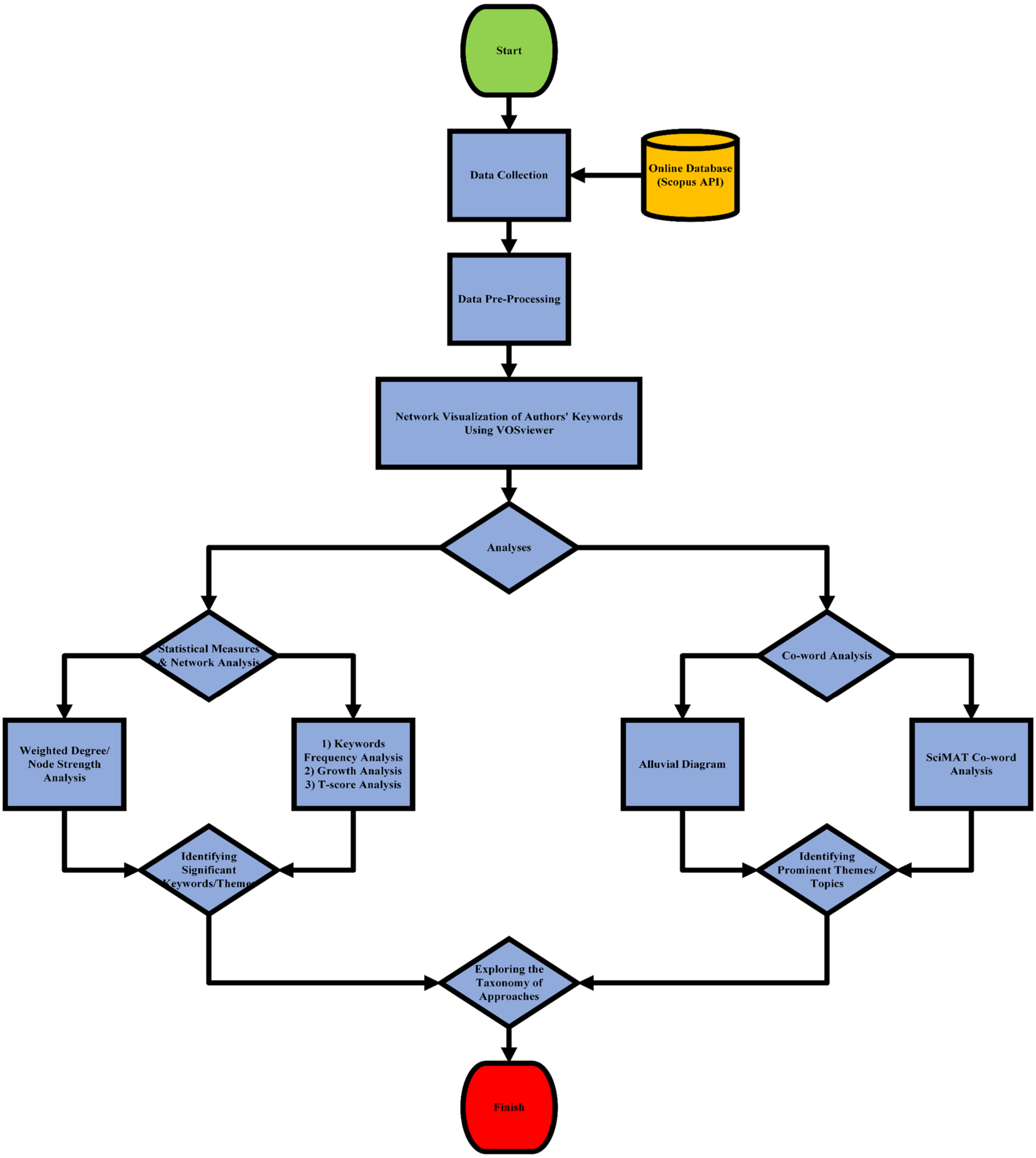

*Figure 10. The Research Methodology*

# 3-1- Data Collection and Pre-processing

As mentioned earlier, this study used the author keywords of the papers published in the social network anonymization field as the dataset. The articles were gathered from the open-source Elsevier Developer Portal via the Scopus API (Elsevier 2022), offering a detailed assortment of features associated with the retrieved papers, such as article title, journal title, year of publication, authors, authors' affiliations, authors' keywords, the number of citations, references, etc.

The search query "social network anonymization" and related searches, such as "online social network anonymization," "graph anonymization," "social network privacy preservation," "graph privacy preservation," and "social graph anonymization," were employed to extract all relevant articles in this area of research.

From the collected articles, we found that 315 papers were published in the field of social network anonymization over the past 16 years (2007-2022). The information displayed in Fig. 11 suggests a consistent rise in the quantity of research papers published in this domain from 2007 to 2019, reflecting a growing interest from researchers. However, from 2020 to 2022, the number of publications declined in the field of social network anonymization.

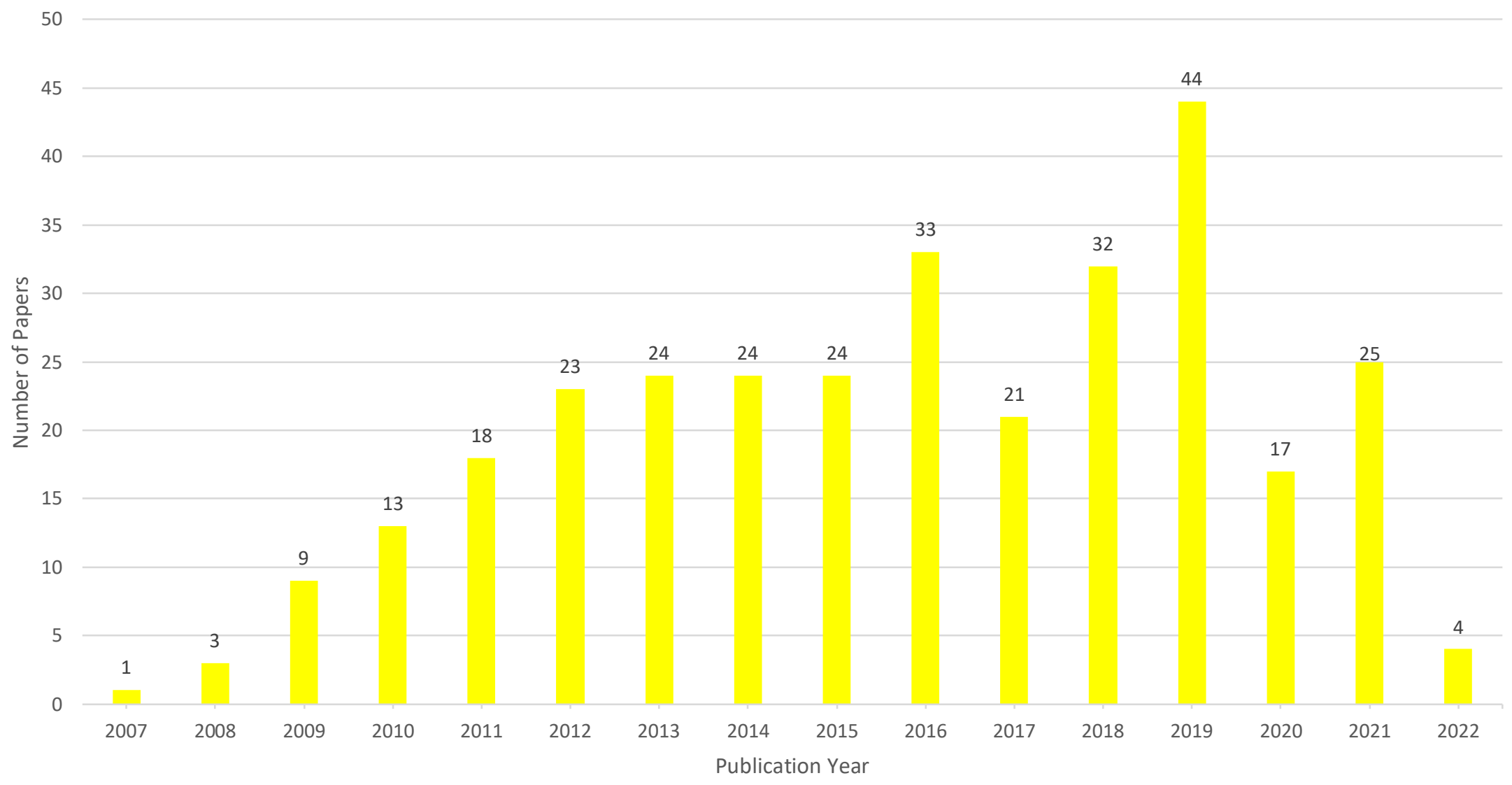

*Figure 11. Number of Papers Per Year*

These papers have been published in 193 different journals and conferences. Some of these publication venues, specifically those with a higher volume of published papers, are ranked and visually represented in Fig. 12.

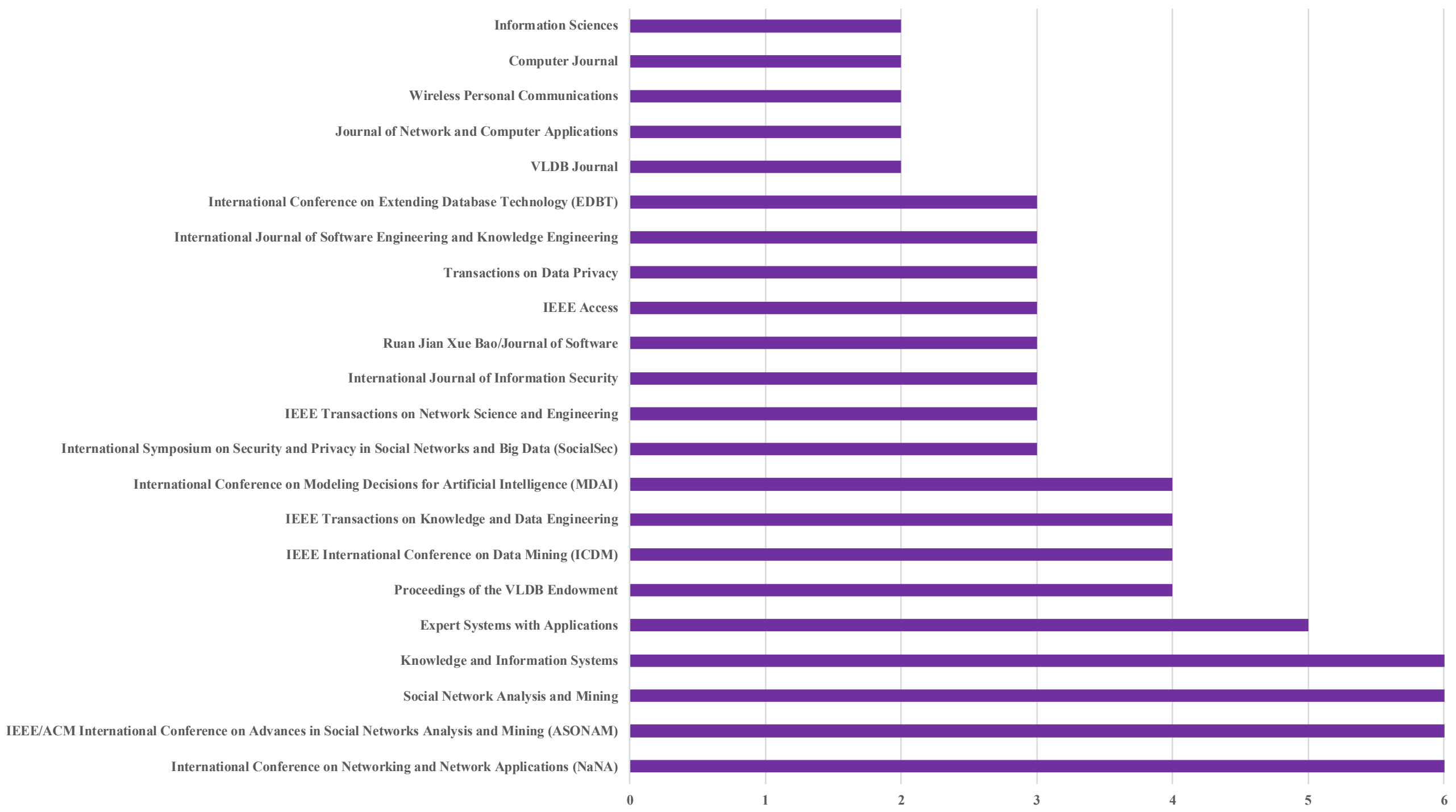

*Figure 12. Top-ranked Publication Venues for the Social Network Anonymization Field*

Furthermore, Fig. 13 provides a graphical representation of the number of articles published by each country, limiting the scope to those countries that have produced at least five articles. The top contributors are China, the United States of America (USA), India, Spain, and Iran. China stands as the leading publisher, contributing 33% of the total articles sourced for the current research study. Following China, USA has produced 22%, India has published 10%, Spain has contributed 7%, and Iran accounts for 5% of the total published articles.

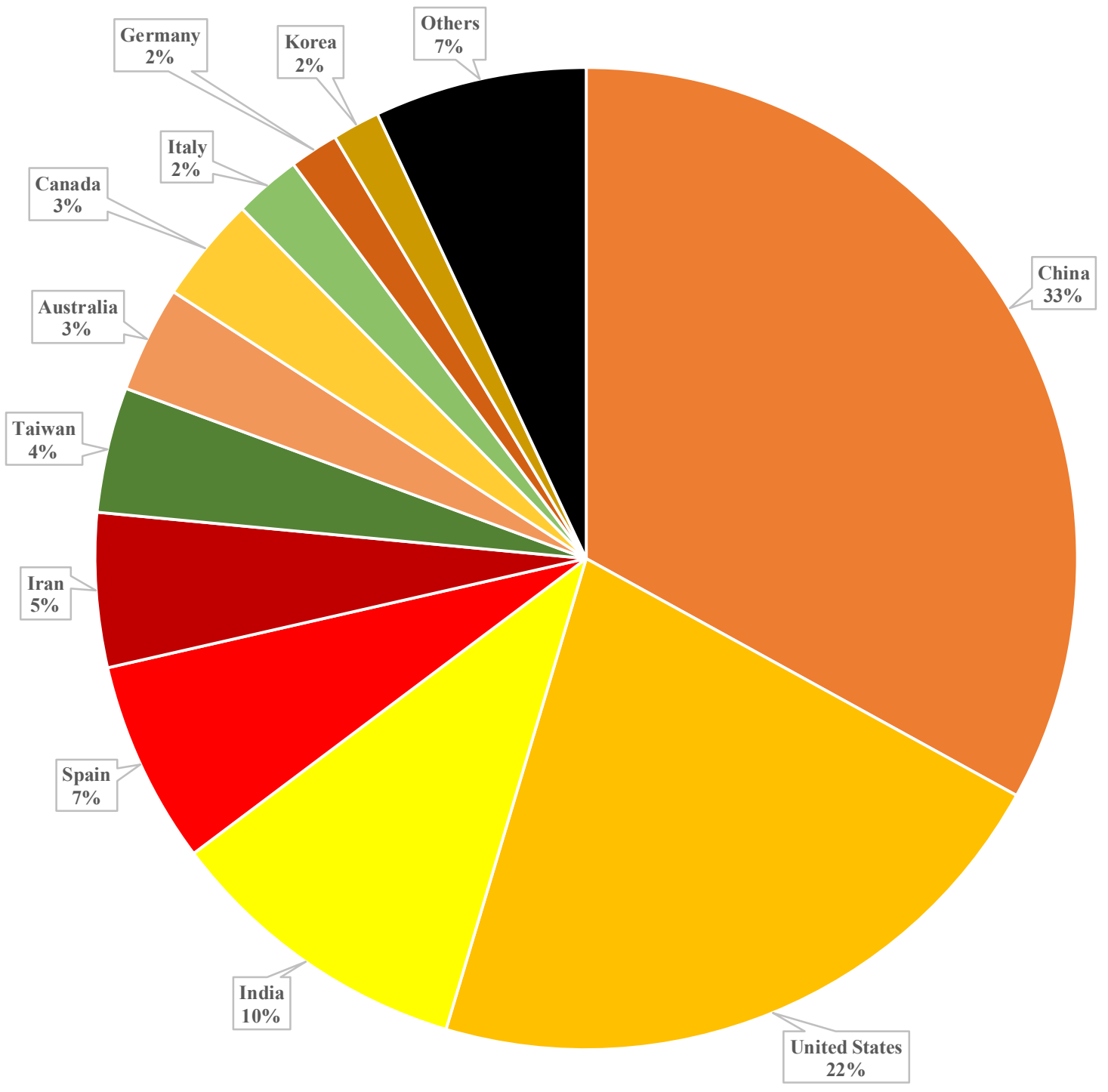

*Figure 13. The Countries with Most Publication in Social Network Anonymization Field*

In the current study, the research period was segmented into four consecutive subperiods: 2007–2010, 2011–2014, 2015–2018, and 2019–2022. Also, the number of publications in each subperiod is demonstrated in Fig. 14. An R-squared value of 0.57 indicates that the regression model explains 57% of the variation in the number of publications. This means that there are other factors that affect the number of publications besides the time period alone. The remaining 43% of the variation may be due to other factors, such as changes in the research landscape, funding availability, or other external factors.

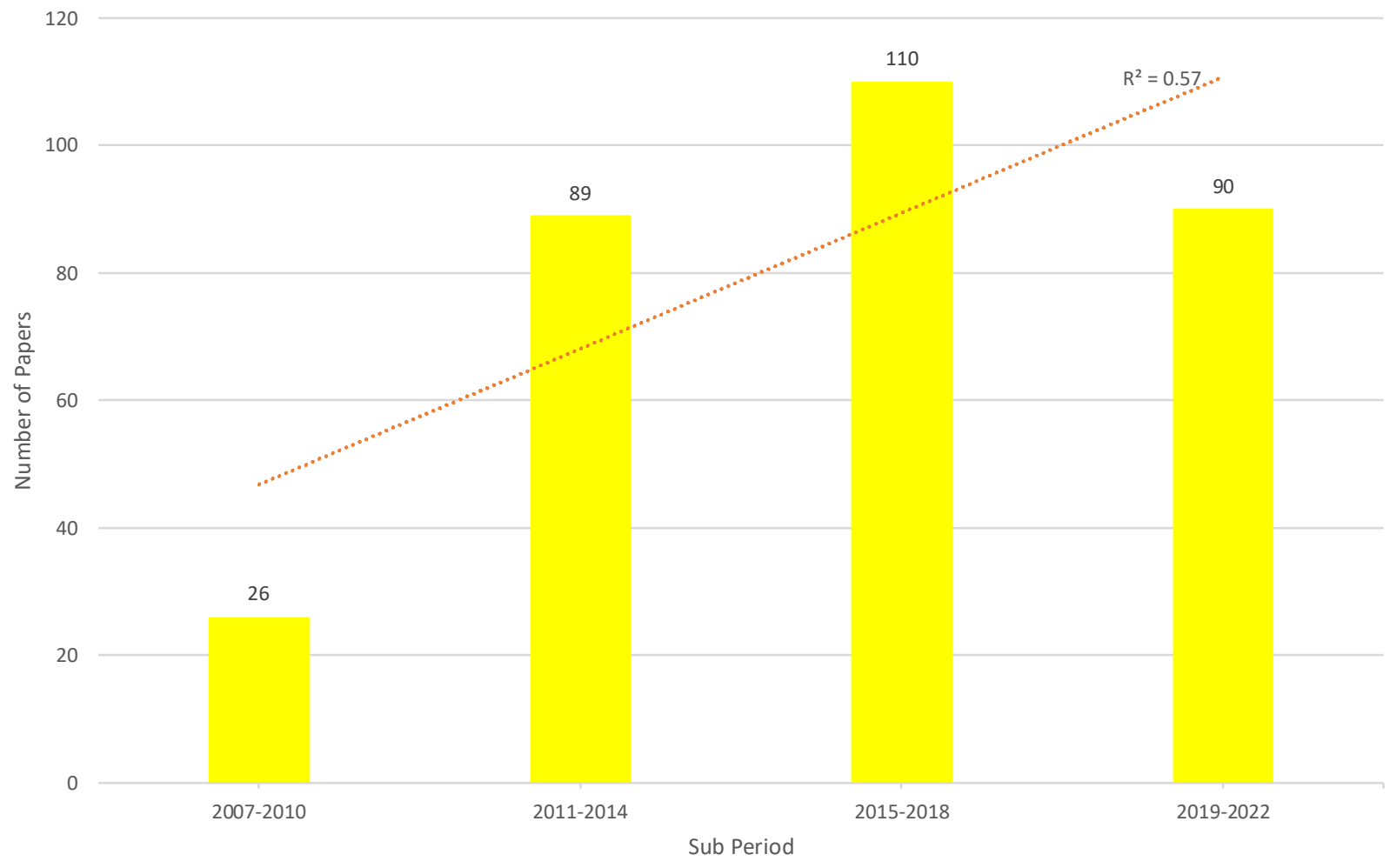

*Figure 14. Number of Papers in Each Subperiod and the Fitted Regression Line*

With respect to the data pre-processing stage, it is worth mentioning that the following processes were performed to improve the quality and clarity of the keywords in the dataset, making it more suitable for analysis:

- Removing duplicates: Duplicate keywords were removed to reduce the noise in the dataset.
- Stemming: Words were reduced to their root form to group different variations of the same word.
- Lemmatization: Words were converted to their base form to capture their core meaning.
- Removing special characters: Special characters were removed to improve the readability of the dataset.
- Missing values: For papers lacking specific authors' keywords, indexed keywords supplied by the Elsevier Scopus API (Elsevier 2022) were used as substitutes.
- Misspelled words correction: Some papers contained misspelled keywords, like "anonimisation". To address this problem, we developed Python scripts to identify keywords with the most similar spelling and replace the incorrect ones. Thus, the incorrectly spelled keyword "anonymisation" was replaced with the correct term, "anonymization."

Following the application of the above-mentioned pre-processing steps, the dataset's overall attributes are as follows: the total count of keywords in the dataset amounts was determined to be 1547, out of which only 361 keywords are unique. Therefore, each paper contains an average of 4.91 keywords.

Additionally, the most frequently used keyword in the dataset appeared 244 times, while the least frequent ones appeared only once. These statistics provide a general idea about the distribution of keywords in the dataset and help to understand the characteristics of the research in the field of social network anonymization. The information will be used to identify the dominant themes and topics in the field and gain insights into the areas that have received more attention from researchers.

## 3-2- Authors’ Keywords Network Visualization

The frequency information mentioned earlier was utilized in constructing a keyword-keyword network using the VOSviewer software. This network represents keywords as nodes and establishes edges based on the frequency of co-occurrence between pairs of keywords. By examining this network, it becomes possible to identify relationships between keywords and gain a comprehensive understanding of their connections. Moreover, the network incorporates clusters of keywords that depend on the strength of their relationship (Ali et al., 2023; Van Eck & Waltman, 2010). Additionally, the frequency information aids in ranking the keywords according to their popularity, offering valuable insights into the most relevant topics within the field of social network anonymization.

Following the construction of the keyword-keyword network, an analysis was conducted to unveil concealed patterns and relationships. This analysis involved the presentation of network diagrams, density visualizations in the form of heatmaps, and overlay visualization. By utilizing the VOSviewer tool, it was possible to identify the most important and influential keywords, examine the relationships between keywords, determine the time period during which each keyword was utilized, and comprehend the overall structure of the keyword-keyword network.

In the following figures, the co-occurrence mapping of the keywords for the four defined subperiods is depicted.

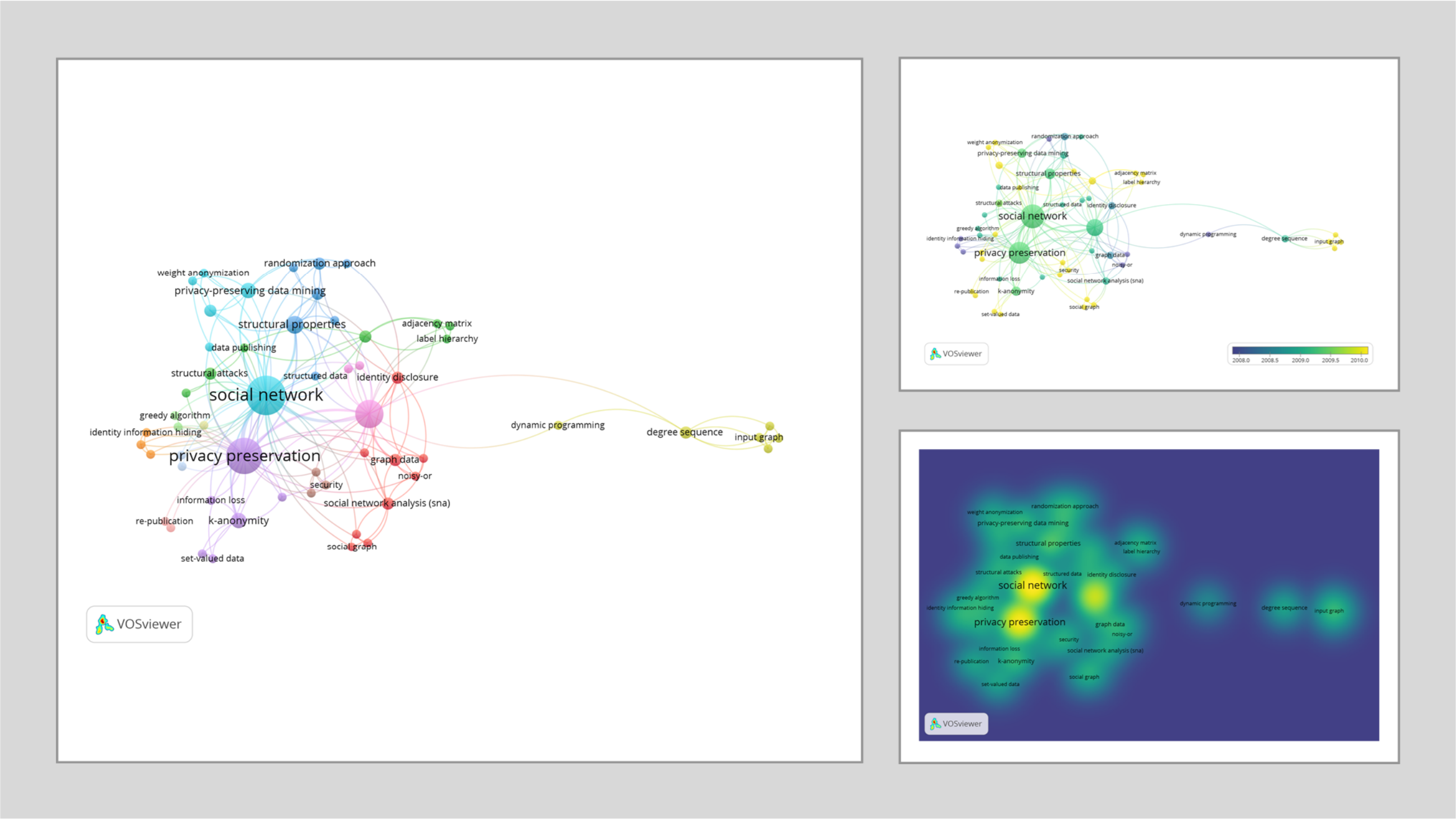

*Figure 15. Co-occurrence Mapping of 2007-2010 Authors’ Keywords*

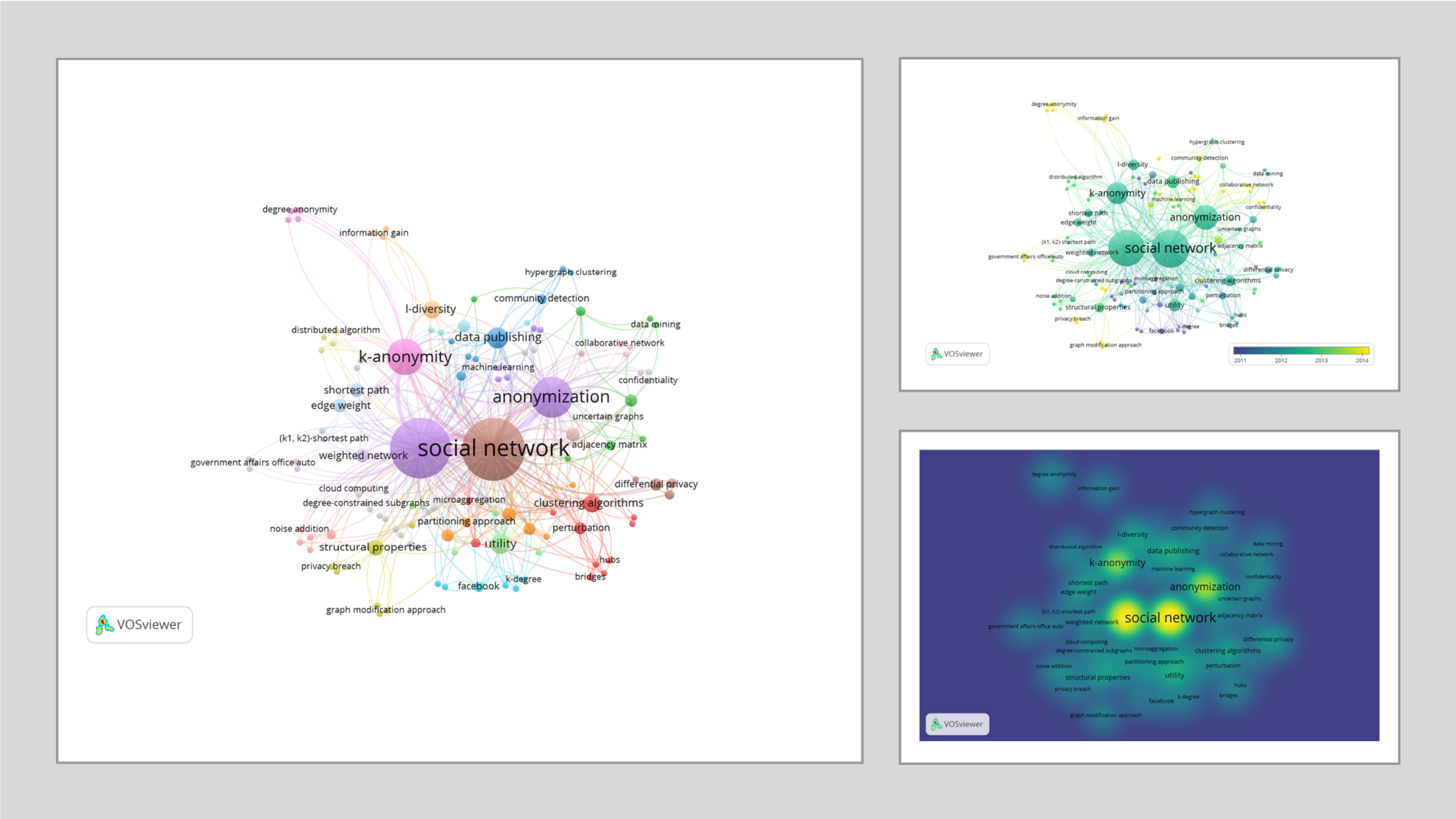

*Figure 16. Co-occurrence Mapping of 2011-2014 Authors' Keywords*

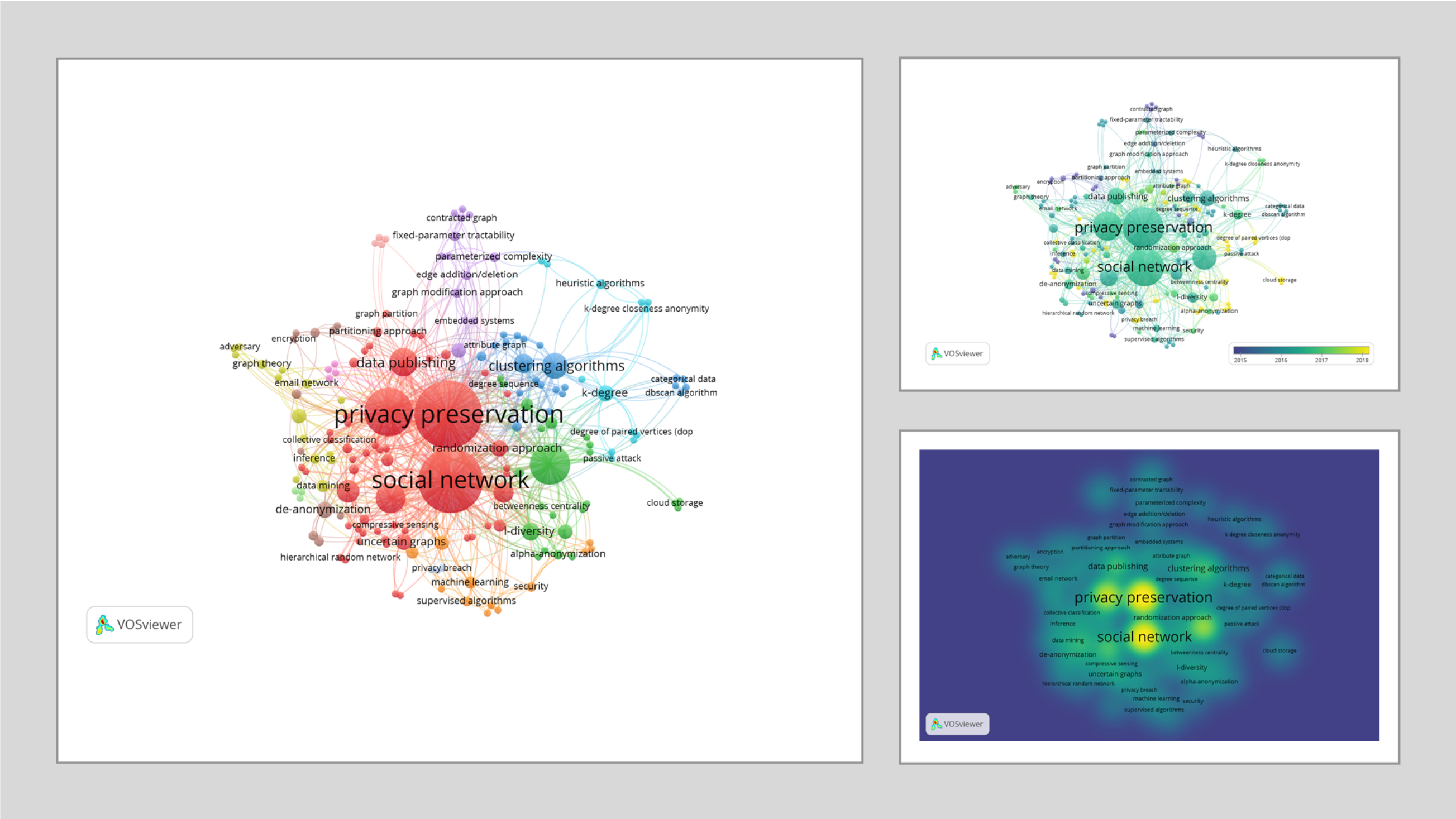

*Figure 17. Co-occurrence Mapping of 2015-2018 Authors' Keywords*

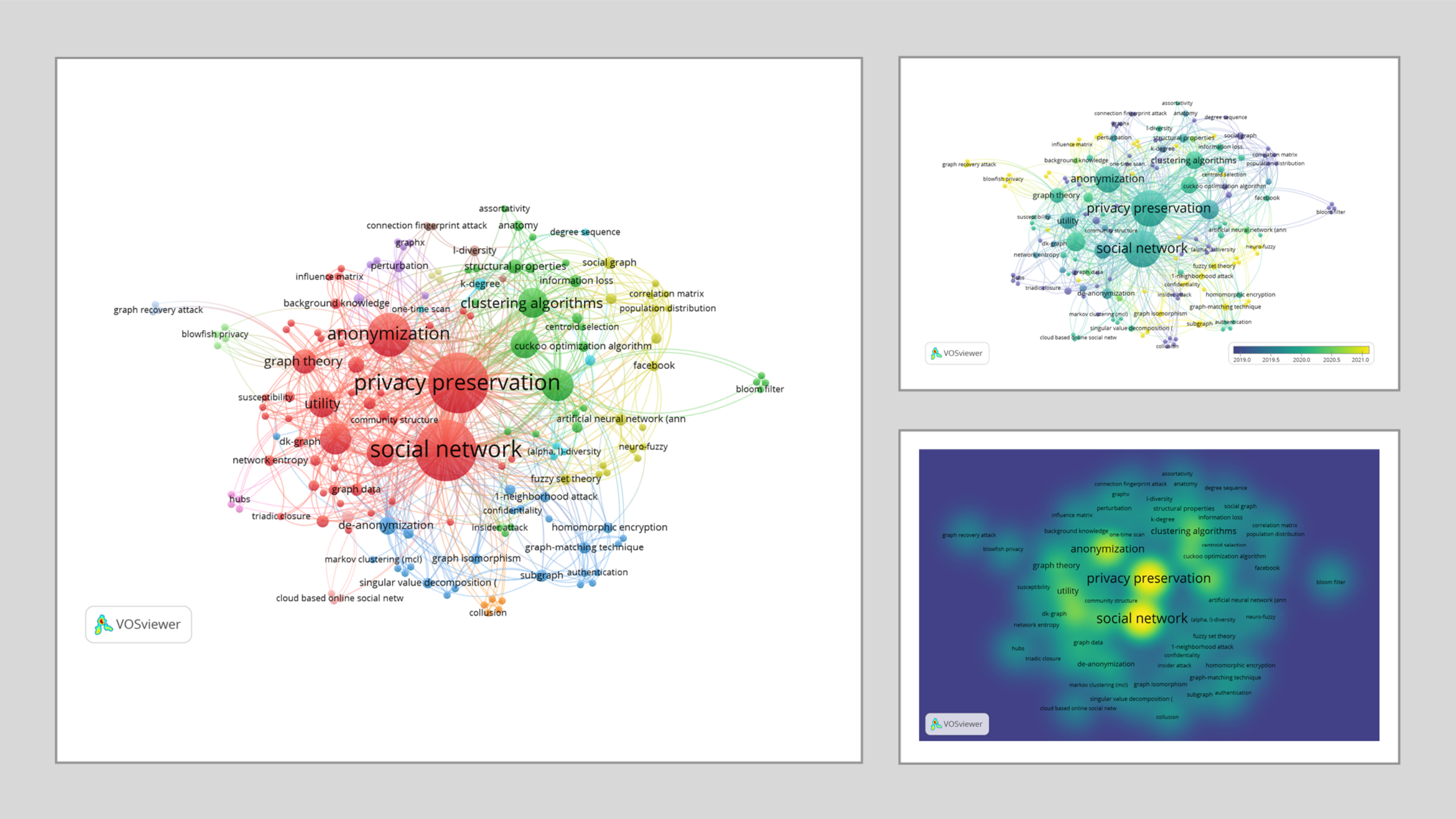

*Figure 18. Co-occurrence Mapping of 2019-2022 Authors' Keywords*

In the network in Fig. 15 for the 2007-2010 subperiod, "Social Network", "Privacy Preservation", and "Anonymization" are influential keywords that are positioned near the center with larger nodes. This implies that these keywords have persistently been significant keywords in the field of social network anonymization. The "Structural Properties," "K-anonymity," "Generalization Approach," and "Social Network Analysis (SNA)" keywords are closely related to the mentioned influential keywords, demonstrating the co-occurrence of these keywords in the social network anonymization field.

Furthermore, in the 2011-2014 network illustrated in Fig. 16, "Social Network," "Privacy Preservation," "Anonymization," and "K-anonymity" emerge as prominent keywords. Again, these keywords are situated close to the center with larger nodes, indicating their significant influence. It is worth mentioning that "Data Publishing," "Clustering Algorithms," "Structural Properties," "Utility," and "L-diversity" keywords have a strong association with the aforementioned influential keywords. This illustrates their co-occurrence and interrelation in the field of social network anonymization.

In the 2015-2018 network shown in Fig. 17, once again, "Social Network," "Privacy Preservation," "Anonymization," and "K-anonymity" stand out as influential keywords. Moreover, a number of keywords, such as "Utility," "Data Publishing," "Clustering Algorithms," "Generalization Approach," "Differential Privacy," "Graph Modification Approach," "Background Knowledge," "Machine Learning," "Uncertain Graphs," and "L-diversity" exhibit a strong connection with the influential nodes. This illustrates that the co-occurrence of these keywords was of considerable interest to researchers during the 2014-2018 subperiod.

In the network of the last subperiod 2019-2022 in Fig. 18, keywords "Social Network", "Privacy Preservation," "Anonymization," "K-anonymity," "Clustering Algorithms," "Generalization Approach," "Differential Privacy," "Utility," "Data Publishing," and "Graph Theory" are influential nodes. Additionally, keywords "Graph Modification Approach," "Structural Properties," "Information Loss," "Genetic Algorithm (GA)," "Randomization Approach," "Perturbation Techniques,"

“Background Knowledge,” “K-degree,” “Graph Matching Technique,” “Uncertain Graphs,” “Optimization,” “Fuzzy Set Theory,” and “Artificial Neural Network (ANN)” demonstrate a strong link with the influential nodes mentioned earlier, indicating their interrelation in the field of social network anonymization.

The co-occurrence mapping of all the authors’ keywords used in the 2007-2022 period is provided in Fig. 19.

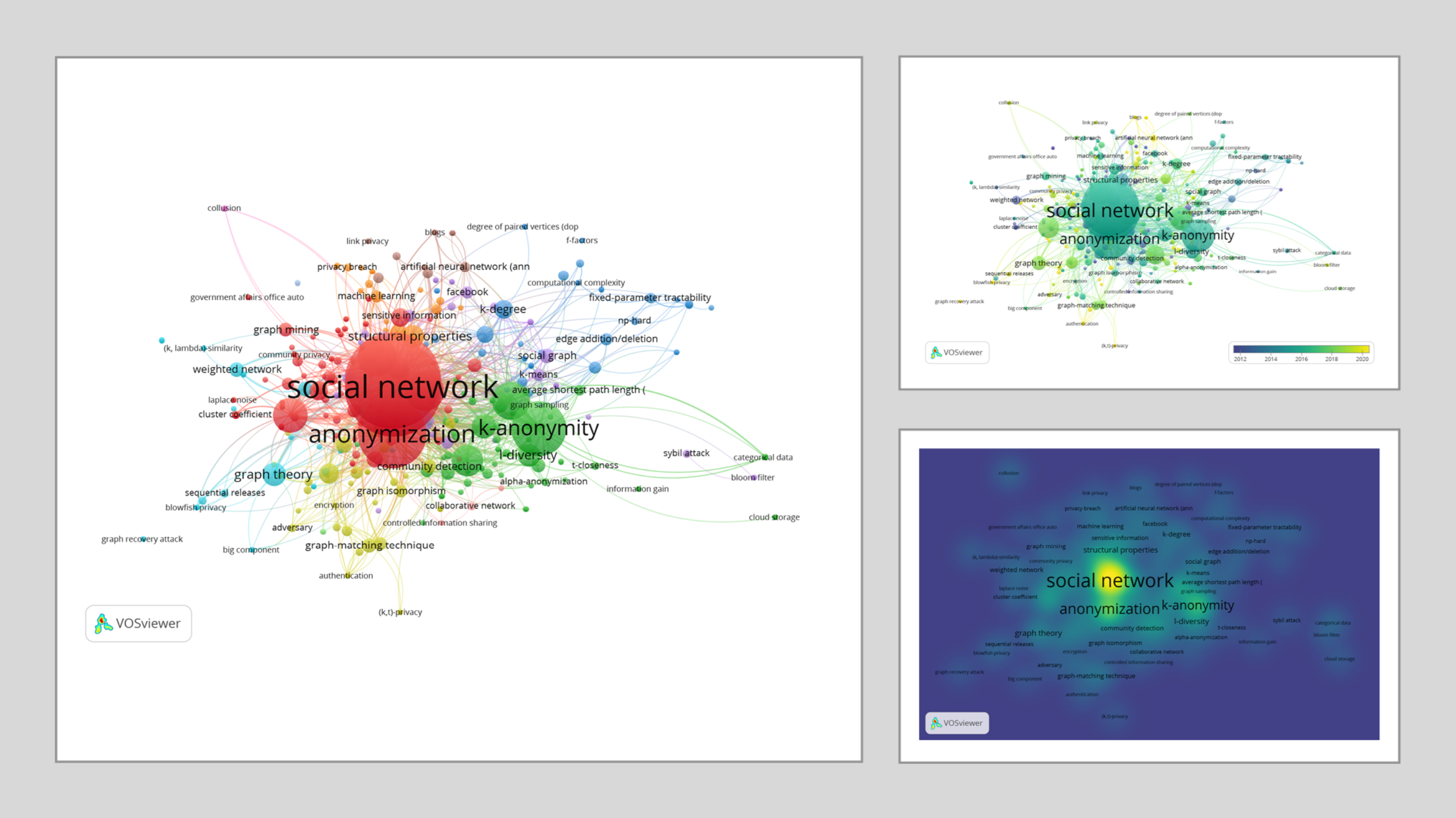

*Figure 19. Co-occurrence Mapping of 2007-2022 Authors’ Keywords*

As shown in this figure, keywords “Social Network”, “Privacy Preservation”, “Anonymization”, “K-anonymity”, “Data Publishing”, “Clustering Algorithms”, “Generalization Approach”, “Differential Privacy”, “Graph Modification Approach”, “Uncertain Graphs”, “Information Loss”, “Structural Properties”, “Randomization Approach”, “L-diversity”, “Graph Theory”, “Background Knowledge”, “Social Network Analysis (SNA)”, “K-degree”, “Machine Learning”, and “Perturbation Techniques” are among the most influential keywords used by researchers in the social network anonymization field. Moreover, the overlay visualization provides key insights into the period of usage for particular keywords by researchers in the specified field. For instance, the most recent keywords utilized by researchers include “Graph-matching Technique”, “Artificial Neural Network (ANN)”, “Optimization”, “Particle Swarm Optimization Algorithm (PSO)”, “Cuckoo Optimization Algorithm (COA)”, “Firefly Algorithm (FA)”, “Hierarchical Clustering”, “Blowfish Privacy”, and “Assortativity”.

## 3-3- Statistical Measures & Network Analysis

In the following section, statistical analyses and network analysis, including the frequency analysis of the keywords, their relative growth, the T-score values, and weighted degree/node strength, are provided on the research dataset during the successive subperiods.

### 3-3-1- Most Significant Keywords

Here, we provide an explanation of the statistical and network analysis techniques utilized for analyzing the authors' keywords. These methods encompass growth analysis, T-score, and weighted degree/node strength.

The growth analysis technique is employed to determine the expansion of various topics within the research area (Amiri et al., 2021). By examining the increases or decreases in the occurrences of specific keywords, we can discern the growth patterns of related topics across consecutive subperiods. The growth measure is computed based on Eq. (1), where the variable $f_i$ represents the frequency of a specific keyword during the $i$-th subperiod:

$$Growth = \sum_{i=1}^{n-1} \frac{(f_{i+1} - f_i)}{f_i} \qquad (1)$$

In addition, the T-score analysis technique assists in enhancing the evaluation of the relative importance of keywords. This standardized score calculates the probability of a specific keyword occurring. By utilizing the mean and standard deviation values, the T-score analysis converts the frequencies into a standardized metric using Eq. (2). In this equation, $X$ represents the value to be converted, $\mu$ denotes the mean, $\sigma$ represents the standard deviation, and $n$ is the number of samples (with n-1 degrees of freedom). The T-score value mathematically signifies the probability of occurrence, which can be derived from the t-tables (Kim, 2015). A keyword with a high T-score indicates a lower probability of its occurrence, thereby suggesting its lower significance. Conversely, a lower T-score implies a higher probability of occurrence, indicating greater significance. In this study, a significance level of α=0.05 was considered, where a T-score with a p-value less than α (i.e., p-value < α) is deemed statistically significant.

$$T_score = \frac{X - \mu}{\left(\sigma / \sqrt{n}\right)} \qquad (2)$$

Moreover, the weighted node strength is a network analysis measure to quantify the importance of nodes in a network. Node strength is calculated as the sum of the weights of all the edges connected to a node. In a weighted network, the strength of a node is determined by the sum of the weights of the edges connected to that node. The formula for calculating the weighted node strength of a node $i$ in a network is:

$$WNS_i = \sum_{i \neq j} w_{ij} \qquad (3)$$

where $WNS_i$ is the weighted node strength of node $i$; and $w_{ij}$ is the weight of the edge connecting node $i$ to node $j$. The summation is over all nodes $j$ that are directly connected to node $i$. A keyword with a high $WNS$ means that it is more important compared to other keywords.

As previously mentioned, in this work, the study period was divided into four consecutive subperiods. The top 30 frequent authors' keywords in the whole studying period extracted from the dataset are provided in Table 1. Also, the relative growth, T-score, and weighted node strength values of each frequent keyword are reported in Table 2.

We report these metrics not only to assess the impact and importance of the keywords thus far but also to predict which keywords are more likely to become future trends. By analyzing these metrics, we aimed to identify keywords that exhibit promising patterns and have the potential to gain prominence in the future. To do so, we computed the designated metrics for each commonly used keyword and proposed Eq. 4 to assess the significance of each keyword, sorting the frequently appearing keywords in descending order of importance. It is worth mentioning that the most significant frequent keyword is the one with the highest relative growth, highest weighted node strength, and lowest T-score values. Therefore, if we want to include all these metrics in a single formula to determine the impact or importance of each frequent keyword, we should consider the absolute value of α and p-value (i.e., $|\,\alpha - (p_value)|$) instead of the T-score value. Also, the normalized value (NV) of each metric is calculated and used in Eq. (4):

$$Importance_i = (W_1 * NV(Growth_i)) + (W_2 * NV(|\alpha - p_value|_i)) + (W_3 * NV(WNS_i)) \quad (4)$$

In the above formula, the user-defined importance weights, denoted as $W_1, W_2, and\ W_3$, should be assigned to each calculated metric. However, for this study, we treated the metrics as equally important and, therefore, assigned an importance weight of 1/3 to each metric.

Table 1 presents the distribution of each frequent keyword in the 2007-2022 period, and Fig. 20 illustrates the frequency distribution of the commonly used keywords over the four determined subperiods. This figure provides a more insightful representation of the growth of the unique keywords presented in Table 1. Specifically, a large portion of the bar plots for each keyword pertains to the most recent subperiods, indicating that they are currently experiencing increased usage and positive growth.

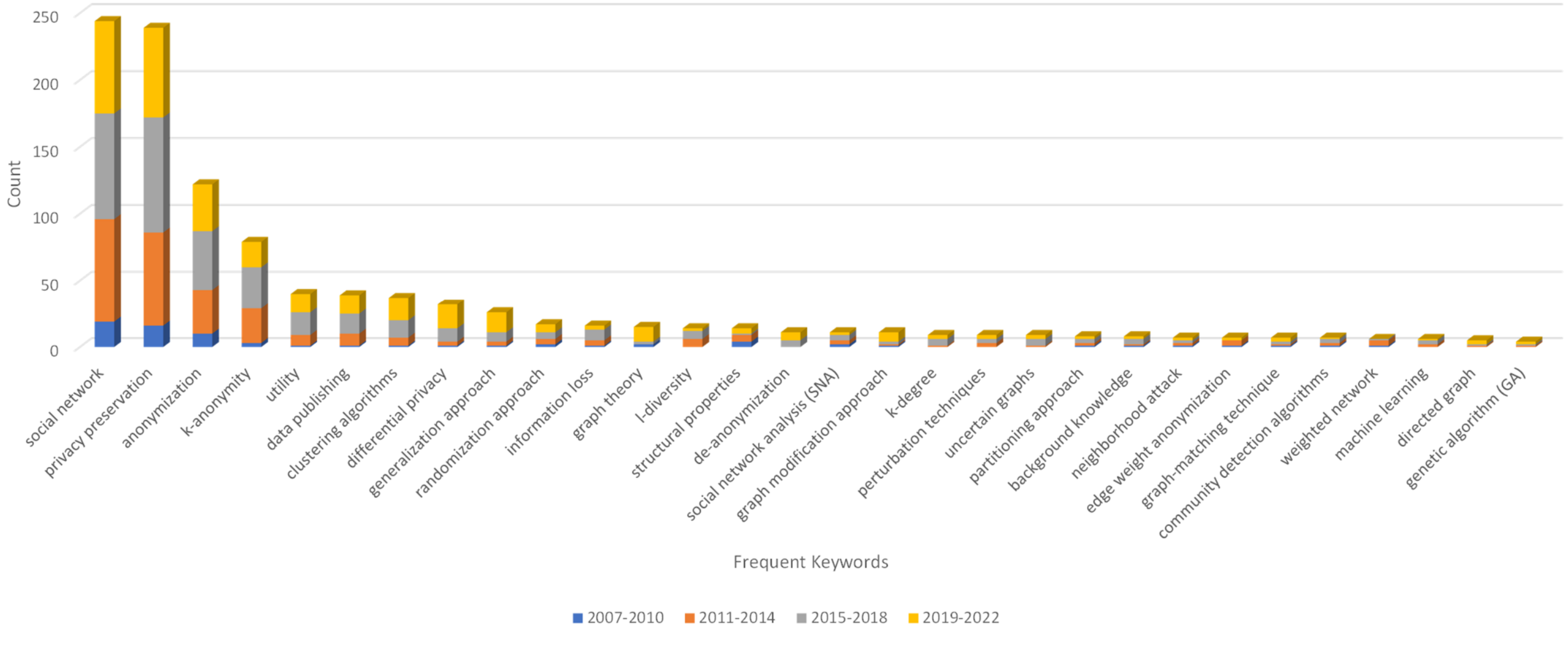

*Figure 20. Analysis of Frequency Distribution of Frequent Keywords Across Four Subperiods*

Furthermore, Table 2 presents the growth, T-score, weighted node strength, and importance factor associated with the frequent keywords. Notably, the growth, T-score, and weighted node strength values for each keyword hold more significance compared to their frequency alone. This is because our goal is to assess the impact and relevance of keywords and topics up to the present time as well as to predict which ones are more likely to become future trends. It is noteworthy that all the frequent keywords listed in Table 1 exhibit statistically significant and positive growth, while none indicate a statistically significant decline according to the data presented in Table 2.

The keywords with the highest relative significant growth rate, T-score, and weighted node strength in Table 2 are “Social Network”, “Privacy Preservation”, “Anonymization”, “K-anonymity”, “Data Publishing”, “Utility”, “Clustering Algorithms”, “Differential Privacy”, and “Generalization Approach”. Therefore, these keywords are considered pertinent to the social network anonymization field. Furthermore, given their high relative importance factor, it is more likely that we will encounter them in future works within this domain.

Additionally, some keywords did not exhibit noticeable growth or high weighted node strength compared to the ones previously mentioned. However, their corresponding T-score significance values suggest that they still represent important topics. These keywords, as shown in Table 2, include “Structural Properties”, “L-diversity”, “Information Loss”, “Graph Modification Approach”, “Randomization Approach”, “Social Network Analysis (SNA)”, “Perturbation Techniques”, “Graph-Matching Techniques”, “Neighborhood Attack”, “Community Detection Algorithms”, “Uncertain Graphs”, “Machine Learning”, and “Genetic Algorithm (GA)”.

*Table 1. The Number of Frequent Keywords in Each Year*

| Frequent Keywords | social network | privacy preservation | anonymization | k-anonymity | data publishing | utility | clustering algorithms | differential privacy | generalization approach | structural properties | l-diversity | information loss | graph modification | graph theory | randomization approach | social network analysis (SNA) | de-anonymization | edge weight anonymization | k-degree | perturbation techniques | graph-matching | neighborhood attack | community detection | background knowledge | partitioning approach | uncertain graphs | weighted network | machine learning | genetic algorithm | directed graph |
|---|---|---|---|---|---|---|---|---|---|---|---|---|---|---|---|---|---|---|---|---|---|---|---|---|---|---|---|---|---|---|
| 2007 | 1 | 1 | 0 | 0 | 0 | 0 | 0 | 0 | 0 | 0 | 0 | 0 | 0 | 0 | 0 | 0 | 0 | 0 | 0 | 0 | 0 | 0 | 0 | 0 | 0 | 0 | 0 | 0 | 0 | 0 |
| 2008 | 1 | 1 | 2 | 0 | 0 | 1 | 0 | 0 | 0 | 1 | 0 | 0 | 0 | 0 | 1 | 1 | 0 | 0 | 0 | 0 | 0 | 0 | 1 | 1 | 0 | 0 | 0 | 0 | 0 | 0 |
| 2009 | 8 | 6 | 4 | 2 | 0 | 0 | 1 | 1 | 0 | 1 | 0 | 1 | 0 | 0 | 1 | 0 | 0 | 0 | 0 | 0 | 0 | 0 | 0 | 0 | 1 | 0 | 0 | 0 | 0 | 0 |
| 2010 | 9 | 8 | 4 | 1 | 1 | 0 | 0 | 0 | 1 | 2 | 0 | 0 | 1 | 2 | 0 | 1 | 0 | 1 | 0 | 0 | 1 | 1 | 0 | 0 | 0 | 0 | 1 | 0 | 0 | 0 |
| 2011 | 18 | 16 | 7 | 7 | 2 | 2 | 1 | 1 | 1 | 0 | 2 | 0 | 0 | 0 | 0 | 0 | 0 | 1 | 1 | 1 | 0 | 1 | 0 | 0 | 0 | 0 | 1 | 0 | 0 | 0 |
| 2012 | 21 | 18 | 8 | 7 | 1 | 1 | 2 | 1 | 1 | 3 | 1 | 3 | 0 | 0 | 1 | 2 | 0 | 1 | 0 | 2 | 1 | 0 | 0 | 0 | 2 | 1 | 1 | 0 | 1 | 0 |
| 2013 | 21 | 18 | 10 | 5 | 3 | 3 | 3 | 1 | 1 | 0 | 1 | 1 | 0 | 0 | 0 | 1 | 0 | 1 | 0 | 0 | 0 | 0 | 1 | 0 | 0 | 0 | 2 | 1 | 0 | 0 |
| 2014 | 17 | 18 | 8 | 7 | 3 | 2 | 0 | 0 | 0 | 2 | 2 | 0 | 1 | 0 | 3 | 0 | 0 | 1 | 0 | 0 | 0 | 1 | 1 | 1 | 0 | 0 | 0 | 1 | 0 | 1 |
| 2015 | 17 | 21 | 10 | 4 | 3 | 6 | 5 | 1 | 3 | 0 | 1 | 3 | 0 | 0 | 0 | 1 | 2 | 0 | 0 | 0 | 1 | 0 | 1 | 2 | 1 | 2 | 0 | 0 | 0 | 1 |
| 2016 | 23 | 26 | 12 | 16 | 3 | 3 | 3 | 2 | 1 | 1 | 3 | 2 | 1 | 1 | 0 | 2 | 0 | 0 | 2 | 1 | 0 | 0 | 1 | 0 | 2 | 0 | 1 | 2 | 1 | 0 |
| 2017 | 16 | 16 | 7 | 3 | 6 | 4 | 2 | 4 | 1 | 0 | 1 | 1 | 1 | 1 | 4 | 0 | 1 | 0 | 2 | 1 | 0 | 1 | 0 | 2 | 0 | 2 | 0 | 1 | 0 | 0 |
| 2018 | 23 | 23 | 15 | 8 | 3 | 4 | 3 | 3 | 2 | 0 | 1 | 2 | 0 | 0 | 1 | 1 | 2 | 0 | 1 | 1 | 1 | 1 | 1 | 0 | 0 | 1 | 0 | 0 | 0 | 0 |
| 2019 | 35 | 32 | 16 | 12 | 8 | 7 | 7 | 6 | 6 | 2 | 1 | 1 | 1 | 5 | 2 | 1 | 3 | 0 | 2 | 2 | 0 | 0 | 0 | 1 | 1 | 3 | 0 | 0 | 0 | 2 |
| 2020 | 13 | 14 | 9 | 1 | 4 | 4 | 3 | 6 | 3 | 1 | 0 | 2 | 2 | 2 | 2 | 1 | 2 | 0 | 1 | 0 | 1 | 0 | 1 | 0 | 0 | 0 | 0 | 0 | 1 | 1 |
| 2021 | 19 | 18 | 8 | 6 | 1 | 3 | 6 | 5 | 5 | 1 | 1 | 0 | 2 | 3 | 2 | 0 | 1 | 2 | 0 | 1 | 1 | 1 | 0 | 0 | 1 | 0 | 0 | 1 | 1 | 1 |
| 2022 | 2 | 3 | 2 | 0 | 1 | 0 | 1 | 1 | 1 | 0 | 0 | 0 | 2 | 1 | 0 | 0 | 0 | 0 | 0 | 0 | 1 | 1 | 0 | 1 | 0 | 0 | 0 | 0 | 0 | 0 |

*Table 2. The Frequent Keywords' Related Growth, T-score, Weighted Node Strength, and Importance Factor*

| Frequent Keywords | 2007-2010 | 2011-2014 | 2015-2018 | 2019-2022 | Total | Growth | T-score ($\|\alpha - p\ value\|$) | Weighted Node Strength | Importance Factor |
|---|---|---|---|---|---|---|---|---|---|
| **social network** | 19 | 77 | 79 | 69 | 244 | 1 | 0.99992 | 1 | 0.99997 |
| **privacy preservation** | 16 | 70 | 86 | 67 | 239 | 0.88295 | 0.99992 | 0.97007 | 0.95098 |
| **anonymization** | 10 | 33 | 44 | 35 | 122 | 0.56618 | 0.99994 | 0.51186 | 0.69266 |
| **k-anonymity** | 3 | 26 | 31 | 19 | 79 | 0.47703 | 0.99398 | 0.33436 | 0.60179 |
| **data publishing** | 1 | 9 | 15 | 14 | 39 | 0.43732 | 0.99474 | 0.17750 | 0.53652 |
| **utility** | 1 | 8 | 17 | 14 | 40 | 0.40764 | 0.99672 | 0.18988 | 0.53142 |
| **clustering algorithms** | 1 | 6 | 13 | 17 | 37 | 0.34049 | 0.99285 | 0.17750 | 0.50361 |
| **differential privacy** | 1 | 3 | 10 | 18 | 32 | 0.2794 | 0.98098 | 0.12899 | 0.46312 |
| **generalization approach** | 1 | 3 | 7 | 15 | 26 | 0.24946 | 0.97596 | 0.14035 | 0.45526 |
| **structural properties** | 4 | 5 | 1 | 4 | 14 | 0.29382 | 0.97657 | 0.59855 | 0.44342 |
| **l-diversity** | 0 | 6 | 6 | 2 | 14 | 0.24295 | 0.98728 | 0.0650 | 0.43175 |
| **information loss** | 1 | 4 | 8 | 3 | 16 | 0.1993 | 0.97637 | 0.07533 | 0.417 |
| **graph modification** | 1 | 1 | 2 | 7 | 11 | 0.20499 | 0.9655 | 0.05056 | 0.40702 |
| **graph theory** | 2 | 0 | 2 | 11 | 15 | 0.34165 | 0.80343 | 0.06914 | 0.40474 |
| **randomization approach** | 2 | 4 | 5 | 6 | 17 | 0.15716 | 0.96331 | 0.04850 | 0.38966 |
| **social network analysis (SNA)** | 2 | 3 | 4 | 2 | 11 | 0.10629 | 0.98594 | 0.05572 | 0.38265 |
| **de-anonymization** | 0 | 0 | 5 | 6 | 11 | 0.23688 | 0.83892 | 0.05159 | 0.3758 |
| **edge weight anonymization** | 1 | 4 | 0 | 2 | 7 | 0.22777 | 0.86029 | 0.02373 | 0.3706 |
| **k-degree** | 0 | 1 | 5 | 3 | 9 | 0.20955 | 0.85537 | 0.04024 | 0.36839 |
| **perturbation techniques** | 0 | 3 | 3 | 3 | 9 | 0.13666 | 0.92578 | 0.04231 | 0.36825 |
| **graph-matching** | 1 | 1 | 2 | 3 | 7 | 0.11388 | 0.96167 | 0.02683 | 0.36746 |
| **neighborhood attack** | 1 | 2 | 2 | 2 | 7 | 0.09111 | 0.96167 | 0.02579 | 0.35953 |
| **community detection** | 1 | 2 | 3 | 1 | 7 | 0.08352 | 0.96167 | 0.02270 | 0.35596 |
| **background knowledge** | 1 | 1 | 4 | 2 | 8 | 0.15944 | 0.84769 | 0.03921 | 0.34878 |
| **partitioning approach** | 1 | 2 | 3 | 2 | 8 | 0.0987 | 0.84769 | 0.01857 | 0.32166 |
| **uncertain graphs** | 0 | 1 | 5 | 3 | 9 | 0.20955 | 0.66096 | 0.02889 | 0.2998 |
| **weighted network** | 1 | 4 | 1 | 0 | 6 | 0.1025 | 0.71472 | 0.02373 | 0.28032 |
| **machine learning** | 0 | 2 | 3 | 1 | 6 | 0.08352 | 0.71472 | 0.01960 | 0.27261 |
| **genetic algorithm** | 0 | 1 | 1 | 2 | 4 | 0.09111 | 0.59031 | 0.02063 | 0.23402 |
| **directed graph** | 0 | 1 | 1 | 3 | 5 | 0.13666 | 0.44513 | 0.01547 | 0.19909 |

# 3-4- Co-word Analysis

In this section, we describe the co-word analysis (Callon et al., 1986), which builds upon the earlier co-citation analysis method (Small, 1973; Small & Griffith, 1974). Unlike citation-based methodologies, the co-word analysis allows us to explore the relationships and connections between keywords, as well as identify patterns amongst them within the research context. In other words, the co-word analysis explores the frequency with which pairs of keywords appear simultaneously in the literature, enabling the identification of relationships between clusters of keywords or themes (Courtial, 1994; Law et al., 1988) and tracking their developmental trends (Coulter et al., 1998; Lee & Jeong, 2008). The analysis of keyword co-occurrence patterns can reveal the intellectual structure of a particular research field and the connections among its various themes. This method allows for a deeper understanding of the relationships between different topics within the field being studied (Ronda-Pupo & Guerras-Martin, 2012). Over the years, the co-word analysis method has been enhanced and refined by incorporating innovative techniques, such as co-word clustering (Callon et al., 1991), social network analysis (Ding et al., 2001), and strategic diagrams (Stegmann & Grohmann, 2003). Of particular interest is the strategic diagram, which utilizes measures of density and centrality to map the dynamics of themes and topics within a research field. The strategic diagram serves as a valuable tool in visualizing and analyzing the interplay and evolution of various themes, providing valuable insights into the intellectual landscape of the research field.

The SciMAT software, developed by (Cobo et al., 2012), was used to perform co-word analysis using an algorithm to identify themes across the four different subperiods. The software creates networks of keywords, and edges represent the co-occurrence of keyword pairs in the analyzed documents. The edge's weight indicates the importance of the relationship in the entire set of documents related to the research field being studied. The results of the analysis are then used to create strategic diagrams that show how thematic areas evolved throughout each subperiod. The SciMAT software follows three main stages:

- Extract clusters of keywords for each subperiod.
- Investigate the evolution of the extracted clusters over time, with the aim of identifying the primary themes of the research field, their origins, and the connections among them.
- Analyze the performance of the identified themes within each subperiod using quantitative measures such as the number of documents, average citations, and h-index.

Each stage is further explained in the following sub-sections.

## 3-4-1- Process of Detecting Themes

In this section, we discuss the SciMAT software that was applied to analyze the keywords from papers related to social network anonymization and detect the themes. To evaluate the effectiveness and quality of the identified themes and thematic areas, a quantitative and impact analysis was conducted for each subperiod. This analysis involved examining the quantity of documents linked to each theme, encompassing both core documents and secondary documents. Core documents are defined as those that contain at least two keywords appearing within the network of a particular theme, whereas secondary documents are those that have only one keyword associated with the theme's network. It is important to note that both core and secondary documents can potentially belong to multiple thematic networks (Cobo et al., 2011). Additionally, the quantity of documents, citations, average citation counts, and h-index value of each identified theme were tracked to evaluate their quality and impact.

In the following subsections, the themes are depicted visually. Then, the evolution of the themes is discussed in relation to their performance measures.

### 3-4-2- Visualization of Social Network Anonymization Themes

To explore themes related to the social network anonymization field in different time periods, two types of strategic diagrams are illustrated using the SciMAT software. These diagrams show the spheres' sizes, with the first diagram representing the number of citations received and the second one representing the number of core documents published for each theme (Appendix A displays the core documents relating to each theme within each subperiod). It is worth mentioning that, based on their placement within the strategic diagram, there are four distinct types of themes (Cahlik, 2000; Callon et al., 1991; Coulter et al., 1998; Courtial, 1994; He, 1999):

- The themes related to motors located in the upper-right quadrant of the strategic diagram are highly central and dense, indicating their significance for the organization and development of the research field. In other words, these themes are well-established and fundamental to the field.
- The upper-left quadrant of the strategic diagram contains the specialized and peripheral themes, which have high density, indicating their advanced development. However, their low centrality implies that these themes have little significance or importance for the research field.
- Themes that are either emerging or disappearing are located in the lower-left quadrant of the strategic diagram. These themes exhibit low density and centrality, suggesting that they are underdeveloped and not very significant for the research field.
- The lower-right quadrant of the strategic diagram contains basic, transversal, and general themes that are significant for the research field but are not yet well-developed.

The strategic diagrams and their related tables with quantitative measures for each subperiod are subsequently presented.

During the first subperiod, from 2007 to 2010, a total of 26 documents was extracted and analyzed. The strategic diagrams are presented in Fig. 21.

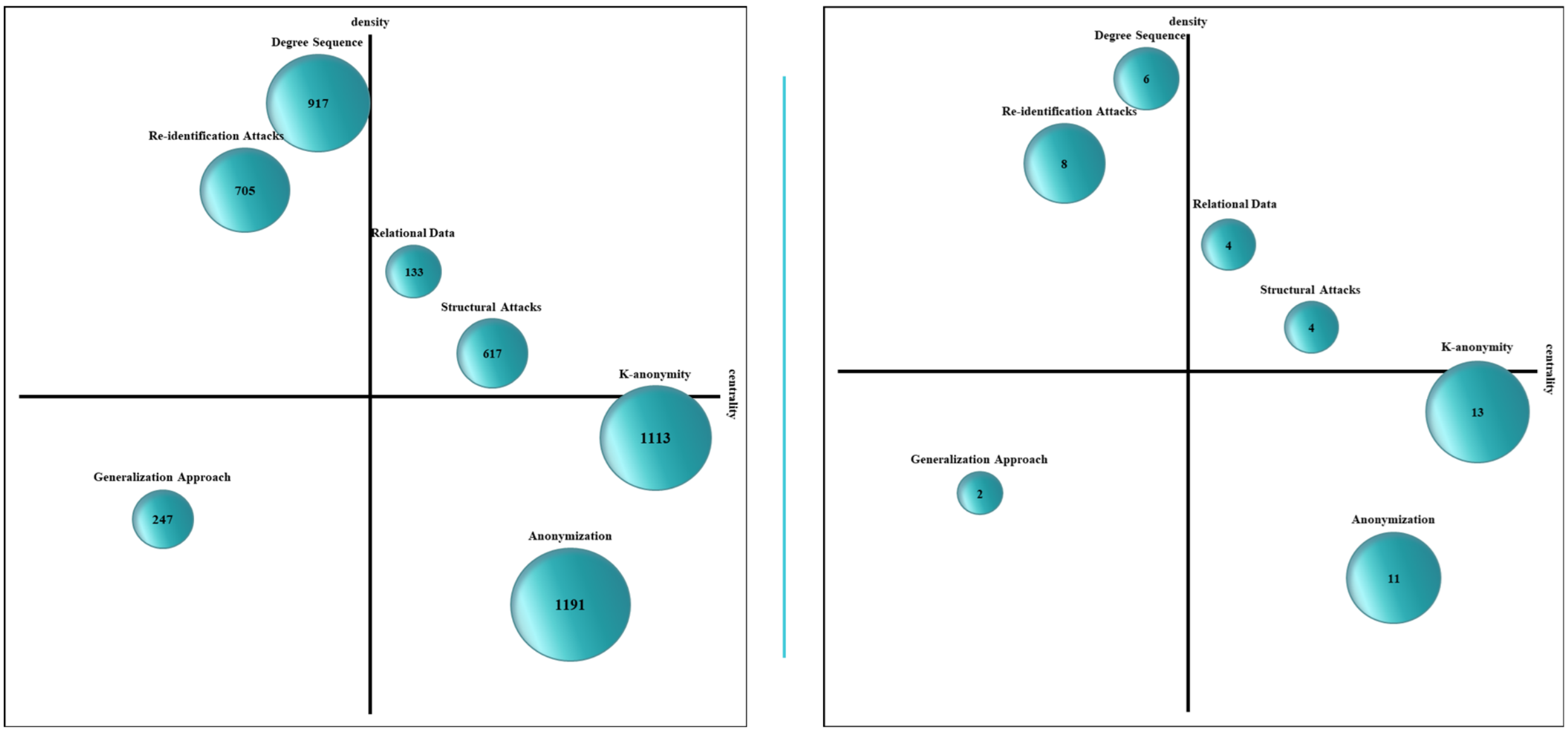

(a) (b)

*Figure 21. The Strategic Diagrams Designed for the Initial Subperiod of 2007-2010 Utilizing Two Parameters for Analysis: a) Number of Citations Received by These Documents, and b) Total Count of Core Documents*

Table 3 provides the quantitative measures related to the extracted themes.

*Table 3. The Metrics Used to Evaluate the Performance of the Themes During the Subperiod of 2007-2010*

| Theme Name | Core Documents Count | Core Documents h-index | Core Documents Citations | Core Documents Average Citations |
|---|---|---|---|---|
| Relational Data | 4 | 3 | 133 | 33.25 |
| Re-identification Attacks | 8 | 6 | 705 | 88.12 |
| K-anonymity | 13 | 8 | 1,113 | 85.62 |
| Degree Sequence | 6 | 5 | 917 | 152.83 |
| Structural Attacks | 4 | 3 | 617 | 154.25 |
| Anonymization | 11 | 9 | 1,191 | 108.27 |
| Generalization Approach | 2 | 2 | 247 | 123.5 |

Based on the obtained strategic map and the quantitative measures, we provide the following explanation for each theme during the 2007-2010 subperiod:

"Structural Attacks" and "Relational Data" can be considered motor themes. The significance of these themes lies in their high centrality and density, suggesting that they play a vital role in shaping and advancing the social network anonymization research domain. The "Structural Attacks" has 4 core documents, an h-index of 3, and a high total citation count of 617. Having the highest average citation count per document (154.25), this theme can be said to have had a significant impact and was highly visible in the field during this subperiod. Despite having fewer core documents, the theme exhibits high average citations per document, emphasizing its importance and influence in the field during this subperiod. In addition, the "Relational Data", with 4 core documents, an h-index of 3, and 133 total citations, has a relatively lower impact compared to some other themes. The average citation count per document (33.25) is also lower, indicating moderate visibility in the field.

The "Degree Sequence" and "Re-identification Attacks" are specialized and peripheral themes, which have high density, suggesting advanced development. However, their low centrality implies a lesser significance in the overall research field. Nevertheless, they still have a substantial impact due to their high total citation counts and average citations per document.

The “Generalization Approach” is among the emerging or disappearing themes in this subperiod since this theme exhibits low density and centrality. With only 2 core documents and an h-index of 2, it can be an emerged theme, so it is underdeveloped and not very significant for the research field in this subperiod. However, its relatively high average citation count per document indicates some visibility during this subperiod.

“Anonymization” and “K-anonymity” are among the basic, transversal, and general themes in this subperiod. They have high significance for the research field but are not yet well-developed. Both themes have a considerable number of core documents, high h-index values, and high total citation counts. Their placement in the lower-right quadrant suggests that they have the potential to grow in importance and influence the field’s development in the future.

Fig. 22 depicts the strategic diagram of the second subperiod of 2011-2014.

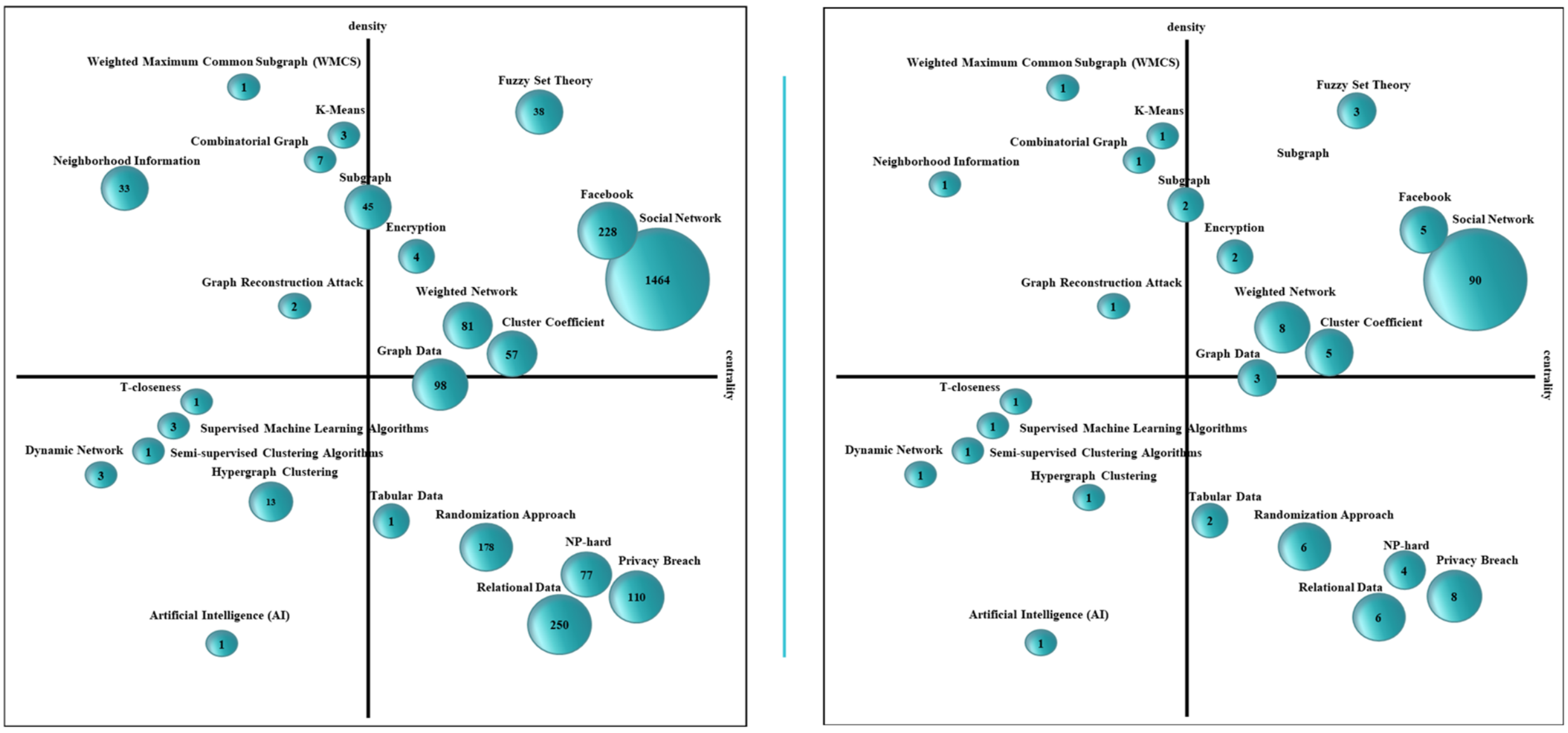

(a) (b)

*Figure 22. The Strategic Diagrams Designed for the Initial Sub Period of 2011-2014 Utilizing Two Parameters for Analysis: a) Number of Citations Received by These Documents, and b) Total Count of Core Documents*

Also, the quantitative measures related to the extracted themes are provided in Table 4.

*Table 4. The Metrics Used to Evaluate the Performance of the Themes During the Subperiod of 2011-2014*

| Theme Name | Core Documents Count | Core Documents h-index | Core Documents Citations | Core Documents Average Citations |
|---|---|---|---|---|
| Facebook | 5 | 5 | 228 | 45.6 |
| Fuzzy Set Theory | 3 | 3 | 38 | 12.67 |
| Social Network | 90 | 20 | 1,464 | 16.27 |
| Cluster Coefficient | 5 | 3 | 57 | 11.4 |
| Randomization Approach | 6 | 5 | 178 | 29.67 |
| Relational Data | 6 | 4 | 250 | 41.67 |
| Weighted Network | 8 | 3 | 81 | 10.12 |
| Privacy Breach | 8 | 5 | 110 | 13.75 |
| NP-hard | 4 | 3 | 77 | 19.25 |
| Encryption | 2 | 1 | 4 | 2 |
| Graph Data | 3 | 3 | 98 | 32.67 |
| Subgraph | 2 | 2 | 45 | 22.5 |
| Tabular Data | 2 | 1 | 1 | 0.5 |
| Weighted Maximum Common Subgraph (WMCS) | 1 | 1 | 1 | 1 |
| K-means | 1 | 1 | 3 | 3 |
| Neighborhood Information | 1 | 1 | 33 | 33 |
| Combinatorial Graph | 1 | 1 | 7 | 7 |
| Graph Reconstruction Attack | 1 | 1 | 2 | 2 |
| Hypergraph Clustering | 1 | 1 | 13 | 13 |
| Dynamic Network | 1 | 1 | 3 | 3 |
| Semi-Supervised Clustering Algorithms | 1 | 1 | 1 | 1 |
| Supervised Machine Learning Algorithms | 1 | 1 | 3 | 3 |
| T-closeness | 1 | 1 | 1 | 1 |
| Artificial Intelligence (AI) | 1 | 1 | 4 | 4 |

Based on the information presented in Fig. 22 and Table 4, the subsequent discoveries were made:

“Social Network”, “Facebook”, “Fuzzy Set Theory”, “Cluster Coefficient”, “Encryption”, and “Weighted Network” are considered motor themes because they are highly central and dense. The “Social Network” theme is the most prominent theme with the highest core document count (90) and h-index (20), which indicates its central role in the research landscape. “Facebook” is another motor theme, highlighting the importance of using this social network as the experimental dataset in social network anonymization research. “Fuzzy Set Theory”, “Cluster Coefficient”, “Encryption”, and “Weighted Network” have lower core document counts and citations compared to “Social Network” and “Facebook”, but their placement in the upper-right quadrant indicates their importance in the research field during this period.

“Weighted Maximum Common Subgraph (WMCS)”, “K-Means”, “Combinatorial Graph”, “Graph Reconstruction Attack”, and “Neighborhood Information” are considered as specialized and peripheral themes. These themes have fewer core documents and lower citation counts, which reflects their specialized nature and limited impact during this period.

“T-closeness”, “Supervised Machine Learning Algorithms”, “Semi-supervised Clustering Algorithms”, “Hypergraph Clustering”, “Dynamic Network”, and “Artificial Intelligence (AI)” are emerging or disappearing themes. Most of these themes have only one core document and low citation counts, indicating their limited influence and visibility during this subperiod.

“Tabular Data”, “Randomization Approach”, “NP-hard”, “Privacy Breach”, and “Relational Data” are considered basic, transversal, and general themes, which have a varying range of core document counts, h-index values, and citation counts. “Tabular Data”, “Relational Data”, and “Privacy Breach” can be considered basic themes in this field because they convey information about using anonymization methods on structured and relational datasets. Also, placing the “NP-hard” and “Randomization Approach” themes in the lower-right quadrant suggests their potential to grow in importance and influence the field’s development in the future.

Moreover, it is worth nothing that themes on the border lines, i.e., "Graph Data" and "Subgraph", have mixed characteristics, suggesting they might be transitioning between different levels of significance and development within the field. Hence, with respect to the quantitative values of both "Graph Data" and "Subgraph" themes from Table 4, they can be considered motor themes because they are fundamental and crucial themes for the development of the social network anonymization field.

Fig. 23 illustrates the strategic map of the extracted themes for the 2015-2018 subperiod.

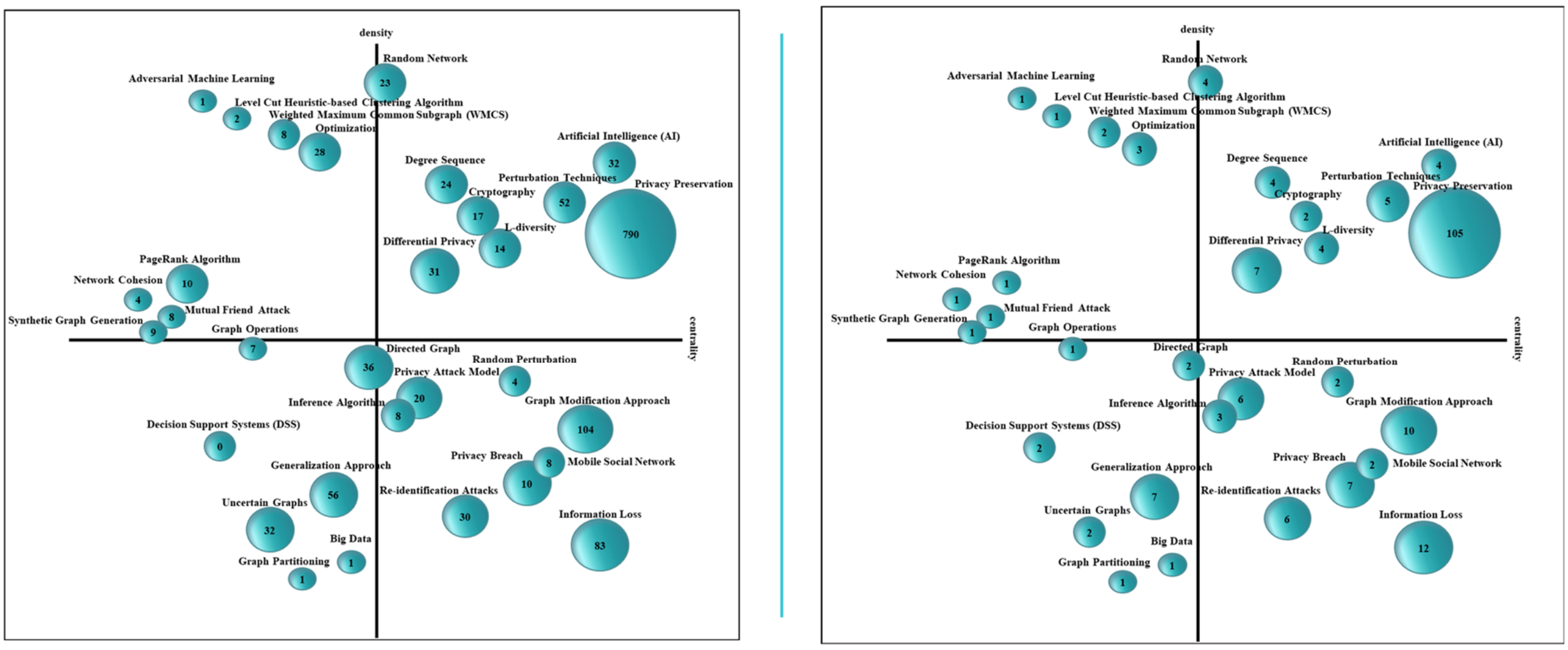

(a) (b)

*Figure 23. The Strategic Diagrams Designed for the Initial Sub Period of 2015-2018 Utilizing Two Parameters for Analysis: a) Number of Citations Received by These Documents, and b) Total Count of Core Documents*

In addition, the quantitative measures related to the themes of this subperiod are shown in Table 5.

*Table 5. The Metrics Used to Evaluate the Performance of the Themes During the Subperiod of 2015-2018*

| Theme Name | Core Documents Count | Core Documents h-index | Core Documents Citations | Core Documents Average Citations |
|---|---|---|---|---|
| Random Network | 4 | 2 | 23 | 5.75 |
| Perturbation Techniques | 5 | 2 | 52 | 10.4 |
| Privacy Preservation | 105 | 13 | 790 | 7.52 |
| Degree Sequence | 4 | 3 | 24 | 6 |
| Privacy Breach | 7 | 2 | 10 | 1.43 |
| Generalization Approach | 7 | 5 | 56 | 8 |
| Artificial Intelligence (AI) | 4 | 4 | 32 | 8 |
| Privacy Attack Model | 6 | 3 | 20 | 3.33 |
| Graph Modification Approach | 10 | 4 | 104 | 10.4 |
| Information Loss | 12 | 4 | 83 | 6.92 |
| L-diversity | 4 | 2 | 14 | 3.5 |
| Re-identification Attacks | 6 | 3 | 30 | 5 |
| Differential Privacy | 7 | 4 | 31 | 4.43 |
| Optimization | 3 | 3 | 28 | 9.33 |
| Weighted Maximum Common Subgraph (WMCS) | 2 | 2 | 8 | 4 |
| Cryptography | 2 | 2 | 17 | 8.5 |
| Random Perturbation | 2 | 1 | 4 | 2 |
| Inference Algorithm | 3 | 2 | 8 | 2.67 |
| Directed Graph | 2 | 2 | 36 | 18 |
| Level Cut Heuristic-based Clustering Algorithm | 1 | 1 | 2 | 2 |
| Adversarial Machine Learning | 1 | 1 | 1 | 1 |
| Mobile Social Network | 2 | 1 | 8 | 4 |
| Decision Support Systems (DSS) | 2 | 0 | 0 | 0 |
| Graph Operations | 1 | 1 | 7 | 7 |
| Synthetic Graph Generation | 1 | 1 | 9 | 9 |
| Mutual Friend Attack | 1 | 1 | 8 | 8 |
| Network Cohesion | 1 | 1 | 4 | 4 |
| PageRank Algorithm | 1 | 1 | 10 | 10 |
| Uncertain Graphs | 2 | 2 | 32 | 16 |
| Graph Partitioning | 1 | 1 | 1 | 1 |
| Big Data | 1 | 1 | 1 | 1 |

Based on the information given, the subsequent observations should be highlighted:

"Privacy Preservation", "Artificial Intelligence (AI)", "Perturbation Techniques", "L-diversity", "Differential Privacy", "Cryptography", "Degree Sequence", and "Random Network" are placed in the upper-right quadrant and are considered motor themes. "Privacy Preservation" is the dominant theme with the highest core document count (105) and h-index (13), reflecting its central role in the research during this period. Also, "Artificial Intelligence (AI)", "Perturbation Techniques", "L-diversity", "Differential Privacy", "Cryptography", "Degree Sequence", and "Random Network" have lower core document counts and citations compared to "Privacy Preservation", but they are essential to the research field in this period due to their high centrality and density.

"Optimization", "Adversarial Machine Learning", "Level Cut Heuristic-based Clustering Algorithm", "Weighted Maximum Common Subgraph (WMCS)", "PageRank Algorithm", "Network Cohesion", "Mutual Friend Attack", and "Synthetic Graph Generation" are situated in the upper-left quadrant and are considered specialized and peripheral themes that are advanced but less significant in this subperiod. The "Optimization" theme, with 3 core documents and an average of 9.33 citations per document, is a developed and specialized theme. However, "Adversarial Machine Learning", "Level Cut Heuristic-based Clustering Algorithm", "Weighted Maximum Common Subgraph (WMCS)", "PageRank Algorithm", "Network Cohesion", "Mutual Friend Attack", and "Synthetic Graph Generation" themes exhibit advanced development but have limited impact on the overall research field during this subperiod, as evidenced by their lower core document counts and citation numbers.

"Graph Operations", "Decision Support Systems (DSS)", "Generalization Approach", "Directed Graph", "Uncertain Graphs", "Big Data", and "Graph Partitioning" themes are located in the lower-left quadrant and can be considered as emerging or disappearing themes. They have limited influence and visibility, with most having only one or two core documents and low citation counts. While their low density and centrality indicate that they might be emerging in the field, they are underdeveloped and not very significant as of yet.

The lower-right quadrant contains the themes "Random Perturbations," "Graph Modification Approach," "Information Loss," "Re-identification Attacks," "Privacy Breach," "Mobile Social Network," "Privacy Attack Model," and "Inference Algorithm". Among these, "Random Perturbations" has 2 core documents, an h-index of 2, and an average of 2 citations per document. "Graph Modification Approach" has 10 core documents, an h-index of 4, and an average of 10.4 citations per document. "Information Loss" has 12 core documents, a h-index of 4, and an average of 6.92 citations per document. “Re-identification Attacks” has 6 core documents, a h-index of 3, and an average of 5 citations per document. “Privacy Breach” has 7 core documents, a h-index of 2, and an average of 1.43 citations per document. These themes are important but not yet well-developed and can be considered basic, transversal, and general themes. However, “Mobile Social Network”, “Privacy Attack Model”, and “Inference Algorithm” themes have a varying range of core document counts, h-index values, and citation counts, but their high centrality and low density suggest potential growth in importance.

Moreover, the strategic diagram of the last subperiod, 2019-2022, is depicted in Fig. 24.

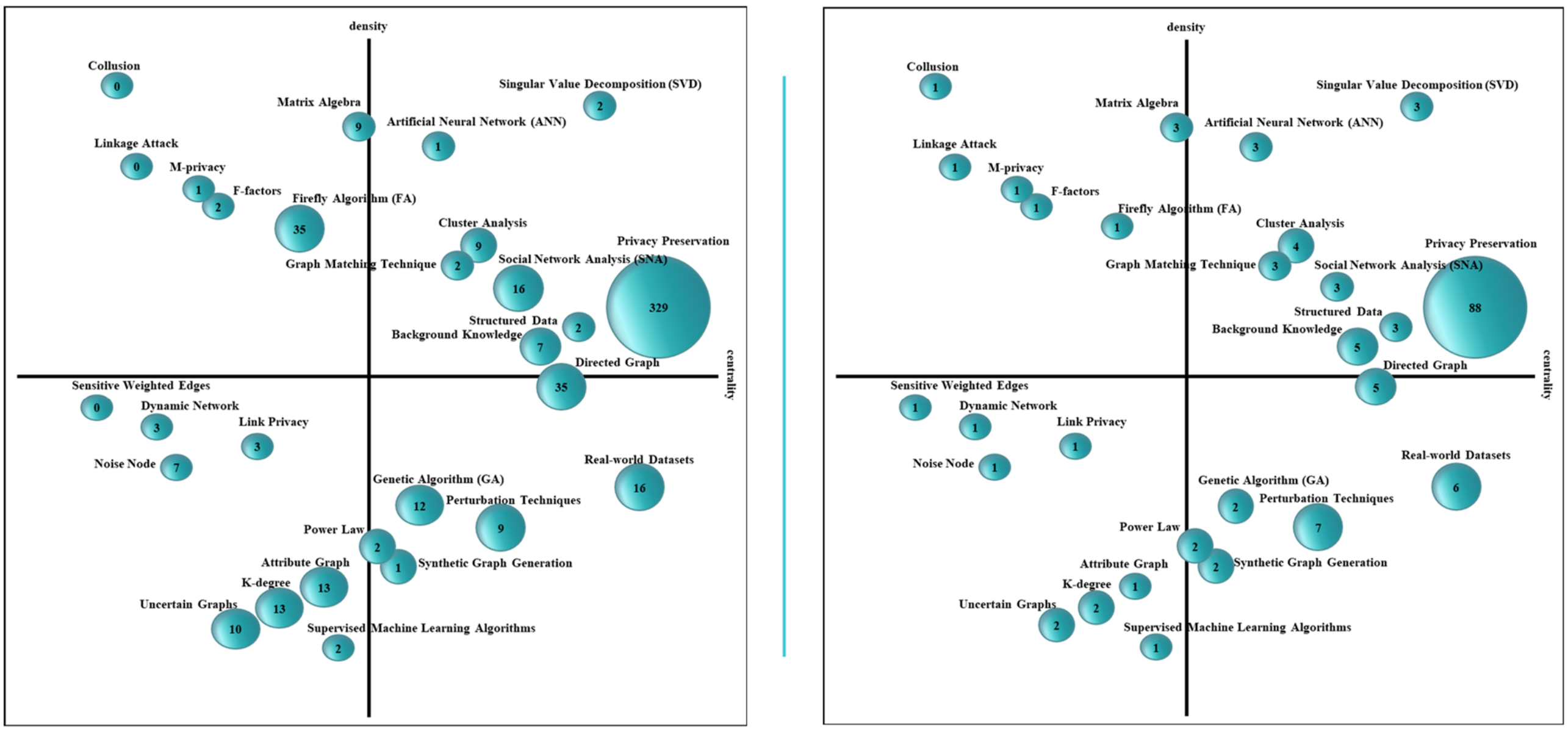

(a) (b)

*Figure 24. The Strategic Diagrams Designed for the Initial Sub Period of 2019-2022 Utilizing Two Parameters for Analysis: a) Number of Citations Received by These Documents, and b) Total Count of Core Documents*

In addition, the quantitative measures of the themes in the last subperiod are provided in Table 6. Since this is the final subperiod, the average number of citations is significantly lower than the previous subperiods. This is primarily due to the relationship between this factor and the number of papers. In the last subperiods, the citation count decreases due to the limited number of documents in the field.

*Table 6. The Metrics Used to Evaluate the Performance of the Themes During the Subperiod of 2015-2018*

| Theme Name | Core Documents Count | Core Documents h-index | Core Documents Citations | Core Documents Average Citations |
|---|---|---|---|---|
| Singular Value Decomposition (SVD) | 3 | 1 | 2 | 0.67 |
| Privacy Preservation | 88 | 9 | 329 | 3.74 |
| Artificial Neural Network (ANN) | 3 | 1 | 1 | 0.33 |
| Graph Matching Technique | 3 | 1 | 2 | 0.67 |
| Matrix Algebra | 3 | 2 | 9 | 3 |
| Perturbation Techniques | 7 | 2 | 9 | 1.29 |
| Cluster Analysis | 4 | 2 | 9 | 2.25 |
| Social Network Analysis (SNA) | 3 | 2 | 16 | 5.33 |
| Structured Data | 3 | 1 | 2 | 0.67 |
| Background Knowledge | 5 | 2 | 7 | 1.4 |
| Real-World Datasets | 6 | 3 | 16 | 2.6 |
| Directed Graph | 5 | 3 | 35 | 7 |
| Genetic Algorithm (GA) | 2 | 2 | 12 | 6 |
| Collusion | 1 | 0 | 0 | 0 |
| Power Law | 2 | 1 | 2 | 1 |
| Synthetic Graph Generation | 2 | 1 | 1 | 0.5 |
| Firefly Algorithm (FA) | 1 | 1 | 35 | 35 |
| F-factors | 1 | 1 | 2 | 2 |
| M-privacy | 1 | 1 | 1 | 1 |
| Linkage Attack | 1 | 0 | 0 | 0 |
| Uncertain Graphs | 2 | 2 | 10 | 5 |
| K-degree | 2 | 1 | 13 | 6.5 |
| Noise Node | 1 | 1 | 7 | 7 |
| Link Privacy | 1 | 1 | 3 | 3 |
| Dynamic Network | 1 | 1 | 3 | 3 |
| Sensitive Weighted Edges | 1 | 0 | 0 | 0 |
| Attribute Graph | 1 | 1 | 13 | 13 |
| Supervised Machine Learning Algorithms | 1 | 1 | 2 | 2 |

Hence, based on the information provided in Fig. 24 and Table 6, it can be observed that:

"Privacy Preservation", "Background knowledge", "Artificial Neural Network (ANN)", "Graph Matching Technique", "Cluster Analysis", "Singular Value Decomposition (SVD)", "Structured Data", and "Social Network Analysis (SNA)" themes are located in the upper-right quadrant, which is designated for motor themes. In this regard, "Privacy Preservation" with 88 core documents and 329 citations, can be considered as the most crucial and well-established in the field, focusing on preserving the personal information and sensitive relationships of social network users in a social network.

"Collusion", "Matrix Algebra", "Linkage Attack", "M-privacy", "F-factors", and "Firefly Algorithm (FA)" themes are situated in the upper-left quadrant, representing niche areas of research with high specialization but a limited influence on the broader field. In other words, these themes represent specific areas of research that have a high specialization, indicating that they require in-depth knowledge to work on them. However, their influence on the broader field is limited, as they have lower centrality. For instance, the "Firefly Algorithm (FA)" is a specialized optimization technique that may be applied to certain social network anonymization problems but is not central to the entire field. "M-privacy" is an example of a specific privacy model that is important in certain contexts but does not apply to all social network privacy preservation scenarios.

The "Sensitive Weighted Edges", "Dynamic Network", "Link Privacy", "Noise Node", "Attribute Graph", "K-degree", "Uncertain Graphs", and "Supervised Machine Learning Algorithms" themes are placed in the lower-left quadrant, which may represent emerging research directions or fading interests in the field. For example, "Link Privacy", "Noise Node", "Uncertain Graphs", "Dynamic Network", and "Supervised Machine Learning Algorithms" may represent emerging themes in the social network anonymization field as the researchers can use them to introduce novel methods or techniques in the domain. On the other hand, "Attribute Graph", "K-degree", and "Sensitive Weighted Edges" could be disappearing themes if their importance has diminished due to advances in other themes.

The "Genetic Algorithm (GA)", "Directed Graph", "Real-world Datasets", "Perturbation Techniques", "Power Law", and "Synthetic Graph Generation" themes are in the lower-right quadrant, which have a broader impact on the research field but are not yet well-developed. They can be considered essential building blocks or methods that are applicable across different areas of the research field. For instance, “Perturbation Techniques” are general methods used to anonymize graph data by introducing noise or altering the graphs in some way. These techniques can be applied across a wide range of social network privacy preservation problems. Similarly, “Directed Graph” and “Genetic Algorithm (GA)” are fundamental concepts that can be used in various contexts within the field.

Based on the conducted analyses of this section, it is worth noting that the basic and motor themes in the social network anonymization domain achieve the highest citation scores and impacts. By recognizing these basic themes, researchers can better comprehend the foundational concepts, principles, and building blocks that support this research area. This understanding aids in establishing a robust foundation, which is essential for grasping more advanced and specialized topics. Moreover, basic themes frequently act as a launchpad for new research, allowing researchers to spot potential research gaps and opportunities for further investigation, innovation, and cooperation. Additionally, being familiar with the motor themes offers valuable insights into the field’s primary challenges and prospects, assisting researchers in developing new research questions and hypotheses.

### 3-4-3- Evolution of Social Network Anonymization Themes

After examining how the number of common keywords changes throughout various subperiods. Subsequently, we investigated the development of thematic areas to track the progression of themes.

As depicted in Fig. 25, there is a variation in the constancy of keywords across each subperiod. Although several keywords consistently appear, some newly appear or fade away within each subperiod. In addition, specific keywords like “T-closeness” exhibit uniqueness by being exclusive to particular subperiods. Despite this, a number of keywords are consistently present across all the analyzed subperiods, such as “Social Network”, “Privacy Preservation”, “Anonymization”, “K-anonymity”, “Generalization Approach”, “Differential Privacy”, “Graph Modification Approach”, “Randomization Approach”, and “Clustering Algorithms”.

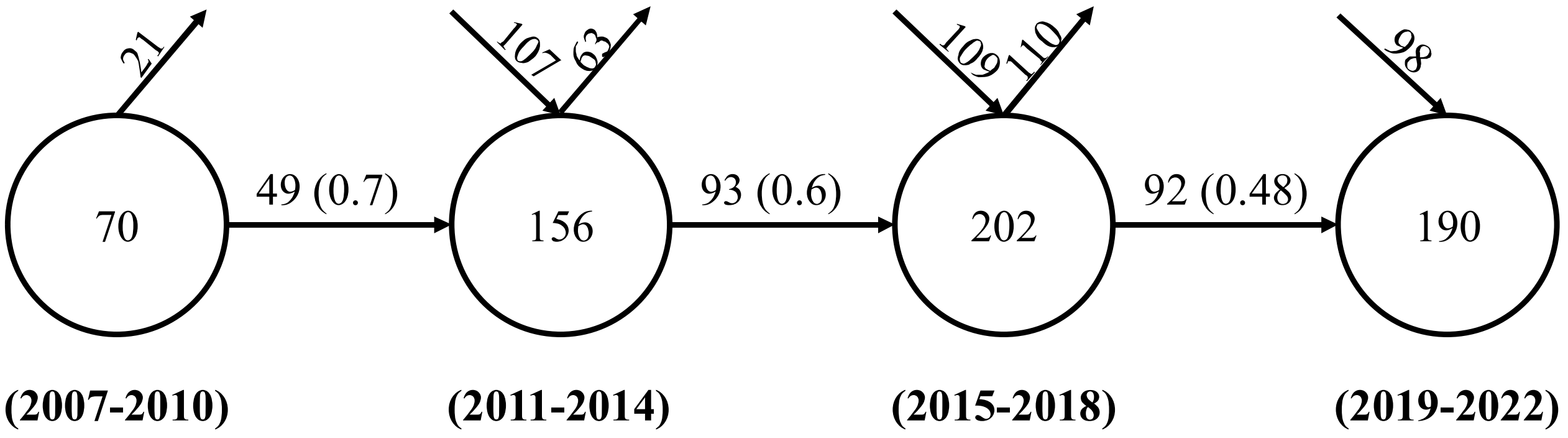

*Figure 25. Overlap Mapping: The Proportion of Keywords That Are Shared Between Consecutive Subperiods, Both Incoming and Outgoing*

It is worth noting that, in overlapping map, the inclusion similarity index refers to the degree of overlap or similarity in the set of keywords between two subperiods, with a higher value indicating a greater degree of similarity. Taking this into consideration and based on the information presented in Fig. 25, the number of shared keywords between the (2007-2010) and (2011-2014) subperiods is 49. This indicates that these two subperiods share some common research topics or themes in the social anonymization field. Also, the inclusion similarity index of 0.7 between the (2007-2010) and (2011-2014) subperiods suggests a relatively high degree of overlap/similarity in the set of shared keywords. This implies a certain level of continuity or similarity in research themes or topics between these two subperiods.

Furthermore, the number of shared keywords increases between the (2011-2014) and (2015-2018) subperiods, which suggests that there is more overlap in research topics or themes between these two subperiods. However, the inclusion similarity index between the (2011-2014) and (2015-2018) subperiods decreases to 0.6, indicating a lower degree of overlap or similarity in the set of shared keywords between these two subperiods. This signifies a decreasing level of continuity or similarity in research themes or topics between these two subperiods.

Interestingly, the number of shared keywords between the (2015-2018) and (2019-2022) subperiods is slightly lower at 92, which may indicate a slight decrease in the level of overlap or commonality in research topics between these two subperiods. In addition, the inclusion similarity index between the (2015-2018) and (2019-2022) subperiods decreases even further to 0.48, which suggests a reduction in the degree of overlap or similarity in the set of shared keywords between these two subperiods. This shows a further decrease in the level of continuity or similarity in research themes or topics between these two subperiods.

Overall, the shared keywords and inclusion similarity index values between the subperiods suggest a decreasing level of continuity or similarity in research themes or topics in the social anonymization field. This reduction between the subperiods may indicate a shift in research focus or divergence in scientific disciplines within the social anonymization field, which could have implications for interdisciplinary collaboration and knowledge transfer within this field.

In the ensuing subsections, we present a comprehensive analysis of the development of themes within the realm of social network anonymization by utilizing the evolution map and alluvial diagrams.

- **Evolution Map**

SciMAT's evolution map analysis was used to examine the evolution of themes in the social network anonymization field over time. This map illustrates the primary topics and themes within the field and how they have transformed or progressed across different periods. The creation of this evolution map relies on the analysis of co-occurring networks of keywords, enabling us to identify the most pertinent research topics and their relationships. Therefore, through employing this map, we gained a better understanding of the field's development, identified emerging topics or research gaps, and explored potential future research directions.

Fig. 26 illustrates the evolution of themes in the social network anonymization field. In this evolution map:

- A solid line indicates that either both themes have the same name or one of the themes shares a keyword with the other.
- A dashed line signifies that the themes share keywords that are not identical to the names of the themes.
- The width of the link between themes is directly proportional to the inclusion similarity index.

- The size of the spheres corresponds to the number of published core documents for each theme, meaning that a larger sphere indicates a higher number of published core documents.

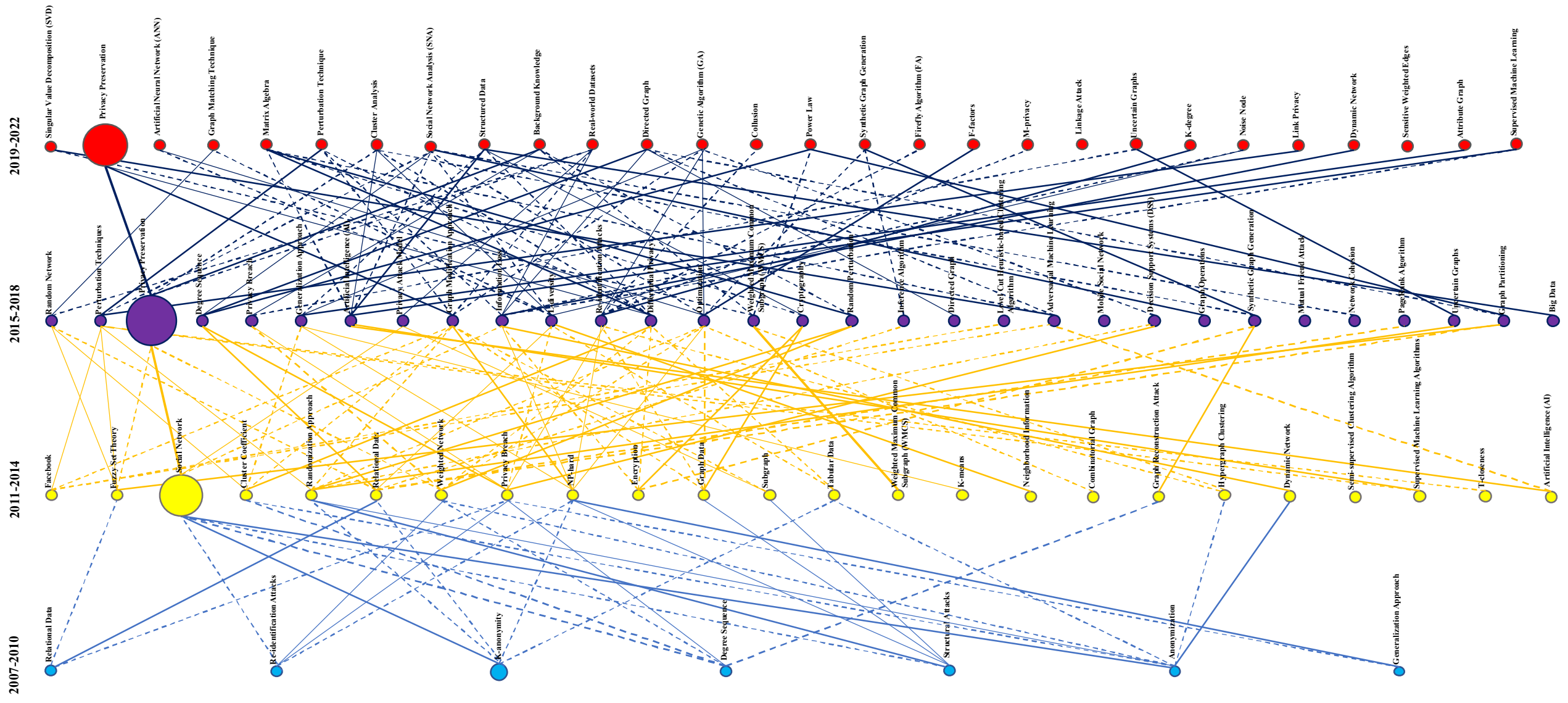

*Figure 26. The Evolution Map of the Social Network Anonymization Field*

This evolution map illustrates that during each subperiod, some newly emerged keywords represent new research topics or unexplored areas of study in the social network anonymization field. These keywords may indicate the introduction of new methods or techniques, novel research areas, or the participation of new authors in the field. For example, in the subperiod spanning 2011-2014, ten newly emerged keywords include “Artificial Intelligence (AI)”, “T-closeness”, “Supervised Machine Learning Algorithms”, “Semi-supervised Clustering Algorithm”, “Combinatorial Graph”, “Neighbourhood Information”, “K-means”, “Weighted Maximum Common Subgraph (WMCS)”, “Encryption”, and “Facebook”.

The appearance of “Artificial Intelligence (AI)”, “Supervised Machine Learning Algorithms”, and “Semi-supervised Clustering Algorithm” as newly emerged keywords suggest that researchers are starting to explore the application of AI and machine learning techniques to address challenges of social network privacy. These algorithms could be used to identify patterns in social network data that reveal sensitive information, such as location data or personal relationships, and then obfuscate that information to protect individual privacy.

One newly emerged keyword, “T-closeness”, is an anonymization model for relational data that aims to protect the privacy of social network users by ensuring that the distribution of data values in the anonymized dataset is similar to the distribution in the original dataset. This model has superiority over the two well-known K-anonymity and L-diversity anonymization models because it can deal with various types of attacks, such as homogeneity and background knowledge.

Another newly emerged keyword, “Combinatorial Graph” refers to a graph structure that can be used to model complex relationships between nodes in a social network. This keyword highlights the importance of using mathematical and computational tools to understand and solve the social network anonymization problem.

Moreover, the “Weighted Maximum Common Subgraph (WMCS)” is a graph-matching algorithm that can be used to identify the largest subgraph that is common to two or more graphs. This algorithm is used in social network anonymization to identify and preserve significant structures (those with high total weights) when applying anonymization techniques.

“Neighbourhood Information” is a newly emerged keyword that refers to the study of the relationships between the nodes and their neighbors in a social network. This information can be used to identify patterns in social network data and understand the structure of social networks. In the context of social network anonymization, the researchers can use this information to develop new anonymization methods that consider the relationships between nodes in a social network.

Also, “K-means” is a clustering algorithm that groups data points together based on their similarity. In the context of social network anonymization, K-means can be used to group nodes together based on their similarity (e.g., attributes or connections) to protect individual privacy.

Besides, “Encryption” is employed as a privacy protection mechanism for social network data to limit unauthorized access and keep users’ information secure. However, since using encryption alone is insufficient, it must be paired with anonymization techniques to guarantee complete anonymity.

“Facebook” is a newly emerged keyword that likely refers to the growing interest in using this platform to study ways of social network anonymization to protect users’ privacy.

Furthermore, in the 2015-2018 subperiod, the newly emerged keywords contain “Big Data”, “Mutual Friend Attack”, “Graph Operations”, “Mobile Social Network”, and “Directed Graph”.

“Big Data” refers to the huge and complex sets of information that are generated by users in social networks and various electronic devices like smartphones, wearable tech, or home automation systems. These large datasets create significant challenges when it comes to keeping personal information private and anonymous.

In this subperiod, “Mutual Friend Attack” emerged, which is a type of attack that aims to re-identify social network users by exploiting information about their mutual friends or connections.

“Graph Operations” refers to the various computational and mathematical techniques used to analyze and manipulate social network graphs, such as algorithms for graph clustering, graph contraction, graph aggregation, graph pruning, and graph obfuscation.

The emergence of the “Directed Graph” in this subperiod shows a trend in studying the social network anonymization methods on this type of graph where relationships or interactions have a directional nature.

In the final subperiod, from 2019 to 2022, “Linkage Attack” emerged as a new theme. This novel keyword suggests that researchers are increasingly concentrating on addressing this type of attack in social network anonymization studies. During a linkage attack, an attacker could exploit the structural properties or patterns present in the anonymized network data, linking them to known properties or patterns in external data sources to re-identify a target.

It is worth noting that in different subperiods of the evolution map, there are some connections between clusters or individual nodes that are interesting to discuss and help understand the relationships and interactions among various research topics or themes over time.

The “Graph Modification Approach” was first introduced during the 2007-2010 period as a node of the “Re-identification Attacks” theme. The edge weight between these two keywords (representing the strength of the relationship between those keywords, calculated based on their co-occurrence in the articles) in the theme suggests that there is a significant overlap in the research conducted on these topics. Therefore, these two keywords have a significant tendency to appear together, indicating that graph modification approaches are frequently studied and used as a way to prevent re-identification attacks in the context of social network anonymization. This can help researchers understand the importance of developing robust graph modification techniques to enhance privacy protection in social networks. The strong relationship between these two keywords also highlights the ongoing challenge of balancing the need for privacy with the utility of the data as re-identification attacks continue to evolve and become more sophisticated. As a result, researchers working in this area should pay close attention to advances in both graph modification techniques and re-identification attack strategies to ensure that they are aware of the latest developments and can adapt their methods accordingly.

Furthermore, in the 2011-2014 subperiod, the “Graph Modification Approach” keyword appeared in the “NP-hard” theme. The edge weight between these keywords indicates a relatively low strength of association between them within the social network anonymization field. This means that although these terms are related, they may not frequently co-occur in the published papers, or their relationship is not as strong as other keyword pairs in the domain. In 2015-2018, the “Graph Modification Approach” keyword gained prominence within the social anonymization field, leading to the formation of a new theme centered around it with the same label. The edge weight between the “Graph Modification Approach” and “Edge Addition/Deletion” keywords suggests that the combination of graph modification approaches and edge addition/deletion techniques is a central theme in the field of social network anonymization. Researchers in this area focused on developing and improving methods that involve altering the structure of social network graphs, particularly by adding or removing edges between nodes, to protect user privacy while preserving the utility of the data for further analysis. Finally, in 2019-2022, the “Graph Modification Approach” keyword is placed in the “Privacy Preservation” theme, which reveals a growing interest in using this approach to protect privacy in social networks. The strong co-occurrence of this keyword with the “Structural Properties” and “Anonymization” keywords is also interesting. This strong co-occurrence suggests that the “Graph Modification Approach” is a promising technique for privacy preservation and social network anonymization, particularly in the context of preserving structural properties while maintaining data usefulness.

The evolution map of the social network anonymization field shows that in 2007-2010, “Differential Privacy” was placed in the degree sequence theme, which is a mathematical property of networks. This suggests that the concept of differential privacy is linked to mathematical approaches to social network anonymization by limiting the risk of re-identification.

Furthermore, the keyword “Differential Privacy” has strong connections to other keywords, such as “Synthetic Graph Generation”, “Privacy Preserving Data Mining (PPDM)”, “Input Graph”, and “Power Law”. The strong connection between “Differential Privacy” and these other keywords in the evolution map suggests that researchers in the social network anonymization field during the 2007-2010 period were actively exploring the use of differential privacy techniques to create synthetic graphs, develop privacy-preserving data mining algorithms, and analyze the structural properties of input graphs (like degree sequences and power-law distributions) without compromising privacy.

During the 2011-2014 subperiod, the evolution map of the social network anonymization field shows the "Differential Privacy" keyword as a part of the "Cluster Coefficient" cluster. This placement, along with its strong association to keywords like "Divide and Conquer Algorithm", "Laplace Noise", "Random Network", "Cluster Coefficient", and "Graph Mining" emphasizes the concentration of research and the interconnections among these keywords during that period. The prominent link between "Differential Privacy" and other keywords mentioned in the evolution map indicates that researchers in the social network anonymization domain were concentrating on incorporating differential privacy into graph mining activities while investigating the connection between privacy and structural characteristics like cluster coefficients. For instance, the scientists developed divide-and-conquer algorithms for effective graph analysis on random networks and employed Laplace noise to ensure privacy while analyzing the mentioned networks.

During the 2015-2018 period of the evolution map, the "Differential Privacy" keyword emerged as a theme. This signifies the increasing prominence of differential privacy as a central research topic in the mentioned subperiod. The keyword also exhibits strong connections to other keywords, such as "Topological Information", "Persistent Homology", "Laplace Noise", "Matrix Algebra", "Random Matrix", "Correlation Matrix", "Personalization", "Social Recommendations", and "Shortest Path", which highlights the interdisciplinary nature of research during this period. The strong connections between "Differential Privacy" and the above-mentioned keywords suggest that researchers were working on various aspects of differential privacy in social networks. Such methods include applying differential privacy to protect topological information, exploring advanced mathematical techniques like persistent homology, using matrix algebra and random matrices to analyze network properties, and incorporating privacy-preserving techniques into personalized social recommendations and shortest-path algorithms.

In the 2019-2022 subperiod, the "Differential Privacy" keyword is placed in the "Privacy Preservation" theme, indicating the maturation of differential privacy as a well-established technique in the context of privacy preservation in social networks. Its strong connection to various other keywords, such as "Privacy Preservation", "Structural Privacy", "Social Network", and "Anonymization", underscores the continued relevance and focus on these topics during this period. The strong connections between "Differential Privacy" and these keywords in this subperiod suggest that researchers in the social network anonymization field were focusing on integrating differential privacy with other privacy preservation techniques to achieve better structural privacy in social networks. This includes exploring new methods and techniques based on differential privacy for anonymizing social networks, as well as studying the trade-offs between privacy and utility in various anonymization approaches.

Based on the provided evolution map and during the 2007-2010 subperiod, the "Generalization Approach" keyword is placed in its own theme. This indicates that during this time, generalization approaches were a distinct and significant research focus within the field. However, based on the strategy map diagram of the mentioned subperiod, the "Generalization Approach" theme had limited importance and connectivity with other research themes, suggesting that it might have been underdeveloped or overlooked in comparison to other themes during that time. The strong connection between "Generalization Approach" and other keywords, such as "Eigenvector Centrality", "Split Algorithms", and "Decision Support Systems (DSS)", highlights the relationships and interdisciplinary nature of research during this period. The strong connection between "Generalization Approach" and the mentioned keywords in the evolution map suggests that researchers in the social network anonymization field were exploring the use of generalization approaches to anonymize social network data while considering network properties like eigenvector centrality. They also developed and applied various split algorithms, such as union-split algorithms, to improve the efficiency of generalization processes, specifically in the context of graph partitioning and distributed privacy-preserving techniques. Besides, the strong relationship between the "Generalization Approach" and "Decision Support Systems (DSS)" demonstrates that researchers conducted some studies to investigate how they can use the generalization approaches to ensure privacy preservation in decision support systems.

In the 2011-2014 subperiod, the "Generalization Approach" keyword is placed in the "NP-hard" theme. This placement indicates that during this period, the research focus related to generalization approaches in social network anonymization was strongly connected to NP-hard problems and optimization challenges. The connections to the "NP-hard", "Optimization", and "Information Loss" keywords highlight the relationships between these topics during this period. The connections between "Generalization Approach" and the keywords mentioned in this subperiod suggest that researchers in the social network anonymization field were focusing on the challenges associated with solving NP-hard problems related to generalization approaches. They were also exploring optimization techniques to minimize information loss while preserving privacy and dealing with the computational complexities of these problems.

In the 2015-2018 subperiod, the "Generalization Approach" keyword is once again placed in the "Generalization Approach" theme, suggesting that this research theme remained relevant and active during this period. The "Generalization Approach" keyword is also connected to various other keywords, including "K-means", "Node Degree", and "Suppression". The connection to "K-means" indicates that researchers explored the use of clustering algorithms, like K-means, for generalization approaches in social network anonymization. The connection to "Node Degree" suggests that researchers may have been focusing on preserving the degree distribution of nodes in the network while generalizing it for privacy protection. The connection to "Suppression" implies that researchers might have been exploring the use of suppression techniques to protect the users' attribute privacy by suppressing their sensitive data in the network.

During the 2019-2022 subperiod in the evolution map of the social network anonymization field, the "Generalization Approach" keyword is placed in the "Genetic Algorithm (GA)" theme. This indicates that during this period, researchers explored the use of genetic algorithms to improve the efficiency and effectiveness of generalization approaches in social network anonymization. The "Generalization Approach" keyword is also connected to other keywords, including "Particle Swarm Optimization Algorithm (PSO)" and "Hybrid Algorithms". The connection to the "Particle Swarm Optimization Algorithm (PSO)" suggests that researchers used swarm intelligence-based optimization techniques to improve generalization approaches in social network anonymization. The connection to "Hybrid Algorithms" indicates that researchers combined genetic algorithms and other optimization techniques (e.g., PSO) to create more powerful and flexible algorithms for social network anonymization. As a result, the use of genetic algorithms, swarm intelligence, and hybrid algorithms highlight the interdisciplinary nature of research in this field and the importance of leveraging techniques from diverse fields to tackle complex problems.

While the "Uncertain Graphs" keyword was not present in the evolution map of the 2007-2010 subperiod, it does appear the 2011-2014 subperiod for the first time and is placed in the "Randomization Approach" theme. This indicates that during this period, the researchers tried to explore new ways of addressing the challenges associated with applying an uncertain graph approach in social network anonymization.

In the 2015-2018 subperiod, the "Uncertain Graphs" keyword is presented as a theme. This indicates that during this period, researchers investigated the challenges and opportunities associated with using uncertain graphs as an approach to anonymize the social network and focused on this topic as a distinct and important research theme. Additionally, the "Uncertain Graphs" keyword has a strong connection to the "Maximizing Variance" keyword. Variance maximization involves maximizing the variance of the uncertain graph by selecting or perturbing its edges or weights. This can help to improve the accuracy of analyses performed on the graph while preserving privacy.

During the 2019-2022 subperiod, the "Uncertain Graphs" keyword is once again placed in the "Uncertain Graphs" theme. This suggests that during this period, researchers continued to focus on uncertain graphs as an important and distinct research theme in social network anonymization. The "Uncertain Graphs" keyword is also connected to the "Node Characteristics" and "Triadic Closure" keywords in the theme. The focus on node characteristics and triadic closures highlights the importance of understanding the structural properties and dynamics of social networks to propose novel uncertain graph approaches for anonymizing social networks effectively.

Furthermore, based on the proposed evolution map of the social network anonymization field, we aim to identify connections between clusters or individual nodes across different subperiods. This can help researchers to understand the relationships and interactions among different research topics and themes over time.

As shown in the evolution map, the "K-anonymity" theme emerged for the first time in the 2007-2010 subperiod. This means that the researchers tried to apply the previously-defined anonymization models used for relational and structured datasets to the graph data, which is unstructured data. They proposed various social network anonymization models to deal with the structural attacks based on the K-anonymity model along with different graph structural properties, such as degree sequence, graph isomorphism, graph automorphism, and centrality criterion. Subsequently, between 2011 and 2014, researchers focused on defeating the shortcomings of K-anonymity-based models in combating re-identification and neighborhood attacks. They introduced the "L-diversity" anonymization model to social network anonymization, which led to the development of various models. Furthermore, during this time, researchers also started to propose social network anonymization models based on the "T-closeness" anonymization model to tackle the limitations of L-diversity-based models.

The proposed evolution map highlights a significant connection pattern involving "Generalization Approach", "NP-hard", "Optimization", "Heuristics", and "Meta-heuristics". Fig. 26 demonstrates a link between the "Generalization Approach" theme in the 2007-2010 subperiod and the "NP-hard" theme in the 2011-2014 subperiod. The "NP-hard" theme encompasses keywords like "Optimization," "Generalization Approach," and "Information Loss," and there is a robust relationship among these keywords within this cluster. That is, in this subperiod, the researchers' focus shifted to tackling NP-hard problems related to generalization approaches during that time. This connection suggests that as researchers explored generalization approaches in the earlier subperiod, they discovered that some problems in the field were NP-hard, which means they are computationally difficult to solve in an efficient manner. An example of this is identifying the optimal node clusters in the generalization approach that result in minimal information loss. Consequently, the research emphasis transitioned to discovering optimization methods, heuristics, and meta-heuristics to tackle these intricate issues more effectively.

In the following subperiod of the evolution map, 2015-2018, the “NP-hard” keyword is associated with the “Re-identification Attacks” theme. This association emphasizes the intricate nature of some re-identification problems and underscores the necessity of creating robust and efficient anonymization methods to counter these attacks. The evolution map illustrates the importance of comprehending the computational obstacles inherent in re-identification attacks and the unceasing efforts within the anonymization discipline to address these challenges. Besides, in this subperiod, the keyword “Optimization” arises as a distinct theme in the context of social network anonymization. This theme demonstrates the growing importance of optimization techniques in addressing privacy and anonymization challenges in social networks. Within the optimization theme, several related keywords are connected, such as “Genetic Algorithm (GA),” “Heuristics,” and “Combinatorial Optimization.” These keywords represent different optimization methodologies that have been employed in the field to improve anonymization techniques. Additionally, these optimization-related keywords have external connections with other relevant keywords from other themes. For instance, “Clustering Algorithms,” “Graph Modification Approach,” and “K-anonymity” are linked to the optimization theme. These external links signify the interdisciplinary nature of social network anonymization research and reveal that optimization techniques are being integrated with other approaches to enhance privacy protection. The emergence of the “Optimization” theme, along with its connections to other keywords and themes, highlights the increasing role of optimization techniques in the field of social network anonymization. This trend highlights the ongoing efforts to develop more effective and efficient methods for protecting users’ privacy in social networks by leveraging various optimization strategies.

In the last subperiod, the “Optimization” keyword appears within the “Artificial Neural Network (ANN)” theme, signifying the increasing use of optimization techniques in the context of ANNs for social network anonymization. The keyword has connections with keywords like “Cuckoo Optimization Algorithm (COA),” “Graph Neural Network (GNN),” “Backpropagation Algorithm,” and “High Degree Nodes,” illustrating the diverse optimization approaches being integrated with neural networks to address privacy challenges in social networks.

Besides, in this subperiod, the emergence of “Genetic Algorithm (GA)” as a theme, along with its connections to related keywords like “Particle Swarm Optimization Algorithm (PSO),” “Hybrid Algorithms,” “Generalization Approach,” and “Edge Addition/Deletion,” highlights the growing importance of optimization techniques in social network anonymization field. The diverse connections between GA and these keywords demonstrate the ongoing efforts to develop more effective privacy-preserving methods by integrating GA with various meta-heuristics, generalization techniques, and graph modification approaches to address privacy challenges in social networks.

Also, the placement of “Combinatorial Optimization” within the “Synthetic Graph Generation” theme, along with its external connections to “Differential Privacy” and “Machine Learning” keywords, illustrates the diverse applications of combinatorial optimization in the field of social network anonymization. Emphasizing its relevance in generating synthetic graphs that maintain user privacy, combinatorial optimization techniques are applied in conjunction with differential privacy mechanisms and machine learning-based approaches to develop more effective privacy-preserving solutions. These interdisciplinary connections highlight the ongoing efforts to enhance privacy protection in social networks by integrating combinatorial optimization with various techniques and methodologies.

The emergence of the “Firefly Algorithm (FA)” as a theme in social network anonymization is notable in this subperiod. As a meta-heuristic algorithm, it has the potential for complex optimization problems. The “Firefly Algorithm (FA)” theme is connected to other important concepts, such as “Fuzzy Clustering” and “Identity Disclosure.” The connection between “Fuzzy Clustering” and the “Firefly Algorithm (FA)” theme implies that the algorithm was applied in optimizing fuzzy clustering techniques to enhance the effectiveness of social network anonymization. Also, the connection between “Identity Disclosure” and the “Firefly Algorithm (FA)” theme suggests that the algorithm was used to mitigate the risk of identity disclosure by optimizing anonymization techniques that balance data utility and privacy preservation. Additionally, it has external links with relevant keywords, including “Optimization,” “K-anonymity,” “Information Loss,” and “Clustering Algorithms.” These connections highlight its relevance in privacy preservation and optimizing social network anonymization techniques.

In the current subperiod, the “Neighborhood Attraction Firefly Algorithm (NAFA)” keyword is associated with the “F-factors” theme in social network anonymization research. This algorithm, which belongs to the metaheuristic optimization category, is utilized to optimize privacy-preserving techniques within the field. Moreover, external connections to other keywords, such as “Graph Modification Approach,” “Structural Properties,” “Utility,” and “Anonymization,” suggest potential applications of the NAFA algorithm in enhancing privacy through graph modification, maintaining the structural properties of social networks during anonymization, optimizing the balance between privacy protection and data utility, and improving the overall anonymization process. Taken together, these connections highlight the importance of the NAFA algorithm in optimizing social network anonymization techniques and protecting user privacy.

Another intriguing pattern in the provided evolution map is the “Artificial Intelligence (AI)” keyword, which first emerged as a theme within the subperiod of 2011-2014. This theme exhibits a connection to the “Machine Learning” keyword, indicating a strong relationship between these two concepts. Additionally, the “Artificial Intelligence (AI)” theme displays external connections to other themes in the field. Specifically, there are links to the keywords “Anonymization”, “K-anonymity”, and “Clustering Algorithms”, suggesting potential areas of intersection between these keywords. This pattern provides valuable insight into the evolving trends and themes within the field, highlighting the emergence of AI as a central concept in the 2011-2014 subperiod. The connections between “Artificial Intelligence (AI)” and “Machine Learning”, as well as other key themes, suggest potential avenues for further exploration and research in the field.

In the subsequent subperiod, 2015-2018, the “Artificial Intelligence (AI)” keyword has already established itself as a prominent theme. Notably, this keyword exhibits connections to several other keywords, including “Adaptive Random Walk”, “Supervised Machine Learning Algorithms”, “Friendship Attack”, and “DBSCAN Algorithm”. These connections suggest potential overlap and intersection between different subthemes within the field. Specifically, the link to “Adaptive Random Walk” highlights its potential utility as an AI-based approach in the context of social network anonymization. The connection to “Supervised Machine Learning Algorithms” points to the ongoing importance of machine learning in developing AI-based privacy-preserving methods. Additionally, the links to “Friendship Attack” and “DBSCAN Algorithm” suggest potential research areas related to privacy and security in the context of social networks.

Moreover, the “Machine Learning” keyword is placed within the “Adversarial Machine Learning” theme, reflecting the growing importance of security and privacy in developing machine learning algorithms for social network anonymization. This keyword exhibits connections to several other keywords, including “Privacy Preservation”, “Supervised Machine Learning”, and “Social

Relationships”. The links between “Machine Learning” and these keywords provide valuable insights into the ongoing trends and themes within the field of social network anonymization. They highlight the importance of protecting social network users’ privacy through anonymization techniques that utilize machine learning algorithms. Additionally, it is crucial to keep user data private when analyzing social networks using machine learning algorithms. The connections also emphasize the importance of supervised learning algorithms in developing machine learning algorithms for social network anonymization.

In the next subperiod, 2019-2022, the “Artificial Intelligence (AI)” keyword is placed within the “Structured Data” theme. This placement suggests a shift towards using AI techniques to analyze and manage structured data in the context of social network anonymization. Within this theme, the “Artificial Intelligence (AI)” keyword has connections to several other keywords, including “Assortativity”, “Sequential Clustering”, and “Node Degree”. These connections suggest potential research areas related to the use of AI techniques for identifying patterns and structures within social network data. Furthermore, the “Artificial Intelligence (AI)” keyword has external connections to several other themes in the field, including “Hierarchical Clustering”, “K-degree”, “Information Loss”, “Anonymization”, “Differential Privacy”, and “Clustering Algorithms”. These connections suggest that there is growing interest in using AI techniques for addressing issues related to privacy and security in the context of social network anonymization, as well as developing new clustering and anonymization algorithms.

Moreover, the term “Machine Learning” is positioned within the “Matrix Algebra” theme. This keyword is associated with concepts like “Adaptive Random Walk”, “Generative Adversarial Network (GAN)”, “Feature Learning”, “Random Projection Algorithm”, and “Random Perturbation” within the specified theme. Additionally, it has external connections to “Differential Privacy”, “Structural Properties”, “Synthetic Graph Generation”, “Graph Generation Model”, and “Randomization Approach”. The connection among machine learning techniques and matrix algebra, adaptive random walks, GANs, feature learning, random projection algorithms, and random perturbation shows that the researchers used them together to protect user privacy while preserving the overall graph structure. These techniques have external connections to concepts like differential privacy, structural properties, synthetic graph generation, graph generation models, and randomization approaches, which together contribute to creating synthetic social network graphs that maintain privacy and data utility for analysis.

In the current subperiod, the “Adversarial Machine Learning” keyword is placed within the “Singular Value Decomposition (SVD)” theme. It connects to “Structural Attacks”, “Matrix Decomposition”, and “Markov Clustering (MCL)” keywords within this theme. The mentioned connections can be interpreted as follows. In the context of social network anonymization, understanding structural attacks helps researchers develop strategies to counteract them using adversarial machine learning approaches, thereby strengthening the privacy of the anonymized network. In adversarial machine learning for social network anonymization, matrix decomposition techniques like SVD can be employed to process and analyze complex graph data while mitigating the impact of adversarial perturbations. Besides, in the context of adversarial machine learning, MCL can help develop robust privacy-preserving methods that can withstand attacks aimed at exploiting community structures in social networks.

Additionally, the “Adversarial Machine Learning” links externally to keywords, such as “Graph Modification Approach”, “Differential Privacy”, “Clustering Algorithms”, “Graph Isomorphism”, “Fuzzy Set Theory”, and “Neuro Fuzzy”. These connections demonstrate the key role of adversarial machine learning in protecting user privacy and maintaining data utility for analysis in social network anonymization.

It is noteworthy that the term "Supervised Machine Learning" emerged as a distinct theme during the 2019-2022 subperiod. This concept is intrinsically linked to the "Health Information" keyword, highlighting its applicability in the privacy-preserving analysis of health-related social networks. Furthermore, it exhibits external connections with keywords like "Clustering Algorithms", "Artificial Neural Network (ANN)", "Anonymization", "K-anonymity", "Graph Neural Network (GAN)", and "Backpropagation Algorithm". These connections demonstrate the interdisciplinary nature of supervised machine learning in addressing the challenges of privacy preservation and data utility within social network anonymization. In other words, these external connections highlight the versatility of supervised machine learning and its ability to contribute to various aspects of social network anonymization, ranging from privacy preservation and data utility to advanced modeling and optimization.

- **Alluvial Diagram**

In this subsection, an alluvial diagram is presented to show the flow or transition of keywords between different categories. In the context of this study on social network anonymization, the alluvial diagram demonstrates how the authors' keywords have changed over time and how they relate to each other. The diagram consists of a series of vertical bars or columns, each representing one of the four subperiods that was defined previously. The columns are connected by horizontal lines, and the width of the lines represents the number of keywords that are shared between adjacent subperiods. The PageRank score of each keyword is calculated in each block, demonstrating the importance of nodes (keywords) within the network (all the keywords in each module, along with their PageRank scores, are presented in Appendix B). By analyzing the PageRank scores of the keywords within each block, we gained insight into the most influential themes and concepts in the social network anonymization field, as well as an understanding the relationships and connections between these themes. It is worth noting that by analyzing the alluvial diagram in Fig. 27, insight into the evolution of the focus of research on social network anonymization over time is gained. It helps us understand which keywords have remained consistently popular throughout all four subperiods that have become less common over time and which have emerged as new focus areas in more recent years. Also, it is helpful in the way that how different keywords are related to each other and how they cluster together into more prominent themes or topics.

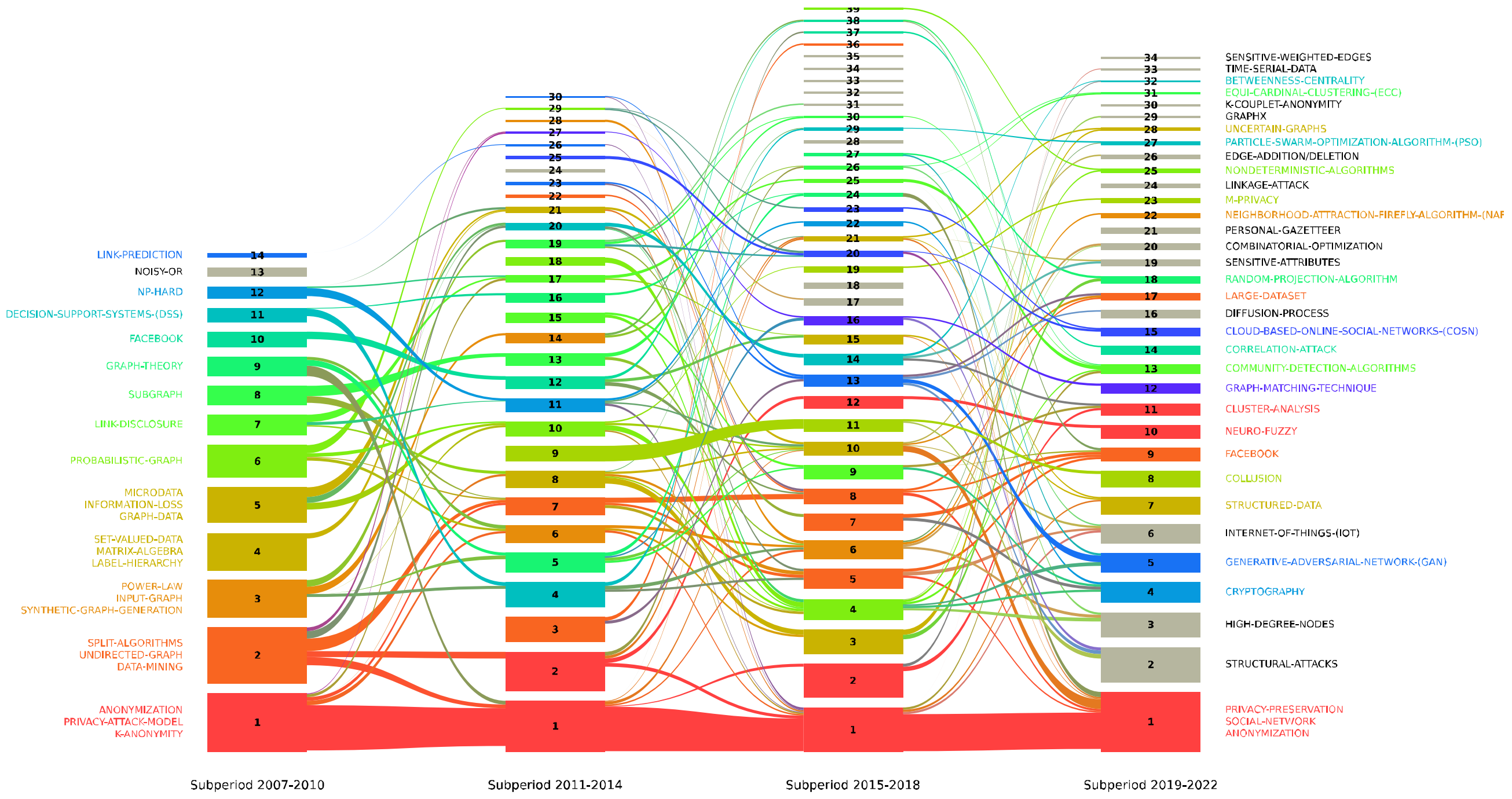

*Figure 27. The Alluvial Diagram of the Social Network Anonymization Field*

From the depicted alluvial diagram, some of the most prominent keywords of the social network anonymization field, along with their PageRank scores, are presented in Table 7. In the following paragraphs, these keywords are analyzed regarding their positions and relations in the alluvial diagram.

*Table 7. Some of the Most Prominent Keywords' Module Placements and PageRank Scores Regarding the Alluvial Diagram*

| Keyword Name | 2007-2010 | | 2011-2014 | | 2015-2018 | | 2019-2022 | |
|---|---|---|---|---|---|---|---|---|
| | Module | PageRank | Module | PageRank | Module | PageRank | Module | PageRank |
| Social Network | 1 | 0.0026 | 1 | 0.0201 | 1 | 0.0152 | 1 | 0.0137 |
| Privacy Preservation | 1 | 0.0121 | 1 | 0.0191 | 1 | 0.0147 | 1 | 0.0152 |
| Anonymization | 1 | 0.0307 | 1 | 0.0150 | 1 | 0.0112 | 1 | 0.0109 |
| K-anonymity | 1 | 0.0197 | 1 | 0.0101 | 1 | 0.00802 | 1 | 0.00616 |
| Clustering Algorithms | 2 | 0.00 | 2 | 0.00987 | 1 | 0.00689 | 1 | 0.00608 |
| Differential Privacy | 3 | 0.00 | 5 | 0.00979 | 24 | 0.00607 | 1 | 0.00719 |
| Generalization Approach | 2 | 0.0166 | 2 | 0.00830 | 21 | 0.00401 | 1 | 0.00436 |
| L-diversity | - | - | 20 | 0.00526 | 14 | 0.00579 | 11 | 0.00503 |
| Graph Modification Approach | 3 | 0.00 | 10 | 0.00516 | 10 | 0.00532 | 1 | 0.00714 |
| Randomization Approach | 7 | 0.00410 | 8 | 0.00750 | 3 | 0.00569 | 18 | 0.00449 |
| K-degree | - | - | 11 | 0.00442 | 15 | 0.00478 | 7 | 0.00310 |
| Perturbation Techniques | - | - | 7 | 0.0102 | 8 | 0.00666 | 1 | 0.00623 |
| Uncertain Graphs | - | - | 1 | 0.00170 | 21 | 0.00361 | 28 | 0.00325 |
| Machine Learning | - | - | 30 | 0.00143 | 13 | 0.00568 | 5 | 0.00797 |
| Genetic Algorithm (GA) | - | - | 1 | 0.00382 | 6 | 0.00485 | 27 | 0.00307 |
| Artificial Intelligence (AI) | - | - | 30 | 0.00249 | 1 | 0.00691 | 7 | 0.00538 |
| Optimization | - | - | 6 | 0.00683 | 6 | 0.00602 | 3 | 0.00864 |

From Fig. 27, it can be understood that the “Social Network” keyword consistently stays in module 1 throughout the four subperiods, with fluctuations in its PageRank score. Analyzing these changes offers insights into the keyword’s evolving significance and relevance in the social network anonymization field. In the 2007-2010 subperiod, “Social Network” was in module 1 out of 14 modules with a 0.00206 PageRank score, indicating its relevance but limited influence and centrality compared to other keywords. From 2011-2014, the keyword remained in 1 out of 30 modules but experienced a substantial increase in its PageRank score to 0.0201, suggesting a rise in importance and influence, making it a more central and prominent keyword. In the 2015-2018 subperiod, the keyword stayed in module 1 out of 39 modules with a slightly lower PageRank score of 0.0152, signifying its continued significance but reduced importance and centrality. In the final subperiod, 2019-2022, the keyword remained in module 1 out of 34 modules, with a further decline in PageRank score to 0.0137, indicating its ongoing relevance but decreasing prominence. The “Social Network” keyword consistently appears in module 1 across all subperiods, highlighting its sustained relevance to a specific cluster of keywords. Its importance and influence within the social network anonymization field have varied, with a notable increase in the second subperiod and a gradual decline in subsequent subperiods; nevertheless, it is the base of the field and remains a significant keyword.

As for the “Privacy Preservation” keyword, the alluvial diagram shows that it consistently remained in module 1 across all four subperiods, with slight fluctuations in its PageRank score. In the 2007-2010 subperiod, “Privacy Preservation” was in module 1 out of 14 modules with a 0.0121 PageRank score, indicating its significance, notable influence, and centrality compared to other keywords. From 2011-2014, the keyword remained in module 1 out of 30 modules, with an increased PageRank score of 0.0191, suggesting a rise in importance and influence, making it an even more central and prominent keyword. In the 2015-2018 subperiod, the keyword stayed in module 1 out of 39 modules with a slightly lower PageRank score of 0.0147, signifying its continued significance but reduced importance and centrality. In the final subperiod, 2019-2022, the keyword remained in module 1 out of 34 modules with a marginally increased PageRank score of 0.0152, indicating a slight resurgence in importance while maintaining its relevance and association with the same cluster of keywords. Consequently, the “Privacy Preservation” keyword consistently appears in module 1 across all subperiods, signifying its enduring relevance to a particular cluster of keywords in the social network anonymization field. The keyword’s importance and influence have experienced minor fluctuations, with an increase in the second subperiod followed by a slight decline and subsequent recovery in the following subperiods. Overall, the “Privacy Preservation” keyword remains a significant theme within the field.

Pertaining to the “Anonymization” keyword, it is apparent that this keyword persistently occupies module 1 across the four subperiods, while its PageRank score exhibits a progressive decline. In the 2007-2010 subperiod, “Anonymization” resided in module 1 out of 14 modules with a PageRank score of 0.0307, signifying that it was a highly consequential and influential theme in the field, showcasing considerable prominence and centrality in comparison to other keywords. In the 2011-2014 subperiod, the keyword maintained its position in module 1 out of 30 modules but with a reduced PageRank score of 0.0150. While this indicates a decrease in importance and influence during this period, it retained its status as a central and pertinent keyword. In the 2015-2018 subperiod, the keyword was in module 1 out of 39 modules, with a further diminished PageRank score of 0.0112, suggesting that the “Anonymization” keyword’s significance and centrality within the domain keep getting less, albeit remaining an essential keyword. In the 2019-2022 subperiod, the keyword persisted in module 1 out of 34 modules, with a marginally lower PageRank score of 0.0109, denoting the ongoing relevance of the “Anonymization” keyword in the field, despite its sustained decline in prominence and influence throughout the examined timeframe. Consequently, the “Anonymization” keyword consistently appears in module 1 across all subperiods, underscoring its lasting relevance and association with a specific group of keywords in the social network anonymization domain. The keyword’s importance and influence have steadily declined, as demonstrated by the descending PageRank scores. However, the “Anonymization” keyword remains a notable and consistently pertinent keyword within the field.

Regarding the “K-anonymity” keyword, the alluvial diagram demonstrates that it consistently presented in module 1 throughout the four subperiods, while its PageRank score undergoes a gradual decline. In the 2007-2010 subperiod, “K-anonymity” was situated in module 1 out of 14 modules with a PageRank score of 0.0197, signifying that it was a highly consequential and influential keyword in the field, showcasing considerable prominence and centrality relative to other keywords. In the 2011-2014 subperiod, the keyword retained its position in module 1 out of 30 modules but with a reduced PageRank score of 0.0101, indicating a decrease in importance and influence during this period, yet it remains a central and pertinent topic. In the 2015-2018 subperiod, the keyword remained in module 1 out of 39 modules, with a further diminished PageRank score of 0.00802, suggesting that the “K-anonymity” keyword’s significance and centrality within the domain continue to wane, albeit remaining an essential topic. In the 2019-2022 subperiod, the keyword persists in module 1 out of 34 modules, with a marginally lower PageRank score of 0.00616, denoting the ongoing relevance of the “K-anonymity” keyword in the field, despite its sustained decline in prominence and influence throughout the examined timeframe. In summary, the “K-anonymity” keyword consistently appears in module 1 across all subperiods, underscoring its lasting relevance and association with a specific group of keywords in the social network anonymization domain. Although the importance and influence of “K-anonymity” have steadily declined, as demonstrated by the descending PageRank scores, it remains a notable and consistently pertinent topic within the field.

Pertaining to the “Clustering Algorithms” keyword, it is clear that across the four subperiods, this keyword undergoes alterations in its module placement and PageRank score. In the 2007-2010 subperiod, the “Clustering Algorithms” keyword was situated in module 2 out of 14 modules with a PageRank score of 0.0, suggesting minimal impact and centrality in the field and a lack of substantial connection to the themes of that subperiod. In the 2011-2014 subperiod, the keyword retained its position in module 2 out of 30 modules but sees a considerable rise in its PageRank score to 0.00987, indicating an increase in importance and influence, thus becoming more central and relevant within its associated cluster. During the 2015-2018 subperiod, the keyword transitioned to module 1 out of 39 modules with a marginally lower PageRank score of 0.00689, signifying a change in its thematic association and aligning with a different cluster of related keywords. Although the keyword’s importance has slightly decreased compared to the previous subperiod, it continues to be a relevant theme in the field. In the 2019-2022 subperiod, the keyword maintained its position in module 1 out of 34 modules, with a marginally lower PageRank score of 0.00608. This suggests that the “Clustering Algorithms” keyword persists as a significant keyword in the field, albeit with a slight reduction in prominence and influence. Hence, this keyword has undergone considerable changes in both module placement and PageRank score throughout the subperiods, illustrating its shifting importance and thematic connections within the social network anonymization domain. The keyword’s significance and influence have grown since the initial subperiod, and despite minor declines in prominence in the latter subperiods, it remains a consistently relevant topic in the field.

With respect to the “Differential Privacy” keyword, significant variations in both module placement and PageRank score are observed across the four subperiods. In the 2007-2010 subperiod, “Differential Privacy” was located in module 3 out of 14 modules with a PageRank score of 0.0, indicating minimal influence and centrality in the field and no substantial connection to the keywords of that subperiod. In the 2011-2014 subperiod, the keyword transitioned to module 5 out of 30 modules, with a notable increase in its PageRank score to 0.00979, suggesting an enhancement in importance and influence and becoming a more central keyword within its associated module. In the 2015-2018 subperiod, the keyword shifted to module 24 out of 39 modules, with a marginally lower PageRank score of 0.00607, signifying a change in its thematic association and alignment with a different cluster of related keywords. In the 2019-2022 subperiod, the keyword moved to module 1 out of 34 modules, with an increased PageRank score of 0.00719, indicating another shift in its thematic association and now being part of a different, potentially more central cluster of keywords. The keyword's prominence and influence have also risen during this period. In sum, the "Differential Privacy" keyword has experienced notable changes in both module placement and PageRank score throughout the subperiods, reflecting its evolving importance and thematic connections within the social network anonymization domain. The keyword's significance and influence have grown since the initial subperiod, and it has undergone shifts in thematic associations, ultimately becoming part of a more central cluster in the final subperiod.

Regarding the "Generalization Approach" keyword, the diagram illustrates that across the four subperiods, the module placement and PageRank score of this keyword underwent variations. In the 2007-2010 subperiod, the keyword was assigned to module 2 out of 14 modules with a PageRank score of 0.0166, signifying its relative importance and influence during this time. In the 2011-2014 subperiod, the keyword remained in module 2 out of 30 modules but with a reduced PageRank score of 0.00830, indicating that although still relevant, its impact and influence have lessened. In the 2015-2018 subperiod, the keyword transitioned to module 21 out of 39 modules with a further decreased PageRank score of 0.00401, suggesting a continued decline in its importance and centrality and association with a different cluster of related keywords. In the 2019-2022 subperiod, the keyword was located in module 1 out of 34 modules with a slightly increased PageRank score of 0.00436, implying a minor resurgence in importance, albeit not as prominent as during the initial subperiod. Overall, the "Generalization Approach" keyword experienced a decline in importance and influence over time, as indicated by the decreasing PageRank scores. The change in module placement in the third subperiod suggests an association with a different cluster of related keywords, with a slight increase in the last subperiod's PageRank score that hints a minor revival of interest.

Concerning the "L-diversity" keyword, the alluvial diagram reveals that across the four subperiods, this keyword emerged and underwent changes in both module placement and PageRank score. In the 2011-2014 subperiod, "L-diversity" appeared for the first time in module 20 out of 30 modules with a PageRank score of 0.00526, indicating its relevance, but not as a central or influential keyword. In the 2015-2018 subperiod, the keyword shifted to module 14 out of 39 modules with a slightly increased PageRank score of 0.00579, suggesting an increase in importance and a change in the thematic association. In the 2019-2022 subperiod, the keyword transitioned to module 11 out of 34 modules with a slightly decreased PageRank score of 0.00503, indicating continued relevance, albeit with a minor decline in prominence and influence and a change in the thematic association. Overall, the "L-diversity" keyword emerged in the second subperiod and experiences variations in both module placement and PageRank score. While not a central or influential theme, it has consistently remained relevant with slight fluctuations in prominence and influence.

With respect to the "Graph Modification Approach," it is apparent that over the four subperiods, the module placement and PageRank score of this keyword were considerably altered. In the subperiod spanning from 2007 to 2010, the term "Graph Modification Approach" was situated in module 3 out of 14 modules, accompanied by a PageRank score of 0.0, signifying minimal influence and centrality in the field and negligible connections to the keywords present during this subperiod. Moving to the subperiod between 2011 to 2014, the keyword migrated to module 10 out of 30 modules and witnesses a considerable augmentation in its PageRank score, reaching 0.00516. This infers that the “Graph Modification Approach” keyword has acquired prominence and sway within the field, evolving into a more central and pertinent keyword inside its affiliated cluster. During the 2015-2018 subperiod, the keyword remained in module 10, but the number of modules in this subperiod is 39 modules. Also, the keyword experienced a slight increase in its PageRank score to 0.00532, indicating that the “Graph Modification Approach” keyword continues to preserve its significance and centrality in the field. In the 2019-2022 subperiod, the keyword transitioned to module 1 out of 34 modules with a heightened PageRank score of 0.00714, suggesting a substantial shift in its thematic association, now existing in a distinct and potentially more central cluster of keywords. The keyword’s prominence and influence also grew during this period. In summation, since the first subperiod, the “Graph Modification Approach” keyword’s relevance and impact have amplified, experiencing shifts in thematic associations and eventually integrating into a more central cluster in the final subperiod.

As depicted in the alluvial diagram, the “Randomization Approach” keyword also experienced a shift in module placement and PageRank score throughout the four subperiods. In the 2007-2010 subperiod, the “Randomization Approach” keyword was positioned in module 7 out of 14 modules, with a PageRank score of 0.00410. During this period, the keyword exhibited moderate influence and centrality in the field, being part of a middle-ranked module, signifying that it was not among the most dominant nor the least significant themes. In the 2011-2014 subperiod, the keyword transitioned to module 8 out of 30 modules, with an increased PageRank score of 0.00750. This suggests that the “Randomization Approach” keyword has gained importance and influence within the field during this period. Its position in module 8, a relatively higher-ranked module, indicates that it has become a more central and relevant topic within its associated cluster. In the 2015-2018 subperiod, the keyword moved to module 3 out of 39 modules, with a slightly decreased PageRank score of 0.00569. This indicates that the “Randomization Approach” keyword experienced a significant shift in its thematic association, moving to a more central cluster of keywords. Although its importance and centrality have somewhat diminished compared to the previous subperiod, it remains a relevant and influential keyword within the field. In the 2019-2022 subperiod, the keyword transitioned to module 18 out of 32 modules, with a decreased PageRank score of 0.00449. This suggests that the “Randomization Approach” keyword has undergone yet another shift in its thematic association, now being part of a relatively lower-ranked cluster of keywords. The keyword’s prominence and influence have also diminished during this period, although it still maintains some relevance within the field. As a result, the “Randomization Approach” keyword’s significance and influence have varied across the subperiods, with the most prominent period being 2011-2014. Despite the decrease in prominence and influence in the last subperiod, the “Randomization Approach” keyword remains a relevant topic within the field.

The alluvial diagram reveals that over the four subperiods, the “K-degree” keyword experienced alterations in both its module placement and PageRank score. During the 2007-2010 subperiod, the “K-degree” keyword had not appeared yet, indicating that it was not a pertinent theme or notion in the social network anonymization domain during that time. In the 2011-2014 subperiod, the “K-degree” keyword surfaced and was situated in module 11 out of 30 modules, accompanied by a PageRank score of 0.00442. This suggests that the keyword evolved into a relevant topic in the field, though it has not yet reached the status of the most central or influential topic, as evidenced by its position in a relatively lower-ranked module. In the 2015-2018 subperiod, the keyword transitioned to module 15 out of 39 modules, with a marginally increased PageRank score of 0.00478. This implies that the "K-degree" keyword became important and influential in the field during this time. Nevertheless, its location in module 15, a middle-ranked module, indicates that it had not yet become one of the most dominant keywords in the field. In the 2019-2022 subperiod, the keyword moved to module 7 out of 32 modules, with a reduced PageRank score of 0.00310. This denotes that the "K-degree" keyword encountered a shift in its thematic association, progressing to a comparatively higher-ranked module. However, the decrease in its PageRank score signals a decline in its prominence and influence within the field. Overall, the "K-degree" keyword made its appearance in the second subperiod and has consistently preserved its relevance within the field since its emergence with minor variations in prominence and influence, despite not ranking among the most central or influential themes.

As depicted in the alluvial diagram, the "Perturbation Techniques" keyword appeared and underwent alterations in both its module placement and PageRank score throughout the four subperiods. In the 2011-2014 subperiod, the "Perturbation Techniques" keyword was presented for the first time and was positioned in module 7 out of 30 modules, with a PageRank score of 0.0102. This suggests that the keyword evolved into a relevant and influential theme in the field during this period. Its location in a comparatively higher-ranked module implies that it became a more central and notable theme within its related cluster. In the 2015-2018 subperiod, the keyword transitioned to module 8 out of 39 modules, with a reduced PageRank score of 0.00666. This implies that the "Perturbation Techniques" keyword experienced a minor decline in importance and influence within the field during this period. Nevertheless, its placement in module 8, a relatively higher-ranked module, denotes that it continues to be a relevant and central theme within its associated cluster. In the 2019-2022 subperiod, the keyword moved to module 1 out of 32 modules, with a marginally decreased PageRank score of 0.00623. This demonstrates that the "Perturbation Techniques" keyword has undergone a substantial shift in its thematic association, advancing to the highest-ranked module. This alteration suggests an enhancement in the keyword's prominence and centrality within the field, incorporating it into a more central cluster of keywords. In sum, the "Perturbation Techniques" keyword materialized in the second subperiod and, since its appearance, has consistently been relevant and influential with its prominence and centrality notably increasing in the final subperiod.

As demonstrated in the alluvial diagram, the "Uncertain Graph" keyword arose and underwent alterations in both its module placement and PageRank score across the four subperiods. In the 2011-2014 subperiod, the "Uncertain Graph" keyword appeared for the first time and was positioned in module 1 out of 30 modules, with a PageRank score of 0.00170. This suggests that the keyword has evolved into a relevant theme in the field during this period. Nevertheless, its relatively low PageRank score implies that it had not yet reached the status of the most central or influential themes in the field. In the 2015-2018 subperiod, the keyword transitioned to module 21 out of 39 modules, with an increased PageRank score of 0.00361. This implies that the "Uncertain Graph" keyword had become significant during this period. However, its location in module 21, a lower-ranked module, denotes that it has not yet become one of the most dominant themes in the field. In the 2019-2022 subperiod, the keyword moved to module 28 out of 32 modules, with a marginally decreased PageRank score of 0.00325. This denotes that the "Uncertain Graph" keyword encountered a shift in its thematic association, progressing to a comparatively lower-ranked module. The slight decrease in its PageRank score signals a decline in its prominence and influence within the field. In short, the "Uncertain Graph" keyword made its appearance in the second subperiod and, although not ranked among the most central or influential themes, has consistently preserved its relevance within the field since its emergence with minor variations in prominence and influence.

As illustrated in the alluvial diagram, the "Machine Learning" keyword appeared and underwent alterations in both its module placement and PageRank score across the four subperiods. In the 2011-2014 subperiod, the "Machine Learning" keyword emerged and was positioned in module 30 out of 30 modules, with a PageRank score of 0.00143. This suggests that the keyword evolved into a relevant theme in the field during this period. Nevertheless, its location in the lowest-ranked module and its relatively low PageRank score imply that it had not yet reached the status of the most central or influential themes in the field. In the 2015-2018 subperiod, the keyword transitioned to module 13 out of 39 modules, with an increased PageRank score of 0.00568. This implies that the "Machine Learning" keyword garnered significance during this period. Its location in module 13, a mid-ranked module, denotes that it became more central and significant within its associated cluster. In the 2019-2022 subperiod, the keyword moved to module 5 out of 32 modules, with an increased PageRank score of 0.00797. This denotes that the "Machine Learning" keyword encountered a considerable shift in its thematic association, progressing to a comparatively higher-ranked module. The increase in its PageRank score signals a rise in its prominence and influence within the field, making it part of a more central cluster of keywords. In summary, the "Machine Learning" keyword made its appearance in the second subperiod, and from its emergence, the keyword consistently gained prominence and influence with substantial enhancements in its centrality and significance in the last two subperiods.

As shown in the alluvial diagram, the "Genetic Algorithm (GA)" keyword first appeared in the second subperiod (2011-2014) within the social network anonymization field and underwent transformation over time, with its prominence and module placement shifting across the subperiods. In the mentioned subperiod, GA made its debut and was positioned in module 1 out of 30 modules, with a 0.00382 PageRank score. Its location in the first module indicates that it gained some relevance or importance during this period, contributing to research in the domain. In the 2015-2018 subperiod, GA was situated in module 6 out of 39 modules, with a 0.00485 PageRank score. The rise in its PageRank score denotes that the importance of this keyword increased, although it now resides in a lower-ranked module. This could suggest that while GA remains pertinent, the field is growing, and alternative research areas or methodologies are gaining prominence. In the 2019-2022 subperiod, GA was positioned in module 27 out of 32 modules, with a 0.00307 PageRank score. The keyword transitioned to a considerably lower-ranked module, and its PageRank score diminished, indicating a decline in its significance within the field during this period. This could be attributed to other approaches or techniques receiving more attention or the research focus shifting to distinct facets of social network anonymization. Overall, the "Genetic Algorithm (GA)" keyword initially surfaced as a relevant notion in the social network anonymization field during the second subperiod and garnered importance in the third subperiod. Nevertheless, its prominence waned in the fourth subperiod due to the emergence of new meta-heuristic algorithms and the evolving concentration within the field.

As depicted in the alluvial diagram, the "Artificial Intelligence (AI)" keyword appeared in the social network anonymization field during the second subperiod (2011-2014) and has since evolved, with its prominence and module placement fluctuating throughout the subperiods. In this subperiod, AI was located in module 30 out of 30 modules with a 0.00249 PageRank score. Its positioning in the final module implies that it was a relatively nascent or less crucial concept in the field during this period, contributing minimally to the research. In the 2015-2018 subperiod, AI was situated in module 1 out of 39 modules with a 0.00691 PageRank score. The substantial increase in its PageRank score and its placement in the top module indicates that the keyword experienced considerable growth in significance and pertinence. This development suggests that AI evolved into a central notion in the social network anonymization field during this period, greatly impacting research. In the 2019-2022 subperiod, AI was positioned in module 7 out of 32 modules with a 0.00538 PageRank score. Although the keyword remained highly ranked, its location in a lower-ranked module and the minor decline in its PageRank score imply that its importance in the field somewhat decreased during this period. This could be attributed to the rise in prominence of other approaches, methods, or research areas or a shift in focus within the field. As a result, the "Artificial Intelligence (AI)" keyword initially surfaced as a relatively insignificant concept in the social network anonymization field during the second subperiod but gained considerable importance in the third subperiod. While its prominence somewhat diminished in the fourth subperiod, AI continues to be a vital concept in the field, reflecting the ever-changing nature of research and AI's impact on social network anonymization.

As displayed in the alluvial diagram, the "Optimization" keyword emerged in the social network anonymization field during the second subperiod (2011-2014) and progressed over time, with its significance and module placement fluctuating throughout the subperiods. In the mentioned subperiod, "Optimization" was situated in module 6 out of 30 modules with a 0.00683 PageRank score. Its location in the sixth module implies that it gained prominence or pertinence during this period, contributing to research in the field. In the 2015-2018 subperiod, "Optimization" was positioned in module 6 out of 39 modules with a 0.00602 PageRank score. Its placement in the same module number, despite an increase in the total number of modules, suggests that the keyword retained its relevance in the field. However, the minor decline in its PageRank score may indicate that its importance in the research area somewhat diminished during this period. In the 2019-2022 subperiod, "Optimization" was situated in module 3 out of 32 modules with a 0.00864 PageRank score. The keyword shifted to a higher-ranked module, and its PageRank score increased, indicating that its significance in the field expanded during this period. This development suggests that "Optimization" has become a more central idea in the research area, potentially influencing the creation of novel methods or approaches in social network anonymization. Overall, the "Optimization" keyword initially appeared as a pertinent concept in the social network anonymization field during the second subperiod and preserved its importance in the third subperiod. In the fourth subperiod, the keyword's prominence grew, reflecting its escalating significance in the field and its potential influence on the research and development of innovative approaches within social network anonymization.

## 3-5- Social Network Anonymization Approaches

In this section, we present a new taxonomy for social network anonymization approaches. This classification is derived from our review of prior surveys discussed in the related work section, combined with the provided bibliometric analyses.

Reviewing the conducted surveys in the social network anonymization field was incredibly beneficial in proposing a new taxonomy. It helped us understand the existing taxonomies and their strengths and weaknesses. Besides, the previous surveys helped us understand the evolution of the field and the trends that have shaped it.

Furthermore, the analyses of the authors' keywords have revealed significant insights into the current trends and key focus areas in the field of social network anonymization. These analyses allowed us to identify the patterns, relationships, and clusters among keywords in the context of the existing literature and research in social network anonymization. Additionally, they provided an understanding of the evolving landscape of various themes, identifying if they are emerging or disappearing, specialized or peripheral, as well as whether they are motor themes or fundamental (basic) themes within this domain.

After interpreting the provided findings, the identified patterns and trends have indeed facilitated our understanding of the prevailing strategies and methodologies within the field. This has effectively allowed us to discern the main approaches utilized in social network anonymization. Hence, in the current study, a taxonomy for social network anonymization is presented. The proposed classification scheme incorporates the discovered trends, central areas of focus, and themes, thereby offering an updated and more comprehension of the field.

In order to accomplish this, we primarily focused on the keywords and themes specifically associated with certain approaches employed in the domain of social network anonymization. Based on the outcomes gleaned from our investigations, we identified the following key techniques in this domain: "Graph Modification Approach", "Generalization Approach", "Differential Privacy", "Uncertain Graphs", and "Cryptography". These keywords and themes serve as the cornerstones of this field. Additionally, our review of existing surveys revealed that numerous studies have combined these methods, leading to the development of hybrid strategies. Consequently, we have included an additional category labelled "Hybrid" to the classification scheme, which is illustrated in Fig. 28.

As depicted in Fig. 28, the primary approaches for social network anonymization comprise "Graph Modification", "Generalization/Clustering", "Differential Privacy", "Uncertain Graphs", "Cryptography", and "Hybrid". We will delve into each of these approaches, along with their respective subcategories, in the subsequent paragraphs.

- **Graph Modification**

This approach involves altering the structure of the social network graph to obscure the identities, relationships, attributes, or all for individuals while preserving the overall network structure. This can be done through various techniques like edge modification (addition, deletion or altering links), node modification (addition, deletion or altering nodes), or a combination of both (Casas-Roma, 2019; Casas-Roma et al., 2017b). Based on the keyword and theme analyses conducted so far, the "Graph Modification" approach can be categorized into four subcategories, including "Random Perturbation", "Constrained Perturbation", "Noise Addition", and "Hybrid".

The "Random Perturbation" inherently comes from randomization and introduces random modification into the social network data and structure. This could be achieved by employing random modification using the mentioned edge modification, node modification, or both. There are several papers in the literature that developed social network anonymization methods based on the "Random Perturbation" method (Casas-Roma, 2014; Dev, 2014; Gong et al., 2021; Guo et al., 2018; Hay et al., 2007; Kumar & Kumar, 2021; Lan, 2015; Liu et al., 2018; Masoumzadeh & Joshi, 2010; Medforth & Wang, 2011; Milani Fard & Wang, 2015; Rousseau et al., 2018; Siddaramappa et al., 2019; Ying & Wu, 2008; Yue et al., 2018; Zhang et al., 2014; Zheleva & Getoor, 2007).

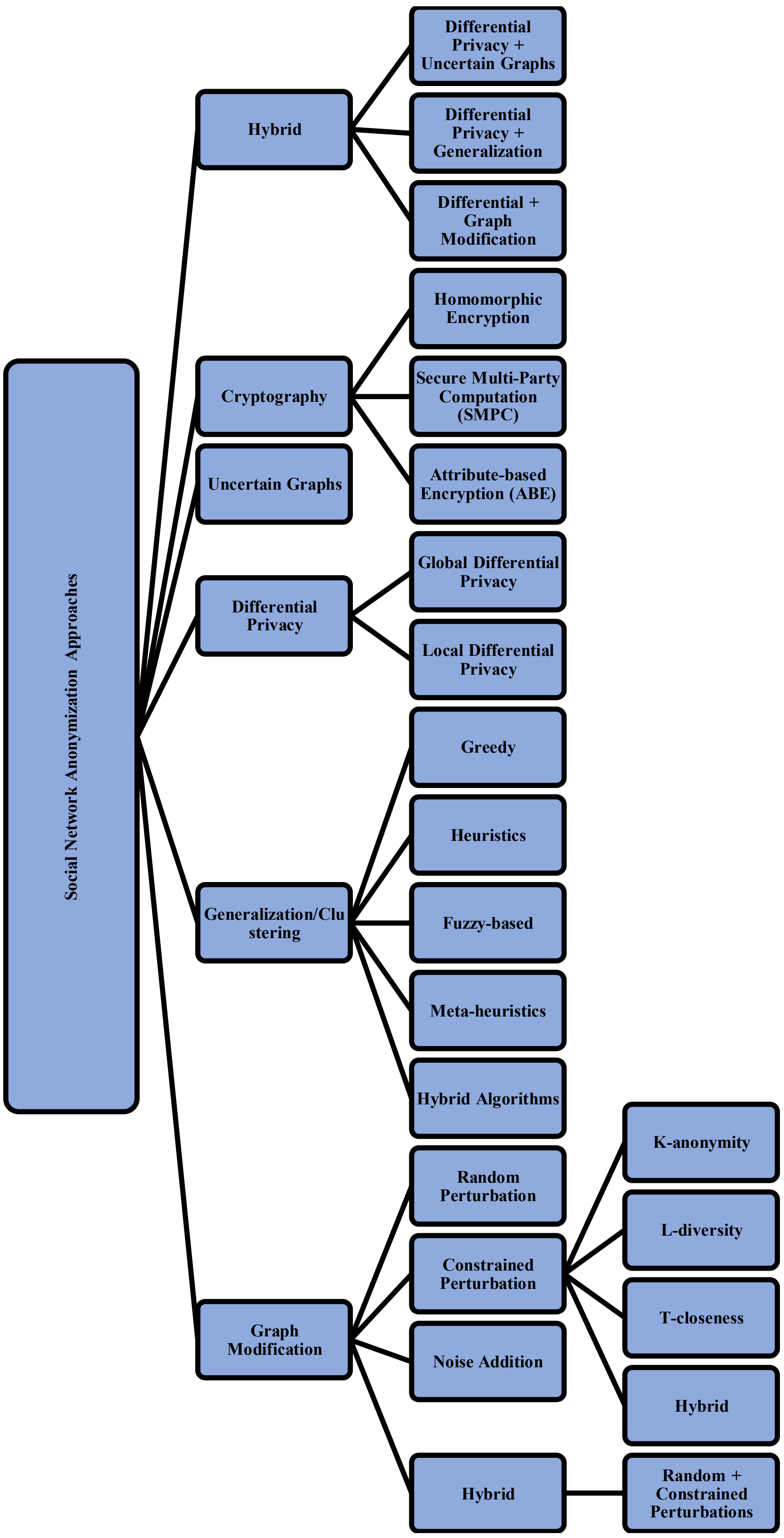

*Figure 28. Social Network Anonymization Approaches*

"Constrained Perturbation" is another variant of the "Graph Modification" approach in social network anonymization, where changes to the graph's structure are restricted by certain rules or constraints.

During constrained perturbation, modifications such like adding, deleting, or changing edges or nodes are performed, but these changes must adhere to predefined constraints to maintain certain structural properties of the original network. After examining the keywords and themes, we identified that various well-established data anonymization models, specifically “K-anonymity”, “L-diversity”, “T-closeness”, and even hybrid models, are widely used in the field of social network anonymization (Alavi et al., 2019; Baktha & Tripathy, 2018; Bensimessaoud et al., 2016; Bhattacharya & Mani, 2015; Bonchi et al., 2014; Casas-Roma et al., 2017a; Casas-Roma et al., 2019; Chakraborty & Tripathy, 2016a, 2016b; Chen et al., 2013; Cheng et al., 2010; Chester et al., 2012; Chester et al., 2011; Clarkson et al., 2010; Djomo & Djotio Ndie, 2021; Elabd et al., 2019; Fu et al., 2018; Hamideh Erfani & Mortazavi, 2019; Hamzehzadeh & Mazinani, 2019; He et al., 2009; Hu, 2015; Huang et al., 2022; Jiao et al., 2014; Kavianpour et al., 2011; Kiabod et al., 2019, 2021; Kumar & Kumar, 2017; Li et al., 2011; Li et al., 2014; Li et al., 2016; Li & Shen, 2010; Liu & Terzi, 2008; Liu et al., 2010; Liu & Li, 2013; Liu et al., 2016; Maag et al., 2014; Macwan & Patel, 2019; Macwan & Patel, 2018; Masoumzadeh & Joshi, 2012; Mauw et al., 2016; Medková, 2020; Minello et al., 2020; Mohapatra & Patra, 2019b; Mortazavi & Erfani, 2020; Ni et al., 2013; Ninggal & Abawajy, 2015; Nobari et al., 2014; Puttaswamy et al., 2009; Qing-jiang et al., 2011; Rajabzadeh et al., 2020; Rajaei et al., 2015a, 2015b; Ren et al., 2022; Safia & Yacine, 2018; Saharkhiz & Shahriari, 2011; Sarah et al., 2018; Sargolzaei et al., 2016; Singh et al., 2018; Skarkala et al., 2012; Srivastava et al., 2016; Stokes & Torra, 2012; Sun et al., 2013; Tai et al., 2011; Tripathy & Panda, 2010; Tsai et al., 2015; Tsai et al., 2012; Wang, Shih, et al., 2013; Wang et al., 2010; Wang, Tsai, et al., 2013; S.-L. Wang et al., 2011; Wang et al., 2014; Y. Wang et al., 2011; W. Wu et al., 2010; Xia, 2018; Xie et al., 2016; Yang et al., 2014; Yu et al., 2019; Yu et al., 2014; Yu et al., 2013; Yu et al., 2012; Yuan et al., 2011; Zhang, Li, et al., 2021; Zhang, Lin, et al., 2021; Zhang et al., 2019; Zhang et al., 2017; Zhang et al., 2016; Zhou & Pei, 2011; Zou et al., 2009). These models are often employed alongside graph modification techniques. Given the inherent constraints these methods have in achieving anonymization, it is justifiable to classify these studies under the “Constrained Perturbation” subcategory.

“Noise Addition” is another class of the “Graph Modification” approach to anonymize social networks. The “Noise Addition” and “Random Perturbation” methods do have similarities, specifically in the context of graph modification for social network anonymization; however, they might differ in the extent of modifications made and the type of randomness introduced. The “Noise Addition” method focuses on introducing a sufficient level of random noise to protect privacy. There is only one study in the literature that specifically used this method to add “noise” or randomness to the data using addition, deletion, or modifying edges or nodes in the graph. In this paper, the authors proposed a privacy-preserving method called seamless privacy for the preservation of the users’ privacy and their sensitive information by adding noise to the output graph (Ding et al., 2013).

Moreover, various research works in the field have devised methods for anonymizing social networks by integrating the previously mentioned techniques. For instance, some researchers merged the “Random Perturbation” and “Constrained Perturbation” strategies to create innovative methods for anonymizing social networks (Gao & Li, 2019a; Hajian et al., 2016; Liu et al., 2017; Liu et al., 2014; Nagaraj et al., 2017; Torra & Salas, 2019; Ying et al., 2009; Zhao & Li, 2019).

- **Generalization/Clustering**

The generalization approach, also recognized as the clustering-based method, works by grouping nodes and edges into clusters, called super-node and super-edge, respectively. Each of these clusters is then anonymized, which publishes collective information regarding the structural properties of its constituent nodes. This method, as outlined by Hay et al., effectively conceals detailed information about individual entities within the network (Hay et al., 2010; Hay et al., 2008).

From the insights derived through our analyses, we classified the "Generalization/Clustering" studies (Babu et al., 2013; Bhagat et al., 2009; Campan & Truta, 2008; Gangarde, Pawar, et al., 2021; Gangarde et al., 2022; Gangarde, Sharma, et al., 2021; Hay et al., 2010; He et al., 2012; Hoang et al., 2019; Jiang, 2015; Kaur & Bhardwaj, 2019; Kaveri & Maheswari, 2015; Langari et al., 2020; Liu & Yang, 2011; Mohapatra & Patra, 2017, 2019a, 2020; Ni et al., 2013; Ros-Martin et al., 2019; Shishodia et al., 2013; Siddula et al., 2018; Siddula et al., 2019; Sihag, 2012; Stokes, 2011, 2013; Su et al., 2021; Thompson & Yao, 2009; Wang et al., 2015; Yazdanjue et al., 2020; Zhang & Qu, 2010) within the social network anonymization literature into five categories. This categorization is based on the type of algorithms these studies have developed to cluster the nodes. The five categories include: "Greedy Algorithms", "Fuzzy-based Algorithms", "Heuristic Algorithms", "Meta-heuristic Algorithms", and "Hybrid Algorithms".

- **Differential Privacy**

As mentioned earlier, Differential Privacy, introduced by Dwork in 2006, is a rigorous mathematical framework for preserving privacy in data analysis and a key approach to social network anonymization. This approach concentrates on the process through which data are released rather than the raw data itself. This is accomplished by introducing a precise amount of randomness into the data publishing process. The central aim of differential privacy is to facilitate the sharing of valuable statistical data insights without compromising the privacy of the individuals within the dataset.

The results of this study indicate that differential privacy has progressively emerged as a prominent approach in the social network anonymization research field since its inception. Additionally, based on the analyses of the keywords and themes, we found that the studies related to the "Differential Privacy" approach can be categorized into "Global Differential Privacy" and "Local Differential Privacy" methods.

The "Global Differential Privacy" adds noise at the time of data queries on the entire dataset (Chen & Zhu, 2015a, 2015b; Gao & Li, 2019b; Huang et al., 2020; A. Li et al., 2020; Macwan & Patel, 2021; Zhu et al., 2019; Zhu et al., 2017). The curator of the social network has access to the original unaltered graph (nodes and edges). When information is requested from this network (like the total number of nodes, average number of connections per user, or other aggregate statistics), noise is added to the result to maintain privacy. This approach allows for more accurate results since noise is optimized across the whole dataset. However, because the original network is unaltered, there may be a higher potential risk to privacy, particularly if the curator is compromised.

On the other hand, "Local Differential Privacy" adds noise to each individual data point (such as individual nodes or edges) before it is added to the database (Gao et al., 2017, 2018; Huang et al., 2021; Ju et al., 2019; Tian et al., 2021; Zhang et al., 2018). This means that the curator never sees the raw data of any individual node or edge. This approach provides more robust privacy protection as it ensures that the information about any specific individual is obfuscated. However, because noise is added to each data point independently, this method may lead to less accurate aggregate statistics or network analyses.

- **Uncertain Graphs**

Our analysis results reveal "Uncertain Graphs" as another approach within the realm of social network anonymization. This approach refers to the technique of introducing uncertainty into the structure of the social network graph to protect user privacy. In a certain graph, the existence of each edge is known with certainty. However, in an uncertain graph, each edge is assigned a probability that indicates the likelihood of its existence. This uncertainty can help to anonymize the data by obscuring the exact relationships between individuals. In other words, the probabilities could be generated to obfuscate the true relationships and make it more difficult for an adversary to re-identify individuals or infer sensitive information. We identified some studies conducted using the "Uncertain Graphs" approach in the social network anonymization field (Boldi et al., 2012; Nguyen et al., 2014; Nguyen et al., 2015; Tian et al., 2018; Xiao et al., 2018; Yan et al., 2017; Yan et al., 2018).

- **Cryptography**

Another keyword or theme that is detected as an approach in the field of social network anonymization is "Cryptography". This approach plays an essential role in preserving privacy and security within social networks, including anonymization. Drawing from the insights gathered through our analyses, we highlight several cryptography-oriented techniques commonly employed for anonymizing social networks:

a) Homomorphic Encryption is a cryptographic technique that enables computations on encrypted data, yielding results as if the operations were performed on the original, unencrypted data. This method has significant implications for anonymizing social networks (P. Li et al., 2020; Zuo et al., 2020). It can enhance user data privacy by allowing calculations on encrypted data, thereby safeguarding users' personal information. Homomorphic encryption also facilitates secure data sharing within the network, allowing sensitive information to remain encrypted even in-use, that is accessible only to authorized entities. Furthermore, it enhances user privacy when interacting with third-party applications on the network by enabling these services to compute encrypted data, preventing them from accessing raw, unencrypted user information.

b) Secure Multi-Party Computation (SMPC) is a cryptographic protocol used in social network anonymization (Yuan et al., 2013, 2015). It enables multiple parties to collectively compute functions on their private inputs without revealing the inputs to each other, preserving individual data privacy. SMPC has various applications in social networks, including privacy-preserving data analysis, collaborative filtering for personalized recommendations, secure integration of anonymized data from multiple sources, and privacy-preserving analysis of social graph structures. These applications allow for data analysis, recommendations, data linkage, and network analysis while maintaining privacy and confidentiality. As a result, the SMPC protocols employ cryptographic techniques to ensure secure computations and simultaneously preserve individual privacy.

c) Attribute-Based Encryption (ABE) is another cryptographic scheme that enables precise access control based on attributes, benefiting social network anonymization by enhancing privacy and access control for sensitive data (Kumaran, 2021). ABE enables data encryption and decryption based on attributes instead of specific identities. Users and data are associated with attributes like roles or affiliations, and encryption policies are defined accordingly. Only users possessing the required attributes can decrypt and access the data. ABE encryption in social networks offers several applications. Selective Data Access allows users with specific attributes or roles to decrypt and access relevant information, ensuring that sensitive data are viewed only by authorized individuals. Privacy-Preserving Sharing enables the secure sharing of sensitive information within a social network by encrypting data based on attributes and restricting decryption and access to users with matching attributes. Anonymous Attribute-Based Access supports controlled data sharing while maintaining user anonymity, ensuring privacy, and granting access based on attributes.

- **Hybrid**

Furthermore, our analyses revealed that numerous studies in the field of social network anonymization have employed a combination of the aforementioned approaches. For instance, some of the authors combined the "Uncertain Graphs" and "Differential Privacy" approaches to preserve the privacy of users in a social network (Hu et al., 2017). Other authors used the "Differential Privacy" approach along with "Generalization/Clustering" to ensure user privacy in social networks (Huang et al., 2020). In another study, the authors tried to use the "Differential Privacy" and "Graph Modification" approaches to propose an efficient social network anonymization model (Shakeel et al., 2021). Besides, "Differential Privacy" and "Cryptography" were combined in another study to achieve a privacy-preserving social network publishing scheme (Zhang et al., 2022).

# 4- Future Research Trends in Social Network Anonymization

Based on the conducted analyses in the current study, we identified and summarized some of the most important research trends in the social network anonymization field.

a) As dynamic social networks continually evolve, they become more complex and interconnected, heightening the need for advanced social network anonymization approaches. These approaches are crucial in preserving users' anonymity in an ever-changing environment and mitigating risks related to cybercrimes. Besides, they should preserve the users' data privacy and ensure the ethical use of AI and machine learning in analyzing dynamic social network data, preventing misuse. Hence, developing novel anonymization methods for dynamic social networks is vital for a safer, more secure, and more inclusive online environment.

b) In today's world of interconnected digital networks, there is an urgent need for the development of new and novel social network anonymization approaches, especially those based on innovative meta-heuristic algorithms. As the complexity and diversity of online interactions increase, traditional anonymization techniques might not be sufficient to maintain privacy and security. Meta-heuristic algorithms, due to their inherent flexibility and adaptability, can offer a potential solution. They can be designed to automatically adapt to the changing topologies of dynamic social networks, effectively anonymizing data in real-time and ensuring that users' identities and personal information remain protected, regardless of how the network evolves. Furthermore, the utilization of meta-heuristic algorithms in anonymization processes can significantly enhance the efficiency and accuracy of privacy-preserving techniques. These algorithms are designed to optimize solutions for complex problems and can help in determining the most effective ways to anonymize data without losing the underlying structure and utility of the data. This is crucial as social networks often serve as rich resources for research and insights, and it is essential to preserve their utility while ensuring privacy using anonymization approaches.

c) The advent of AI and machine learning models in developing new and innovative anonymization approaches has opened up a new frontier in data privacy for social networks. For instance, suppose there is a social network that encompasses a vast user base with diverse demographic attributes. In such a scenario, traditional anonymization methods might struggle to maintain effective privacy protections due to the diversity and volume of data. AI and machine learning models, however, can efficiently manage such diversity and scale by recognizing patterns and dynamically anonymizing user data to uphold user privacy. These AI-driven approaches can provide a more efficient and scalable solution compared to traditional methods, especially when handling high-dimensional datasets with numerous attributes associated with each user. It is worth noting that deep generative models, such as GANs and Variational Autoencoders (VAE), can be beneficial in social network anonymization. These models can be used to create synthetic representations of original social network data. This synthesized data, which mimics the original data's structure and relationships without revealing individual identities, can be used for analysis or research without privacy concerns. Additionally, these deep learning models can be applied to alter the network structure subtly, making it difficult to identify individual nodes (users) while preserving the overall graph's properties. GANs, for instance, can generate synthetic nodes and edges that obscure the original identities but maintain the global network structure. Besides, AI and machine learning have a unique advantage: they can learn and improve over time. They can adapt to new privacy challenges, which is particularly important as social networks and privacy threats continue to evolve. For instance, if a new form of privacy violation emerges, these models can be quickly trained to recognize and counteract it. Overall, the development of anonymization approaches based on AI and machine learning models is essential for all types of social networks.
d) Developing new and novel social network anonymization approaches using hybrid methods, such as combining Differential Privacy with Graph Modification and integrating Uncertain Graphs with Differential Privacy, is becoming increasingly vital. Hybrid approaches have the potential to offer superior privacy protection by harnessing the strengths of different techniques. For example, Differential Privacy provides a mathematical guarantee of privacy by adding a carefully calculated amount of noise to the data. When combined with Graph Modification, which alters the structure of the network to obscure individual identities, it can create a multi-layered approach that enhances overall privacy while still maintaining the utility of the data. Similarly, combining Uncertain Graphs and Differential Privacy could offer another powerful hybrid solution. Uncertain Graphs capture the inherent uncertainty in social relationships, thereby providing an additional layer of anonymization. When coupled with Differential Privacy's mathematical safeguards, this hybrid approach can result in a more comprehensive anonymization strategy, providing robust protection against both current and emerging de-anonymization techniques.
e) Developing social network anonymization approaches based on hyper-heuristics is vital due to their adaptability, efficiency, and flexibility. As high-level problem-solving frameworks, they are able to generate solutions, ensuring effective anonymization that upholds user privacy while preserving data utility dynamically. Their ability to handle large-scale, high-dimensional data can outperform traditional methods, enhancing the quality of anonymization. Moreover, hyper-heuristics can be customized to various types of social networks and anonymization models, providing a universally applicable tool for data privacy.

# 5- Conclusion

In the era of big data, the importance of social network anonymization has become paramount. The rapid digitization of our lives has led to an explosion in the quantity of personal data shared across various platforms, heightening the need for effective anonymization approaches to ensure privacy and security. Within this context, the field of social network anonymization has emerged as a vital area of study and development, aiming to balance the benefits of data sharing with the essential right to privacy.

In this light, our study conducted a comprehensive bibliometric analysis in the burgeoning field of social network anonymization, leveraging a suite of robust analytical tools and methodologies. Our investigation revealed the primary keywords, themes, and topics prevalent in the existing literature, providing a unique insight into the evolution of these themes over time.

Moreover, we have proposed a novel taxonomy of the various approaches currently employed in social network anonymization based on our detailed analysis of collected studies. This taxonomy serves as a valuable tool for academics and practitioners alike, offering a nuanced understanding of the current landscape and guiding navigation through this field.

In anticipation of future trends, we have identified key developments that may shape the trajectory of this domain. These insights, based on our analytical findings, highlight the dynamic nature of social network anonymization and emphasize the necessity for continued research.

In summary, this paper presents a comprehensive snapshot of the present state of social network anonymization and a roadmap for its future progression. Our hope is that this work will serve as a catalyst to foster further exploration and advancements in this critical aspect of online privacy and data protection.

# Appendix A

The core papers of each detected theme by the SciMAT co-word analysis for each subperiod are provided in the following.

# Subperiod 2007-2010

- ## Relational Data
  - THOMPSON, B., YAO, D., The Union-split Algorithm And Cluster-based Anonymization Of Social Networks. 4th International Symposium On Acm Symposium On Information, Computer And Communications Security, Asiaccs'09 : 218-227 (2009). Times cited: 64
  - TRIPATHY, B.K., PANDA, G.K., A New Approach To Manage Security Against Neigborhood Attacks In Social Networks. 2010 International Conference On Advances In Social Network Analysis And Mining, Asonam 2010 : 264-269 (2010). Times cited: 42
  - HE, X., VAIDYA, J., SHAFIQ, B., ADAM, N., ATLURI, V., Preserving Privacy In Social Networks: A Structure-aware Approach. 2009 Ieee/wic/acm International Conference On Web Intelligence, Wi 2009 1: 647-654 (2009). Times cited: 24
  - YANG, C.C., THURAISINGHAM, B.M., Privacy-preserved Social Network Integration And Analysis For Security Informatics. Ieee Intelligent Systems 25:5 88-90 (2010). Times cited: 3
- ## RE-IDENTIFICATION-ATTACKS
  - HAY, M., MIKLAU, G., JENSEN, D., TOWSLEY, D., WEIS, P., Resisting Structural Re-identification In Anonymized Social Networks. Proceedings Of The Vldb Endowment 1:1 102-114 (2010). Times cited: 386
  - ZHELEVA, E., GETOOR, L., Preserving The Privacy Of Sensitive Relationships In Graph Data. 1st Acm Sigkdd International Workshop On Privacy, Security, And Trust In Kdd, Pinkdd 2007 4890 LNCS: 153-171 (2008). Times cited: 196
  - TRIPATHY, B.K., PANDA, G.K., A New Approach To Manage Security Against Neigborhood Attacks In Social Networks. 2010 International Conference On Advances In Social Network Analysis And Mining, Asonam 2010 : 264-269 (2010). Times cited: 42
  - LIU, L., WANG, J., LIU, J., ZHANG, J., Privacy Preservation In Social Networks With Sensitive Edge Weights. 9th Siam International Conference On Data Mining 2009, Sdm 2009 2: 949-960 (2009). Times cited: 41
  - LIU, L., LIU, J., ZHANG, J., WANG, J., Privacy Preservation Of Affinities In Social Networks. Iadis International Conference Information Systems 2010 : 372-376 (2010). Times cited: 17
  - CLARKSON, K.L., LIU, K., TERZI, E., Toward Identity Anonymization In Social Networks. Link Mining: Models, Algorithms, And Applications 9781441965158: 359-385 (2010). Times cited: 15
  - DANEZIS, G., AURA, T., CHEN, S., KICIMAN, E., How To Share Your Favourite Search Results While Preserving Privacy And Quality. 10th International Symposium On Privacy Enhancing Technologies, Pets 2010 6205 LNCS: 273-290 (2010). Times cited: 4

- LI, Y., SHEN, H., On Identity Disclosure In Weighted Graphs. Parallel And Distributed Computing, Applications And Technologies, Pdcat Proceedings : 166-174 (2010). Times cited: 4

- **K-ANONYMITY**
  - LIU, K., TERZI, E., Towards Identity Anonymization On Graphs. 2008 Acm Sigmod International Conference On Management Of Data 2008, Sigmod'08 : 93-106 (2008). Times cited: 562
  - CHENG, J., FU, A.W.-C., LIU, J., K-isomorphism: Privacy Preserving Network Publication Against Structural Attacks. 2010 International Conference On Management Of Data, Sigmod '10 : 459-470 (2010). Times cited: 263
  - CAMPAN, A., TRUTA, T.M., Data And Structural K-anonymity In Social Networks. 2nd Acm Sigkdd International Workshop On Privacy, Security, And Trust In Kdd, Pinkdd 2008 5456 LNCS: 33-54 (2009). Times cited: 98
  - YING, X., PAN, K., WU, X., GUO, L., Comparisons Of Randomization And K-degree Anonymization Schemes For Privacy Preserving Social Network Publishing. 3rd Workshop On Social Network Mining And Analysis, Sna-kdd '09 : - (2009). Times cited: 61
  - BHAGAT, S., CORMODE, G., KRISHNAMURTHY, B., SRIVASTAVA, D., Privacy In Dynamic Social Networks. 19th International World Wide Web Conference, Www2010 : 1059-1060 (2010). Times cited: 37
  - SINGH, L., ZHAN, J., Measuring Topological Anonymity In Social Networks. 2007 Ieee International Conference On Granular Computing, Grc 2007 : 770-774 (2007). Times cited: 28
  - HE, X., VAIDYA, J., SHAFIQ, B., ADAM, N., ATLURI, V., Preserving Privacy In Social Networks: A Structure-aware Approach. 2009 Ieee/wic/acm International Conference On Web Intelligence, Wi 2009 1: 647-654 (2009). Times cited: 24
  - PUTTASWAMY, K.P.N., SALA, A., ZHAO, B.Y., Starclique: Guaranteeing User Privacy In Social Networks Against Intersection Attacks. 2009 Acm Conference On Emerging Networking Experiments And Technologies, Conext'09 : 157-167 (2009). Times cited: 24
  - WANG, S.-L., TSAI, Y.-C., KAO, H.-Y., HONG, T.-P., Anonymizing Set-valued Social Data. 2010 Ieee/acm International Conference On Green Computing And Communications, Greencom 2010, 2010 Ieee/acm International Conference On Cyber, Physical And Social Computing, Cpscom 2010 : 809-812 (2010). Times cited: 7
  - DANEZIS, G., AURA, T., CHEN, S., KICIMAN, E., How To Share Your Favourite Search Results While Preserving Privacy And Quality. 10th International Symposium On Privacy Enhancing Technologies, Pets 2010 6205 LNCS: 273-290 (2010). Times cited: 4
  - YANG, C.C., THURAISINGHAM, B.M., Privacy-preserved Social Network Integration And Analysis For Security Informatics. Ieee Intelligent Systems 25:5 88-90 (2010). Times cited: 3
  - MASOUMZADEH, A., JOSHI, J., Preserving Structural Properties In Anonymization Of Social Networks. Proceedings Of The 6th International Conference On Collaborative Computing: Networking, Applications And Worksharing, Collaboratecom 2010 : - (2010). Times cited: 2

- ZHANG, R., QU, B., Preserving Privacy In Social Networks Against Homogeneity Attack. International Conference On Internet Technology And Applications, Itap 2010 : - (2010). Times cited: 0

- **DEGREE-SEQUENCE**
  - LIU, K., TERZI, E., Towards Identity Anonymization On Graphs. 2008 Acm Sigmod International Conference On Management Of Data 2008, Sigmod'08 : 93-106 (2008). Times cited: 562
  - HAY, M., LI, C., MIKLAU, G., JENSEN, D., Accurate Estimation Of The Degree Distribution Of Private Networks. 9th Ieee International Conference On Data Mining, Icdm 2009 : 169-178 (2009). Times cited: 211
  - THOMPSON, B., YAO, D., The Union-split Algorithm And Cluster-based Anonymization Of Social Networks. 4th International Symposium On Acm Symposium On Information, Computer And Communications Security, Asiaccs'09 : 218-227 (2009). Times cited: 64
  - YING, X., PAN, K., WU, X., GUO, L., Comparisons Of Randomization And K-degree Anonymization Schemes For Privacy Preserving Social Network Publishing. 3rd Workshop On Social Network Mining And Analysis, Sna-kdd '09 : - (2009). Times cited: 61
  - CLARKSON, K.L., LIU, K., TERZI, E., Toward Identity Anonymization In Social Networks. Link Mining: Models, Algorithms, And Applications 9781441965158: 359-385 (2010). Times cited: 15
  - LI, Y., SHEN, H., On Identity Disclosure In Weighted Graphs. Parallel And Distributed Computing, Applications And Technologies, Pdcat Proceedings : 166-174 (2010). Times cited: 4
- **STRUCTURAL-ATTACKS**
  - ZOU, L., CHEN, L., TAMER ÖZSU, M., K-automorphism: A General Framework For Privacy Preserving Network Publication. Proceedings Of The Vldb Endowment 2:1 946-957 (2009). Times cited: 312
  - CHENG, J., FU, A.W.-C., LIU, J., K-isomorphism: Privacy Preserving Network Publication Against Structural Attacks. 2010 International Conference On Management Of Data, Sigmod '10 : 459-470 (2010). Times cited: 263
  - TRIPATHY, B.K., PANDA, G.K., A New Approach To Manage Security Against Neigborhood Attacks In Social Networks. 2010 International Conference On Advances In Social Network Analysis And Mining, Asonam 2010 : 264-269 (2010). Times cited: 42
  - ZHANG, R., QU, B., Preserving Privacy In Social Networks Against Homogeneity Attack. International Conference On Internet Technology And Applications, Itap 2010 : - (2010). Times cited: 0
- **ANONYMIZATION**
  - LIU, K., TERZI, E., Towards Identity Anonymization On Graphs. 2008 Acm Sigmod International Conference On Management Of Data 2008, Sigmod'08 : 93-106 (2008). Times cited: 562

- XIAOWEI, Y., XINTAO, W., Randomizing Social Networks: A Spectrum Preserving Approach. 8th Siam International Conference On Data Mining 2008, Applied Mathematics 130 2: 739-750 (2008). Times cited: 183
- BHAGAT, S., KRISHNAMURTHY, B., CORMODE, G., SRIVASTAVA, D., Classbased Graph Anonymization For Social Network Data. Proceedings Of The Vldb Endowment 2:1 766-777 (2009). Times cited: 159
- WU, W., XIAO, Y., WANG, W., HE, Z., WANG, Z., K-symmetry Model For Identity Anonymization In Social Networks. 13th International Conference On Extending Database Technology: Advances In Database Technology - Edbt 2010 : 111-122 (2010). Times cited: 107
- YING, X., PAN, K., WU, X., GUO, L., Comparisons Of Randomization And K-degree Anonymization Schemes For Privacy Preserving Social Network Publishing. 3rd Workshop On Social Network Mining And Analysis, Sna-kdd '09 : - (2009). Times cited: 61
- BHAGAT, S., CORMODE, G., KRISHNAMURTHY, B., SRIVASTAVA, D., Privacy In Dynamic Social Networks. 19th International World Wide Web Conference, Www2010 : 1059-1060 (2010). Times cited: 37
- SINGH, L., ZHAN, J., Measuring Topological Anonymity In Social Networks. 2007 Ieee International Conference On Granular Computing, Grc 2007 : 770-774 (2007). Times cited: 28
- HE, X., VAIDYA, J., SHAFIQ, B., ADAM, N., ATLURI, V., Preserving Privacy In Social Networks: A Structure-aware Approach. 2009 Ieee/wic/acm International Conference On Web Intelligence, Wi 2009 1: 647-654 (2009). Times cited: 24
- PUTTASWAMY, K.P.N., SALA, A., ZHAO, B.Y., Starclique: Guaranteeing User Privacy In Social Networks Against Intersection Attacks. 2009 Acm Conference On Emerging Networking Experiments And Technologies, Conext'09 : 157-167 (2009). Times cited: 24
- LI, Y., SHEN, H., On Identity Disclosure In Weighted Graphs. Parallel And Distributed Computing, Applications And Technologies, Pdcat Proceedings : 166-174 (2010). Times cited: 4
- MASOUMZADEH, A., JOSHI, J., Preserving Structural Properties In Anonymization Of Social Networks. Proceedings Of The 6th International Conference On Collaborative Computing: Networking, Applications And Worksharing, Collaboratecom 2010 : - (2010). Times cited: 2

# • GENERALIZATION-APPROACH

### ▪ Core Papers

- XIAOWEI, Y., XINTAO, W., Randomizing Social Networks: A Spectrum Preserving Approach. 8th Siam International Conference On Data Mining 2008, Applied Mathematics 130 2: 739-750 (2008). Times cited: 183
- THOMPSON, B., YAO, D., The Union-split Algorithm And Cluster-based Anonymization Of Social Networks. 4th International Symposium On Acm Symposium On Information, Computer And Communications Security, Asiaccs'09 : 218-227 (2009). Times cited: 64

# Subperiod 2011-2014

- **FACEBOOK**
  - SALA, A., ZHAO, X., WILSON, C., ZHENG, H., ZHAO, B.Y., Sharing Graphs Using Differentially Private Graph Models. 2011 Acm Sigcomm Internet Measurement Conference, Imc'11 : 81-97 (2011). Times cited: 158
  - ILYAS, M.U., SHAFIQ, M.Z., LIU, A.X., RADHA, H., A Distributed And Privacy Preserving Algorithm For Identifying Information Hubs In Social Networks. Ieee Infocom 2011 : 561-565 (2011). Times cited: 26
  - DURR, M., MAIER, M., DORFMEISTER, F., Vegas - A Secure And Privacy-preserving Peer-to-peer Online Social Network. 2012 Ase/ieee International Conference On Social Computing, Socialcom 2012 And The 2012 Ase/ieee International Conference On Privacy, Security, Risk And Trust, Passat 2012 : 868-874 (2012). Times cited: 25
  - TRUTA, T.M., CAMPAN, A., RALESCU, A.L., Preservation Of Structural Properties In Anonymized Social Networks. 8th Ieee International Conference On Collaborative Computing: Networking, Applications And Worksharing, Collaboratecom 2012 : 619-627 (2012). Times cited: 14
  - SONG, Y., NOBARI, S., LU, X., KARRAS, P., BRESSAN, S., On The Privacy And Utility Of Anonymized Social Networks. 13th International Conference On Information Integration And Web-based Applications And Services, Iiwas2011 : 246-253 (2011). Times cited: 5
- **FUZZY-SET-THEORY**
  - FARD, A.M., WANG, K., YU, P.S., Limiting Link Disclosure In Social Network Analysis Through Subgraph-wise Perturbation. 15th International Conference On Extending Database Technology, Edbt 2012 : 109-119 (2012). Times cited: 20
  - STOKES, K., TORRA, V., On Some Clustering Approaches For Graphs. 2011 Ieee International Conference On Fuzzy Systems, Fuzz 2011 : 409-415 (2011). Times cited: 15
  - NETTLETON, D.F., Information Loss Evaluation Based On Fuzzy And Crisp Clustering Of Graph Statistics. 2012 Ieee International Conference On Fuzzy Systems, Fuzz 2012 : - (2012). Times cited: 3
- **SOCIAL-NETWORK**
  - ZHOU, B., PEI, J., The K-anonymity And L-diversity Approaches For Privacy Preservation In Social Networks Against Neighborhood Attacks. Knowledge And Information Systems 28:1 47-77 (2011). Times cited: 160
  - SALA, A., ZHAO, X., WILSON, C., ZHENG, H., ZHAO, B.Y., Sharing Graphs Using Differentially Private Graph Models. 2011 Acm Sigcomm Internet Measurement Conference, Imc'11 : 81-97 (2011). Times cited: 158
  - TASSA, T., COHEN, D.J., Anonymization Of Centralized And Distributed Social Networks By Sequential Clustering. Ieee Transactions On Knowledge And Data Engineering 25:2 311-324 (2013). Times cited: 94

- CASAS-ROMA, J., Privacy-preserving On Graphs Using Randomization And Edge-relevance. 11th International Conference On Modeling Decisions For Artificial Intelligence, Mdai 2014 8825: 204-216 (2014). Times cited: 13
- LI, Y., SHEN, H., On Identity Disclosure Control For Hypergraph-based Data Publishing. Ieee Transactions On Information Forensics And Security 8:8 1384-1396 (2013). Times cited: 13
- WANG, Y., XIE, L., ZHENG, B., LEE, K.C.K., Utility-oriented K-anonymization On Social Networks. 16th International Conference On Database Systems For Advanced Applications, Dasfaa 2011 6587 LNCS:PART 1 78-92 (2011). Times cited: 13
- WANG, Y., WU, X., ZHU, J., XIANG, Y., On Learning Cluster Coefficient Of Private Networks. Social Network Analysis And Mining 3:4 925-938 (2013). Times cited: 12
- WATANABE, C., AMAGASA, T., LIU, L., Privacy Risks And Countermeasures In Publishing And Mining Social Network Data. 7th International Conference On Collaborative Computing: Networking, Applications And Worksharing, Coliaboratecom 2011 : 55-66 (2011). Times cited: 12
- LIU, P., LI, X., An Improved Privacy Preserving Algorithm For Publishing Social Network Data. 15th Ieee International Conference On High Performance Computing And Communications, Hpcc 2013 And 11th Ieee/ifip International Conference On Embedded And Ubiquitous Computing, Euc 2013 : 888-895 (2014). Times cited: 11
- BABU, K.S., JENA, S.K., HOTA, J., MOHARANA, B., Anonymizing Social Networks: A Generalization Approach. Computers And Electrical Engineering 39:7 1947-1961 (2013). Times cited: 10
- WANG, Y., WU, X., ZHU, J., XIANG, Y., On Learning Cluster Coefficient Of Private Networks. 2012 Ieee/acm International Conference On Advances In Social Networks Analysis And Mining, Asonam 2012 : 395-402 (2012). Times cited: 9
- SINGH, A., URDANETA, G., VAN STEEN, M., VITENBERG, R., Robust Overlays For Privacy-preserving Data Dissemination Over A Social Graph. 32nd Ieee International Conference On Distributed Computing Systems, Icdcs 2012 : 234-244 (2012). Times cited: 9
- JIAO, J., LIU, P., LI, X., A Personalized Privacy Preserving Method For Publishing Social Network Data. 11th Annual Conference On Theory And Applications Of Models Of Computation, Tamc 2014 8402 LNCS: 141-157 (2014). Times cited: 7
- WANG, R., ZHANG, M., FENG, D., FU, Y., A Clustering Approach For Privacy-preserving In Social Networks. 17th International Conference On Information Security And Cryptology, Icisc 2014 8949: 193-204 (2014). Times cited: 7
- YANG, J., WANG, B., YANG, X., ZHANG, H., XIANG, G., A Secure K-automorphism Privacy Preserving Approach With High Data Utility In Social Networks. Security And Communication Networks 7:9 1399-1411 (2014). Times cited: 7
- SHISHODIA, M.S., JAIN, S., TRIPATHY, B.K., Gasna: Greedy Algorithm For Social Network Anonymization. 2013 Ieee/acm International Conference On Advances In Social Networks Analysis And Mining, Asonam 2013 : 1161-1166 (2013). Times cited: 7

- WANG, S.-L., SHIH, C.-C., TING, I.-H., HONG, T.-P., Edge Selection For Degree Anonymization On K Shortest Paths. Springer Proceedings In Complexity : 55-64 (2013). Times cited: 0
- DING, X., WANG, W., Enabling Dynamic Analysis Of Anonymized Social Network Data. 4th International Conference On Cyber-enabled Distributed Computing And Knowledge Discovery, Cyberc 2012 : 21-26 (2012). Times cited: 0
- ZHOU, B., Answering Vertex Aggregate Queries Using Anonymized Social Network Data. 1st Workshop On Privacy And Security In Online Social Media, Psosm'12 : - (2012). Times cited: 0
- HE, X., VAIDYA, J., SHAFIQ, B., ADAM, N., ATLURI, V., Structure-aware Graph Anonymization. Web Intelligence And Agent Systems 10:2 193-208 (2012). Times cited: 0
- WANG, S.-L., TSAI, Z.-Z., HONG, T.-P., TING, I.-H., TSAI, Y.-C., Anonymizing Multiple K-anonymous Shortest Paths For Social Graphs. 2011 2nd International Conference On Innovations In Bio-inspired Computing And Applications, Ibica 2011 : 195-198 (2011). Times cited: 0
- SAHARKHIZ, A., SHAHRIARI, H.R., A Method For Preserving Privacy In Published Multi-relational Social Networks. International Conference On Knowledge Management And Information Sharing, Kmis 2011 : 341-346 (2011). Times cited: 0
- KAVIANPOUR, S., ISMAIL, Z., MOHTASEBI, A., Effectiveness Of Using Integrated Algorithm In Preserving Privacy Of Social Network Sites Users. International Conference On Digital Information And Communication Technology And Its Applications, Dictap 2011 167 CCIS:PART 2 237-249 (2011). Times cited: 0

## CLUSTER-COEFFICIENT

- CHESTER, S., KAPRON, B., RAMESH, G., SRIVASTAVA, G., THOMO, A., VENKATESH, S., κ-anonymization Of Social Networks By Vertex Addition. 15th East-european Conference On Advances In Databases And Information Systems: Research Communications, Adbis 2011 789: 107-116 (2011). Times cited: 33
- WANG, Y., WU, X., ZHU, J., XIANG, Y., On Learning Cluster Coefficient Of Private Networks. Social Network Analysis And Mining 3:4 925-938 (2013). Times cited: 12
- WANG, Y., WU, X., ZHU, J., XIANG, Y., On Learning Cluster Coefficient Of Private Networks. 2012 Ieee/acm International Conference On Advances In Social Networks Analysis And Mining, Asonam 2012 : 395-402 (2012). Times cited: 9
- PéREZ-SOLà, C., HERRERA-JOANCOMARTí, J., Classifying Online Social Network Users Through The Social Graph. 5th International Symposium On Foundations And Practice Of Security, Fps 2012 7743 LNCS: 115-131 (2013). Times cited: 2
- BAE, D.-H., LEE, J.-M., KIM, S.-W., WON, Y., PARK, Y., Analyzing Network Privacy Preserving Methods: A Perspective Of Social Network Characteristics. Ieice Transactions On Information And Systems E97-D:6 1664-1667 (2014). Times cited: 1

- **RANDOMIZATION-APPROACH**
  - BOLDI, P., BONCHI, F., GIONIS, A., TASSA, T., Injecting Uncertainty In Graphs For Identity Obfuscation. Proceedings Of The Vldb Endowment 5:11 1376-1387 (2012). Times cited: 83
  - BONCHI, F., GIONIS, A., TASSA, T., Identity Obfuscation In Graphs Through The Information Theoretic Lens. 2011 Ieee 27th International Conference On Data Engineering, Icde 2011 : 924-935 (2011). Times cited: 38
  - BONCHI, F., GIONIS, A., TASSA, T., Identity Obfuscation In Graphs Through The Information Theoretic Lens. Information Sciences 275: 232-256 (2014). Times cited: 25
  - XUE, M., KARRAS, P., CHEDY, R., KALNIS, P., PUNG, H.K., Delineating Social Network Data Anonymization Via Random Edge Perturbation. 21st Acm International Conference On Information And Knowledge Management, Cikm 2012 : 475-484 (2012). Times cited: 17
  - CASAS-ROMA, J., Privacy-preserving On Graphs Using Randomization And Edge-relevance. 11th International Conference On Modeling Decisions For Artificial Intelligence, Mdai 2014 8825: 204-216 (2014). Times cited: 13
  - LIU, P., CUI, L., LI, X., A Hybrid Algorithm For Privacy Preserving Social Network Publication. Lecture Notes In Computer Science (including Subseries Lecture Notes In Artificial Intelligence And Lecture Notes In Bioinformatics) 8933: 267-278 (2014). Times cited: 2
- **RELATIONAL-DATA**
  - ZHOU, B., PEI, J., The K-anonymity And L-diversity Approaches For Privacy Preservation In Social Networks Against Neighborhood Attacks. Knowledge And Information Systems 28:1 47-77 (2011). Times cited: 160
  - TAI, C.-H., YU, P.S., YANG, D.-N., CHEN, M.-S., Privacy-preserving Social Network Publication Against Friendship Attacks. 17th Acm Sigkdd International Conference On Knowledge Discovery And Data Mining, Kdd'11 : 1262-1270 (2011). Times cited: 77
  - SHISHODIA, M.S., JAIN, S., TRIPATHY, B.K., Gasna: Greedy Algorithm For Social Network Anonymization. 2013 Ieee/acm International Conference On Advances In Social Networks Analysis And Mining, Asonam 2013 : 1161-1166 (2013). Times cited: 7
  - TRIPATHY, B.K., Anonymisation Of Social Networks And Rough Set Approach. Computational Social Networks: Security And Privacy 9781447140511: 269-309 (2013). Times cited: 4
  - XIANG, K., LUO, W., LU, X., YIN, J., Message Passing Based Privacy Preserve In Social Networks. 2012 4th International Conference On Multimedia And Security, Mines 2012 : 483-487 (2012). Times cited: 2
  - HE, X., VAIDYA, J., SHAFIQ, B., ADAM, N., ATLURI, V., Structure-aware Graph Anonymization. Web Intelligence And Agent Systems 10:2 193-208 (2012). Times cited: 0
- **WEIGHTED-NETWORK**
  - SKARKALA, M.E., MARAGOUDAKIS, M., GRITZALIS, S., MITROU, L., TOIVONEN, H., MOEN, P., Privacy Preservation By K-anonymization Ofweighted

Social Networks. 2012 Ieee/acm International Conference On Advances In Social Networks Analysis And Mining, Asonam 2012 : 423-428 (2012). Times cited: 40

- LIU, X., YANG, X., A Generalization Based Approach For Anonymizing Weighted Social Network Graphs. 12th International Conference On Web-age Information Management, Waim 2011 6897 LNCS: 118-130 (2011). Times cited: 33
- CHEN, K., ZHANG, H., WANG, B., YANG, X., Protecting Sensitive Labels In Weighted Social Networks. 2013 10th Web Information System And Application Conference, Wisa 2013 : 221-226 (2013). Times cited: 5
- WANG, S.-L., TSAI, Z.-Z., TING, I.-H., HONG, T.-P., K-anonymous Path Privacy On Social Graphs. Journal Of Intelligent And Fuzzy Systems 26:3 1191-1199 (2014). Times cited: 1
- NI, W., SUN, F., WENG, G., XU, L., A Hellinger Distance Based Anonymization Method For Weighted Social Networks. 2013 10th Web Information System And Application Conference, Wisa 2013 : 259-264 (2013). Times cited: 1
- TSAI, Y.-C., WANG, S.-L., KAO, H.-Y., HONG, T.-P., Confining Edge Types In K-anonymization Of Shortest Paths. 3rd International Conference On Innovations In Bio-inspired Computing And Applications, Ibica 2012 : 318-322 (2012). Times cited: 1
- WANG, S.-L., SHIH, C.-C., TING, I.-H., HONG, T.-P., Edge Selection For Degree Anonymization On K Shortest Paths. Springer Proceedings In Complexity : 55-64 (2013). Times cited: 0
- WANG, S.-L., TSAI, Z.-Z., HONG, T.-P., TING, I.-H., TSAI, Y.-C., Anonymizing Multiple K-anonymous Shortest Paths For Social Graphs. 2011 2nd International Conference On Innovations In Bio-inspired Computing And Applications, Ibica 2011 : 195-198 (2011). Times cited: 0

# PRIVACY-BREACH

- STOKES, K., TORRA, V., Reidentification And K-anonymity: A Model For Disclosure Risk In Graphs. Soft Computing 16:10 1657-1670 (2012). Times cited: 37
- DURR, M., MAIER, M., DORFMEISTER, F., Vegas - A Secure And Privacy-preserving Peer-to-peer Online Social Network. 2012 Ase/ieee International Conference On Social Computing, Socialcom 2012 And The 2012 Ase/ieee International Conference On Privacy, Security, Risk And Trust, Passat 2012 : 868-874 (2012). Times cited: 25
- FARD, A.M., WANG, K., YU, P.S., Limiting Link Disclosure In Social Network Analysis Through Subgraph-wise Perturbation. 15th International Conference On Extending Database Technology, Edbt 2012 : 109-119 (2012). Times cited: 20
- WATANABE, C., AMAGASA, T., LIU, L., Privacy Risks And Countermeasures In Publishing And Mining Social Network Data. 7th International Conference On Collaborative Computing: Networking, Applications And Worksharing, Coliaboratecom 2011 : 55-66 (2011). Times cited: 12
- BAZGAN, C., NICHTERLEIN, A., Parameterized Inapproximability Of Degree Anonymization. 9th International Symposium On Parameterized And Exact Computation, Ipec 2014 8894: 75-84 (2014). Times cited: 7

- CHEN, K., ZHANG, H., WANG, B., YANG, X., Protecting Sensitive Labels In Weighted Social Networks. 2013 10th Web Information System And Application Conference, Wisa 2013 : 221-226 (2013). Times cited: 5
- NINGGAL, M.I.H., ABAWAJY, J.H., Preserving Utility In Social Network Graph Anonymization. 12th Ieee International Conference On Trust, Security And Privacy In Computing And Communications, Trustcom 2013 : 226-232 (2013). Times cited: 3
- BAE, D.-H., LEE, J.-M., KIM, S.-W., WON, Y., PARK, Y., Analyzing Network Privacy Preserving Methods: A Perspective Of Social Network Characteristics. Ieice Transactions On Information And Systems E97-D:6 1664-1667 (2014). Times cited: 1

## NP-HARD

- CHESTER, S., KAPRON, B.M., SRIVASTAVA, G., VENKATESH, S., Complexity Of Social Network Anonymization. Social Network Analysis And Mining 3:2 151-166 (2013). Times cited: 36
- KAPRON, B., SRIVASTAVA, G., VENKATESH, S., Social Network Anonymization Via Edge Addition. 2011 International Conference On Advances In Social Networks Analysis And Mining, Asonam 2011 : 155-162 (2011). Times cited: 22
- SIHAG, V.K., A Clustering Approach For Structural K-anonymity In Social Networks Using Genetic Algorithm. International Information Technology Conference, Cube 2012 : 701-706 (2012). Times cited: 18
- KAYEM, A.V.D.M., DESHAI, A., HAMMER, S., On Anonymizing Social Network Graphs. 2012 Conference On Information Security For South Africa, Issa 2012 : - (2012). Times cited: 1

## ENCRYPTION

- ALBERTINI, D.A., CARMINATI, B., Relationship-based Information Sharing In Cloud-based Decentralized Social Networks. 4th Acm Conference On Data And Application Security And Privacy, Codaspy 2014 : 297-304 (2014). Times cited: 4
- BRUCE, N., LEE, H.J., Anonymization Algorithm For Security And Confidentiality Of Health Data Set Across Social Network. 5th International Conference On Information And Communication Technology Convergence, Ictc 2014 : 65-70 (2014). Times cited: 0

## GRAPH-DATA

- TAI, C.-H., YU, P.S., YANG, D.-N., CHEN, M.-S., Privacy-preserving Social Network Publication Against Friendship Attacks. 17th Acm Sigkdd International Conference On Knowledge Discovery And Data Mining, Kdd'11 : 1262-1270 (2011). Times cited: 77
- AGGARWAL, C.C., LI, Y., YU, P.S., On The Hardness Of Graph Anonymization. 11th Ieee International Conference On Data Mining, Icdm 2011 : 1002-1007 (2011). Times cited: 15
- LIU, X., WANG, B., YANG, X., Efficiently Anonymizing Social Networks With Reachability Preservation. 22nd Acm International Conference On Information And Knowledge Management, Cikm 2013 : 1613-1618 (2013). Times cited: 6

- **SUBGRAPH**
  - CHESTER, S., GAERTNER, J., STEGE, U., VENKATESH, S., Anonymizing Subsets Of Social Networks With Degree Constrained Subgraphs. 2012 Ieee/acm International Conference On Advances In Social Networks Analysis And Mining, Asonam 2012 : 418-422 (2012). Times cited: 25
  - FARD, A.M., WANG, K., YU, P.S., Limiting Link Disclosure In Social Network Analysis Through Subgraph-wise Perturbation. 15th International Conference On Extending Database Technology, Edbt 2012 : 109-119 (2012). Times cited: 20
- **TABULAR-DATA**
  - YUAN, M., CHEN, L., Semi-edge Anonymity: Graph Publication When The Protection Algorithm Is Available. 17th International Conference On Database Systems For Advanced Applications, Dasfaa 2012 7238 LNCS:PART 1 367-381 (2012). Times cited: 1
  - PADROL, A., MUNTéS-MULERO, V., Graph Anonymization Via Metric Embed-dings: Using Classical Anonymization For Graphs. Intelligent Data Analysis 18:3 365-388 (2014). Times cited: 0
- **WEIGHTED-MAXIMUM-COMMON-SUBGRAPH-(WMCS)**
  - YUAN, M., CHEN, L., YU, P.S., MEI, H., Privacy Preserving Graph Publication In A Distributed Environment. 15th Asia-pacific Web Conference On Web Technologies And Applications, Apweb 2013 7808 LNCS: 75-87 (2013). Times cited: 1
- **K-MEANS**
  - STOKES, K., Graph K-anonymity Through K-means And As Modular Decomposition. 18th Nordic Conference On Secure It Systems, Nordsec 2013 8208 LNCS: 263-278 (2013). Times cited: 3
- **NEIGHBORHOOD-INFORMATION**
  - WANG, G., LIU, Q., LI, F., YANG, S., WU, J., Outsourcing Privacy-preserving Social Networks To A Cloud. 32nd Ieee Conference On Computer Communications, Ieee Infocom 2013 : 2886-2894 (2013). Times cited: 33
- **COMBINATORIAL-GRAPH**
  - BAZGAN, C., NICHTERLEIN, A., Parameterized Inapproximability Of Degree Anonymization. 9th International Symposium On Parameterized And Exact Computation, Ipec 2014 8894: 75-84 (2014). Times cited: 7
- **GRAPH-RECONSTRUCTION-ATTACK**
  - DING, X., WANG, W., WAN, M., GU, M., Seamless Privacy: Privacy-preserving Subgraph Counting In Interactive Social Network Analysis. 2013 5th International Conference On Cyber-enabled Distributed Computing And Knowledge Discovery, Cyberc 2013 : 97-104 (2013). Times cited: 2

- **HYPERGRAPH-CLUSTERING**
    - LI, Y., SHEN, H., On Identity Disclosure Control For Hypergraph-based Data Publishing. Ieee Transactions On Information Forensics And Security 8:8 1384-1396 (2013). Times cited: 13
- **DYNAMIC-NETWORK**
    - ZHANG, W., WANG, X.-R., WANG, J., CHEN, Y.-F., Privacy Preservation In Dynamic Social Networks Based On K-neighborhood Isomorphism. Nanjing Youdian Daxue Xuebao (ziran Kexue Ban)/journal Of Nanjing University Of Posts And Telecommunications (natural Science) 34:5 9-16 (2014). Times cited: 3
- **SEMI-SUPERVISED-CLUSTERING-ALGORITHMS**
    - LI, K., WANG, Z., MA, C., SONG, S., Sns Privacy Protection Based On The Elm Integration And Semi-supervised Clustering. Journal Of Software 8:1 160-167 (2013). Times cited: 1
- **SUPERVISED-MACHINE-LEARNING-ALGORITHMS**
    - LI, C., AGGARWAL, C.C., WANG, J., On Anonymization Of Multi-graphs. 11th Siam International Conference On Data Mining, Sdm 2011 : 711-722 (2011). Times cited: 3
- **T-CLOSENESS**
    - SINGH, A., BANSAL, D., SOFAT, S., An Approach Of Privacy Preserving Based Publishing In Twitter. 7th International Conference On Security Of Information And Networks, Sin 2014 2014-September: 39-42 (2014). Times cited: 1
- **ARTIFICIAL-INTELLIGENCE-(AI)**

    - MAAG, M.L., DENOYER, L., GALLINARI, P., Graph Anonymization Using Machine Learning. 28th Ieee International Conference On Advanced Information Networking And Applications, Ieee Aina 2014 : 1111-1118 (2014). Times cited: 4

# Subperiod 2015-2018

- **RANDOM-NETWORK**
  - CHIASSERINI, C.-F., GARETTO, M., LEONARDI, E., De-anonymizing Clustered Social Networks By Percolation Graph Matching. Acm Transactions On Knowledge Discovery From Data 12:2 - (2018). Times cited: 12
  - CHIASSERINI, C.-F., GARETTO, M., LEONARDI, E., Impact Of Clustering On The Performance Of Network De-anonymization. 3rd Acm Conference On Online Social Networks, Cosn 2015 : 83-94 (2015). Times cited: 10
  - GRINING, K., KLONOWSKI, M., SULKOWSKA, M., How To Cooperate Locally To Improve Global Privacy In Social Networks? On Amplification Of Privacy Preserving Data Aggregation. 16th Ieee International Conference On Trust, Security And Privacy In Computing And Communications, 11th Ieee International Conference On Big Data Science And Engineering And 14th Ieee International Conference On Embedded Software And Systems, Trustcom/bigdatase/icess 2017 : 464-471 (2017). Times cited: 1
  - CHEN, L., ZHU, P., Preserving Network Privacy With A Hierarchical Structure Approach. 12th International Conference On Fuzzy Systems And Knowledge Discovery, Fskd 2015 : 773-777 (2016). Times cited: 0
- **PERTURBATION-TECHNIQUES**
  - MILANI FARD, A., WANG, K., Neighborhood Randomization For Link Privacy In Social Network Analysis. World Wide Web 18:1 9-32 (2015). Times cited: 30
  - SIDDULA, M., LI, L., LI, Y., An Empirical Study On The Privacy Preservation Of Online Social Networks. Ieee Access 6: 19912-19922 (2018). Times cited: 20
  - SHARATH KUMAR, J., MAHESWARI, N., A Survey On Privacy-preserving Techniques For Social Network Data. Asian Journal Of Pharmaceutical And Clinical Research 10: 112-116 (2017). Times cited: 1
  - LAN, L., Preserving Weighted Social Networks Privacy Using Vectors Similarity. 8th International Conference On Biomedical Engineering And Informatics, Bmei 2015 : 789-794 (2016). Times cited: 1
  - WANG, H., LIU, P., LIN, S., LI, X., A Local-perturbation Anonymizing Approach To Preserving Community Structure In Released Social Networks. 12th Eai International Conference On Heterogeneous Networking For Quality, Reliability, Security And Robustness, Qshine 2016 199: 36-45 (2017). Times cited: 0
- **PRIVACY-PRESERVATION**
  - ABAWAJY, J.H., NINGGAL, M.I.H., HERAWAN, T., Privacy Preserving Social Network Data Publication. Ieee Communications Surveys And Tutorials 18:3 1974-1997 (2016). Times cited: 75
  - JI, S., MITTAL, P., BEYAH, R., Graph Data Anonymization, De-anonymization Attacks, And De-anonymizability Quantification: A Survey. Ieee Communications Surveys And Tutorials 19:2 1305-1326 (2017). Times cited: 57

- CASAS-ROMA, J., HERRERA-JOANCOMARTí, J., TORRA, V., A Survey Of Graph-modification Techniques For Privacy-preserving On Networks. Artificial Intelligence Review 47:3 341-366 (2017). Times cited: 51
- CASAS-ROMA, J., HERRERA-JOANCOMARTí, J., TORRA, V., K-degree Anonymity And Edge Selection: Improving Data Utility In Large Networks. Knowledge And Information Systems 50:2 447-474 (2017). Times cited: 44
- LI, X.-Y., ZHANG, C., JUNG, T., QIAN, J., CHEN, L., Graph-based Privacy-preserving Data Publication. 35th Annual Ieee International Conference On Computer Communications, Ieee Infocom 2016 2016-July: - (2016). Times cited: 35
- MILANI FARD, A., WANG, K., Neighborhood Randomization For Link Privacy In Social Network Analysis. World Wide Web 18:1 9-32 (2015). Times cited: 30
- FU, H., ZHANG, A., XIE, X., Effective Social Graph Deanonymization Based On Graph Structure And Descriptive Information. Acm Transactions On Intelligent Systems And Technology 6:4 - (2015). Times cited: 25
- NGUYEN, H.H., IMINE, A., RUSINOWITCH, M., Anonymizing Social Graphs Via Uncertainty Semantics. 10th Acm Symposium On Information, Computer And Communications Security, Asiaccs 2015 : 495-506 (2015). Times cited: 25
- SIDDULA, M., LI, L., LI, Y., An Empirical Study On The Privacy Preservation Of Online Social Networks. Ieee Access 6: 19912-19922 (2018). Times cited: 20
- CAMPAN, A., ALUFAISAN, Y., TRUTA, T.M., Preserving Communities In Anonymized Social Networks. Transactions On Data Privacy 8:1 55-87 (2015). Times cited: 18
- ROUSSEAU, F., CASAS-ROMA, J., VAZIRGIANNIS, M., Community-preserving Anonymization Of Graphs. Knowledge And Information Systems 54:2 315-343 (2018). Times cited: 15
- LIU, Y., JI, S., MITTAL, P., Smartwalk: Enhancing Social Network Security Via Adaptive Random Walks. 23rd Acm Conference On Computer And Communications Security, Ccs 2016 24-28-October-2016: 492-503 (2016). Times cited: 15
- YU, F., CHEN, M., YU, B., LI, W., MA, L., GAO, H., Privacy Preservation Based On Clustering Perturbation Algorithm For Social Network. Multimedia Tools And Applications 77:9 11241-11258 (2018). Times cited: 13
- NINGGAL, M.I.H., ABAWAJY, J.H., Utility-aware Social Network Graph Anonymization. Journal Of Network And Computer Applications 56: 137-148 (2015). Times cited: 13
- LIU, P., BAI, Y., WANG, L., LI, X., Partial κ-anonymity For Privacy-preserving Social Network Data Publishing. International Journal Of Software Engineering And Knowledge Engineering 27:1 71-90 (2017). Times cited: 13

- CHIASSERINI, C.-F., GARETTO, M., LEONARDI, E., De-anonymizing Clustered Social Networks By Percolation Graph Matching. Acm Transactions On Knowledge Discovery From Data 12:2 - (2018). Times cited: 12

International Conference For Internet Technology And Secured Transactions, Icitst 2016 : 80-85 (2017). Times cited: 2

- LIU, X.-Y., LI, J.-J., AN, Y.-Z., ZHOU, D.-H., XIA, X.-F., Efficient Algorithm On Anonymizing Social Networks With Reachability Preservation. Ruan Jian Xue Bao/journal Of Software 27:8 1904-1921 (2016). Times cited: 2
- XIE, Y., ZHENG, M., LIU, L., A Personalized Sensitive Label-preserving Model And Algorithm Based On Utility In Social Network Data Publishing. 2nd International Conference On Human Centered Computing, Hcc 2016 9567: 942-947 (2016). Times cited: 2
- LIN, S.H., LIAO, M.H., Towards Publishing Social Network Data With Graph Anonymization. Journal Of Intelligent And Fuzzy Systems 30:1 333-345 (2016). Times cited: 2
- AGGARWAL, C.C., LI, Y., YU, P.S., On The Anonymizability Of Graphs. Knowledge And Information Systems 45:3 571-588 (2015). Times cited: 2
- GU, Y.-H., LIN, J.-C., GUO, D., Clustering-based Dynamic Privacy Preserving Method For Social Networks. Tongxin Xuebao/journal On Communications 36: - (2015). Times cited: 2
- SINGH, A., BANSAL, D., SOFAT, S., What About Privacy Of My Osn Data?. Cybernetics And Systems 49:1 44-63 (2018). Times cited: 2
- RAJAEI, M., HAGHJOO, M.S., KHANJARI MIYANEH, E., An Anonymization Algorithm for (α,β,γ,δ)-Social Network Privacy Considering Data Utility. Journal Of Universal Computer Science 21:2 268-305 (2015). Times cited: 2
- MONISHA, R., KARTHIK, S., Precision Driven Privacy-preserving Anonymization For Social Data Using Segmentation. 2018 International Conference On Soft-computing And Network Security, Icsns 2018 : - (2018). Times cited: 1
- AN, S., LI, Y., WANG, T., JIN, Y., Contact Graph Based Anonymization For Geosocial Network Datasets. 5th International Conference On Behavioral, Economic, And Socio-cultural Computing, Besc 2018 : 132-137 (2018). Times cited: 1
- ZHANG, M., QIN, S., GUO, F., Satisfying Link Perturbation And K-out Anonymous In Social Network Privacy Protection. 17th Ieee International Conference On Communication Technology, Icct 2017 2017-October: 1387-1391 (2018). Times cited: 1
- BOURAHLA, S., CHALLAL, Y., Social Networks Privacy Preserving Data Publishing. 13th International Conference On Computational Intelligence And Security, Cis 2017 2018-January: 258-262 (2018). Times cited: 1
- TRIPATHY, B.K., BAKTHA, K., Security, Privacy, And Anonymization In Social Networks: Emerging Research And Opportunities. Security, Privacy, And Anonymization In Social Networks: Emerging Research And Opportunities : 1-176 (2018). Times cited: 1
- GARCíA-RECUERO, J. Burdges, Efficient Privacy-preserving Adversarial Learning In Decentralized Online Social Networks. 9th Ieee/acm International Conference On Advances In Social Networks Analysis And Mining, Asonam 2017 : 1132-1135 (2017). Times cited: 1

- SHARATH KUMAR, J., MAHESWARI, N., A Survey On Privacy-preserving Techniques For Social Network Data. Asian Journal Of Pharmaceutical And Clinical Research 10: 112-116 (2017). Times cited: 1
- LI, Y., On Preserving Private Geosocial Networks Against Practical Attacks. 11th International Conference On Intelligent Information Hiding And Multimedia Signal Processing, Iih-msp 2015 : 417-420 (2016). Times cited: 1
- LAN, L., Preserving Weighted Social Networks Privacy Using Vectors Similarity. 8th International Conference On Biomedical Engineering And Informatics, Bmei 2015 : 789-794 (2016). Times cited: 1
- SALAS, J., Sampling And Merging For Graph Anonymization. 13th International Conference On Modeling Decisions For Artificial Intelligence, Mdai 2016 9880 LNAI: 250-260 (2016). Times cited: 1
- GRINING, K., KLONOWSKI, M., SULKOWSKA, M., How To Cooperate Locally To Improve Global Privacy In Social Networks? On Amplification Of Privacy Preserving Data Aggregation. 16th Ieee International Conference On Trust, Security And Privacy In Computing And Communications, 11th Ieee International Conference On Big Data Science And Engineering And 14th Ieee International Conference On Embedded Software And Systems, Trustcom/bigdatase/icess 2017 : 464-471 (2017). Times cited: 1
- GAO, T., LI, F., Preserving Graph Utility In Anonymized Social Networks? A Study On The Persistent Homology. 14th Ieee International Conference On Mobile Ad Hoc And Sensor Systems, Mass 2017 : 348-352 (2017). Times cited: 1
- SRIVASTAVA, GAUTAM, CITULSKY, EVAN, TILBURY, KYLE, ABDELBAR, ASHRAF, AMAGASA, TOSHIYUKI, The Effects Of Ant Colony Optimization On Graph Anonymization. Gstf Journal On Computing (joc) 5:1 91-100 (2016). Times cited: 1
- HUOWEN, J., HUANLIANG, X., HUIYUN, Z., A Novel Approach To Achieving κ-anonymization For Social Network Privacy Preservation Based On Vertex Connectivity. Ieee Advanced Information Technology, Electronic And Automation Control Conference, Iaeac 2015 : 1097-1100 (2016). Times cited: 1
- TIAN, H., LU, Y., LIU, J., YU, J., Betweenness Centrality Based K-anonymity For Privacy Preserving In Social Networks. 16th International Conference On Advances In Mobile Computing And Multimedia, Momm 2018 : 3-7 (2018). Times cited: 0
- XIA, H., An Efficient Algorithm In Anonymous Social Network With Reachability Preservation. 10th International Conference On Intelligent Human-machine Systems And Cybernetics, Ihmsc 2018 1: 383-387 (2018). Times cited: 0
- NAGARAJ, K., SRIDHAR, A., SHARVANI, G.S., Identification Of Network Communities And Assessment Of Privacy Using Hybrid Algorithm. 2nd International Conference On Computational Systems And Information Technology For Sustainable Solutions, Csitss 2017 : - (2018). Times cited: 0
- SAFIA, B., YACINE, C., Privacy Preservation In Social Networks Sequential Publishing. 32nd Ieee International Conference On Advanced Information Networking And Applications, Aina 2018 2018-May: 732-739 (2018). Times cited: 0
- LIU, P., WANG, H., LIN, S., LI, X., Social Network Anonymization Via Local-perturbing Approach. Journal Of Internet Technology 19:1 247-256 (2018). Times cited: 0

- WANG, H., LIU, P., LIN, S., LI, X., A Local-perturbation Anonymizing Approach To Preserving Community Structure In Released Social Networks. 12th Eai International Conference On Heterogeneous Networking For Quality, Reliability, Security And Robustness, Qshine 2016 199: 36-45 (2017). Times cited: 0

## • DEGREE-SEQUENCE

- ROUSSEAU, F., CASAS-ROMA, J., VAZIRGIANNIS, M., Community-preserving Anonymization Of Graphs. Knowledge And Information Systems 54:2 315-343 (2018). Times cited: 15
- WANG, Y., YANG, L., CHEN, X., ZHANG, X., HE, Z., Enhancing Social Network Privacy With Accumulated Non-zero Prior Knowledge. Information Sciences 445-446: 6-21 (2018). Times cited: 5
- PORTELA, J., VILLALBA, L.J.G., SILVA TRUJILLO, A.G., SANDOVAL OROZCO, A.L., KIM, T.-H., Estimation Of Anonymous Email Network Characteristics Through Statistical Disclosure Attacks. Sensors (switzerland) 16:11 - (2016). Times cited: 3
- SALAS, J., Sampling And Merging For Graph Anonymization. 13th International Conference On Modeling Decisions For Artificial Intelligence, Mdai 2016 9880 LNAI: 250-260 (2016). Times cited: 1

## • PRIVACY-BREACH

- LIU, X., LI, M., XIA, X., LI, J., ZONG, C., ZHU, R., Spatio-temporal Features Based Sensitive Relationship Protection In Social Networks. 15th Web Information Systems And Applications Conference, Wisa 2018 11242 LNCS: 330-343 (2018). Times cited: 4
- WEN, G., LIU, H., YAN, J., WU, Z., A Privacy Analysis Method To Anonymous Graph Based On Bayes Rule In Social Networks. 14th International Conference On Computational Intelligence And Security, Cis 2018 : 469-472 (2018). Times cited: 3
- YAN, J., ZHANG, L., SHI, W., HU, J., WU, Z., Uncertain Graph Method Based On Triadic Closure Improving Privacy Preserving In Social Network. 2017 International Conference On Networking And Network Applications, Nana 2017 2018-January: 190-195 (2017). Times cited: 2
- HUOWEN, J., HUANLIANG, X., HUIYUN, Z., A Novel Approach To Achieving κ-anonymization For Social Network Privacy Preservation Based On Vertex Connectivity. Ieee Advanced Information Technology, Electronic And Automation Control Conference, Iaeac 2015 : 1097-1100 (2016). Times cited: 1
- XIA, H., An Efficient Algorithm In Anonymous Social Network With Reachability Preservation. 10th International Conference On Intelligent Human-machine Systems And Cybernetics, Ihmsc 2018 1: 383-387 (2018). Times cited: 0
- HU, Q., JIANG, C., WANG, Y., (k.l)-anonymity For Social Networks Based On K-neighborhood Anonymity. 4th International Conference On Information Science And Cloud Computing, Iscc 2015 18-19-December-2015: - (2015). Times cited: 0
- TSAI, Y.-C., WANG, S.-L., HONG, T.-P., KAO, H.-Y., Extending [k1, K2] Anonymization Of Shortest Paths For Social Networks. 2nd International Conference On Multidisciplinary Social Networks Research, Misnc 2015 540: 187-199 (2015). Times cited: 0

- **GENERALIZATION-APPROACH**
  - NGUYEN, H.H., IMINE, A., RUSINOWITCH, M., Anonymizing Social Graphs Via Uncertainty Semantics. 10th Acm Symposium On Information, Computer And Communications Security, Asiaccs 2015 : 495-506 (2015). Times cited: 25
  - GHATE, R.B., INGLE, R., Clustering Based Anonymization For Privacy Preservation. 2015 International Conference On Pervasive Computing, Icpc 2015 : - (2015). Times cited: 9
  - YIN, D., SHEN, Y., Location- And Relation-based Clustering On Privacy-preserving Social Networks. Tsinghua Science And Technology 23:4 453-462 (2018). Times cited: 6
  - JI, S., DU, T., HONG, Z., WANG, T., BEYAH, R., Quantifying Graph Anonymity, Utility, And De-anonymity. 2018 Ieee Conference On Computer Communications, Infocom 2018 2018-April: 1736-1744 (2018). Times cited: 5
  - JIANG, H.-W., ZHAN, Q.-H., LIU, W.-J., MA, H.-Y., Clustering-anonymity Approach For Privacy Preservation Of Graph Data-publishing. Ruan Jian Xue Bao/journal Of Software 28:9 2323-2333 (2017). Times cited: 5
  - SHYAMALA, C.K., HEMAASHRI, S., An Enhanced Design For Anonymization In Social Networks. Indian Journal Of Science And Technology 9:30 - (2016). Times cited: 5
  - MONISHA, R., KARTHIK, S., Precision Driven Privacy-preserving Anonymization For Social Data Using Segmentation. 2018 International Conference On Soft-computing And Network Security, Icsns 2018 : - (2018). Times cited: 1
- **ARTIFICIAL-INTELLIGENCE-(AI)**
  - LIU, Y., JI, S., MITTAL, P., Smartwalk: Enhancing Social Network Security Via Adaptive Random Walks. 23rd Acm Conference On Computer And Communications Security, Ccs 2016 24-28-October-2016: 492-503 (2016). Times cited: 15
  - CHEN, Z.-G., KANG, H.-S., YIN, S.-N., KIM, S.-R., An Efficient Privacy Protection In Mobility Social Network Services With Novel Clustering-based Anonymization. Eurasip Journal On Wireless Communications And Networking 2016:1 - (2016). Times cited: 8
  - ASIF, W., RAY, I.G., TAHIR, S., RAJARAJAN, M., Privacy-preserving Anonymization With Restricted Search (pars) On Social Network Data For Criminal Investigations. 19th Ieee/acis International Conference On Software Engineering, Artificial Intelligence, Networking And Parallel/distributed Computing, Snpd 2018 : 329-334 (2018). Times cited: 5
  - ZAGHIAN, A., BAGHERI, A., A Combined Model Of Clustering And Classification Methods For Preserving Privacy In Social Networks Against Inference And Neighborhood Attacks. International Journal Of Security And Its Applications 10:1 95-102 (2016). Times cited: 4
- **PRIVACY-ATTACK-MODEL**
  - MAUW, S., TRUJILLO-RASUA, R., XUAN, B., Counteracting Active Attacks In Social Network Graphs. 30th Ifip Wg 11.3 Conference On Data And Applications Security, Dbsec 2016 9766: 233-248 (2016). Times cited: 10

- FU, Y., WANG, W., FU, H., YANG, W., YIN, D., Privacy Preserving Social Network Against Dopv Attacks. 19th International Conference On Web Information Systems Engineering, Wise 2018 11233 LNCS: 178-188 (2018). Times cited: 3
- SU, J., LIU, S., LUO, Z.-Y., SUN, G.-L., Method Of Constructing An Anonymous Graph Based On Information Loss Estimation. Tongxin Xuebao/journal On Communications 37:6 56-64 (2016). Times cited: 3
- ZHAO, Y., WAGNER, I., Poster: Evaluating Privacy Metrics For Graph Anonymization And De-anonymization. 13th Acm Symposium On Information, Computer And Communications Security, Asiaccs 2018 : 817-819 (2018). Times cited: 2
- ZHANG, M., QIN, S., GUO, F., Satisfying Link Perturbation And K-out Anonymous In Social Network Privacy Protection. 17th Ieee International Conference On Communication Technology, Icct 2017 2017-October: 1387-1391 (2018). Times cited: 1
- TRIPATHY, B.K., BAKTHA, K., Security, Privacy, And Anonymization In Social Networks: Emerging Research And Opportunities. Security, Privacy, And Anonymization In Social Networks: Emerging Research And Opportunities : 1-176 (2018). Times cited: 1

## GRAPH-MODIFICATION-APPROACH

- CASAS-ROMA, J., HERRERA-JOANCOMARTí, J., TORRA, V., K-degree Anonymity And Edge Selection: Improving Data Utility In Large Networks. Knowledge And Information Systems 50:2 447-474 (2017). Times cited: 44
- FU, H., ZHANG, A., XIE, X., Effective Social Graph Deanonymization Based On Graph Structure And Descriptive Information. Acm Transactions On Intelligent Systems And Technology 6:4 - (2015). Times cited: 25
- ROUSSEAU, F., CASAS-ROMA, J., VAZIRGIANNIS, M., Community-preserving Anonymization Of Graphs. Knowledge And Information Systems 54:2 315-343 (2018). Times cited: 15
- CASAS-ROMA, J., An Evaluation Of Edge Modification Techniques For Privacy-preserving On Graphs. 12th International Conference On Modeling Decisions For Artificial Intelligence, Mdai 2015 9321: 180-191 (2015). Times cited: 5
- SHI, W., HU, J., YAN, J., WU, Z., LU, L., A Privacy Measurement Method Using Network Structure Entropy. 2017 International Conference On Networking And Network Applications, Nana 2017 2018-January: 147-151 (2017). Times cited: 4
- YUE, R., LI, Y., WANG, T., JIN, Y., An Efficient Adaptive Graph Anonymization Framework For Incremental Data Publication. 5th International Conference On Behavioral, Economic, And Socio-cultural Computing, Besc 2018 : 103-108 (2018). Times cited: 3
- ZHANG, Y., MA, T., CAO, J., TANG, M., K-anonymisation Of Social Network By Vertex And Edge Modification. International Journal Of Embedded Systems 8:2-3 206-216 (2016). Times cited: 3
- TALMON, N., HARTUNG, S., The Complexity Of Degree Anonymization By Graph Contractions. Information And Computation 256: 212-225 (2017). Times cited: 2
- LIU, X.-Y., LI, J.-J., AN, Y.-Z., ZHOU, D.-H., XIA, X.-F., Efficient Algorithm On Anonymizing Social Networks With Reachability Preservation. Ruan Jian Xue Bao/journal Of Software 27:8 1904-1921 (2016). Times cited: 2

- LIU, X., LI, J., ZHOU, D., AN, Y., XIA, X., Preserving The D-reachability When Anonymizing Social Networks. 17th International Conference On Web-age Information Management, Waim 2016 9659: 40-51 (2016). Times cited: 1

- **INFORMATION-LOSS**
  - CASAS-ROMA, J., HERRERA-JOANCOMARTí, J., TORRA, V., K-degree Anonymity And Edge Selection: Improving Data Utility In Large Networks. Knowledge And Information Systems 50:2 447-474 (2017). Times cited: 44
  - YU, F., CHEN, M., YU, B., LI, W., MA, L., GAO, H., Privacy Preservation Based On Clustering Perturbation Algorithm For Social Network. Multimedia Tools And Applications 77:9 11241-11258 (2018). Times cited: 13
  - CASAS-ROMA, J., HERRERA-JOANCOMARTí, J., TORRA, V., Anonymizing Graphs: Measuring Quality For Clustering. Knowledge And Information Systems 44:3 507-528 (2015). Times cited: 11
  - LIU, X., LI, M., XIA, X., LI, J., ZONG, C., ZHU, R., Spatio-temporal Features Based Sensitive Relationship Protection In Social Networks. 15th Web Information Systems And Applications Conference, Wisa 2018 11242 LNCS: 330-343 (2018). Times cited: 4
  - SU, J., LIU, S., LUO, Z.-Y., SUN, G.-L., Method Of Constructing An Anonymous Graph Based On Information Loss Estimation. Tongxin Xuebao/journal On Communications 37:6 56-64 (2016). Times cited: 3
  - ZHANG, Y., MA, T., CAO, J., TANG, M., K-anonymisation Of Social Network By Vertex And Edge Modification. International Journal Of Embedded Systems 8:2-3 206-216 (2016). Times cited: 3
  - GU, Y.-H., LIN, J.-C., GUO, D., Clustering-based Dynamic Privacy Preserving Method For Social Networks. Tongxin Xuebao/journal On Communications 36: - (2015). Times cited: 2
  - SINGH, A., BANSAL, D., SOFAT, S., What About Privacy Of My Osn Data?. Cybernetics And Systems 49:1 44-63 (2018). Times cited: 2
  - LIU, X., LI, J., ZHOU, D., AN, Y., XIA, X., Preserving The D-reachability When Anonymizing Social Networks. 17th International Conference On Web-age Information Management, Waim 2016 9659: 40-51 (2016). Times cited: 1
  - TIAN, H., LU, Y., LIU, J., YU, J., Betweenness Centrality Based K-anonymity For Privacy Preserving In Social Networks. 16th International Conference On Advances In Mobile Computing And Multimedia, Momm 2018 : 3-7 (2018). Times cited: 0
  - LIU, P., WANG, H., LIN, S., LI, X., Social Network Anonymization Via Local-perturbing Approach. Journal Of Internet Technology 19:1 247-256 (2018). Times cited: 0
  - WANG, H., LIU, P., LIN, S., LI, X., A Local-perturbation Anonymizing Approach To Preserving Community Structure In Released Social Networks. 12th Eai International Conference On Heterogeneous Networking For Quality, Reliability, Security And Robustness, Qshine 2016 199: 36-45 (2017). Times cited: 0

- **L-DIVERSITY**
  - CHAKRABORTY, S., TRIPATHY, B.K., Alpha-anonymization Techniques For Privacy Preservation In Social Networks. Social Network Analysis And Mining 6:1 - (2016). Times cited: 8
  - CHAKRABORTY, S., TRIPATHY, B.K., Privacy Preserving Anonymization Of Social Networks Using Eigenvector Centrality Approach. Intelligent Data Analysis 20:3 543-560 (2016). Times cited: 4
  - BAKTHA, K., TRIPATHY, B.K., Alpha Anonymization In Social Networks Using The Lossy-join Approach. Transactions On Data Privacy 11:1 1-22 (2018). Times cited: 2
  - HU, Q., JIANG, C., WANG, Y., (k.l)-anonymity For Social Networks Based On K-neighborhood Anonymity. 4th International Conference On Information Science And Cloud Computing, Iscc 2015 18-19-December-2015: - (2015). Times cited: 0
- **RE-IDENTIFICATION-ATTACKS**
  - CAMPAN, A., ALUFAISAN, Y., TRUTA, T.M., Preserving Communities In Anonymized Social Networks. Transactions On Data Privacy 8:1 55-87 (2015). Times cited: 18
  - SHARAD, K., True Friends Let You Down: Benchmarking Social Graph Anonymization Schemes. 9th Acm Workshop On Artificial Intelligence And Security, Aisec 2016 : 93-104 (2016). Times cited: 4
  - BHATTACHARYA, M., MANI, P., Preserving Privacy In Social Network Graph With K-anonymize Degree Sequence Generation. 9th International Conference On Software, Knowledge, Information Management And Applications, Skima 2015 : - (2016). Times cited: 3
  - BENSIMESSAOUD, S., BADACHE, N., BENMEZIANE, S., DJELLALBIA, A., An Enhanced Approach To Preserving Privacy In Social Network Data Publishing. 11th International Conference For Internet Technology And Secured Transactions, Icitst 2016 : 80-85 (2017). Times cited: 2
  - AGGARWAL, C.C., LI, Y., YU, P.S., On The Anonymizability Of Graphs. Knowledge And Information Systems 45:3 571-588 (2015). Times cited: 2
  - AN, S., LI, Y., WANG, T., JIN, Y., Contact Graph Based Anonymization For Geosocial Network Datasets. 5th International Conference On Behavioral, Economic, And Socio-cultural Computing, Besc 2018 : 132-137 (2018). Times cited: 1
- **DIFFERENTIAL-PRIVACY**
  - AHMED, F., LIU, A.X., JIN, R., Social Graph Publishing With Privacy Guarantees. 36th Ieee International Conference On Distributed Computing Systems, Icdcs 2016 2016-August: 447-456 (2016). Times cited: 11
  - WANG, Y., YANG, L., CHEN, X., ZHANG, X., HE, Z., Enhancing Social Network Privacy With Accumulated Non-zero Prior Knowledge. Information Sciences 445-446: 6-21 (2018). Times cited: 5
  - GAO, T., LI, F., Studying The Utility Preservation In Social Network Anonymization Via Persistent Homology. Computers And Security 77: 49-64 (2018). Times cited: 4

- GAO, T., LI, F., CHEN, Y., ZOU, X., Preserving Local Differential Privacy In Online Social Networks. 12th International Conference On Wireless Algorithms, Systems, And Applications, Wasa 2017 10251 LNCS: 393-405 (2017). Times cited: 4
- CHEN, L., ZHU, P., Preserving The Privacy Of Social Recommendation With A Differentially Private Approach. Ieee International Conference On Smart City, Smartcity 2015 : 780-785 (2015). Times cited: 4
- ZHU, T., YANG, M., XIONG, P., XIANG, Y., ZHOU, W., An Iteration-based Differentially Private Social Network Data Release. Computer Systems Science And Engineering 33:2 61-69 (2018). Times cited: 2
- GAO, T., LI, F., Preserving Graph Utility In Anonymized Social Networks? A Study On The Persistent Homology. 14th Ieee International Conference On Mobile Ad Hoc And Sensor Systems, Mass 2017 : 348-352 (2017). Times cited: 1

## OPTIMIZATION

- HARTUNG, S., NICHTERLEIN, A., NIEDERMEIER, R., SUCHý, O., A Refined Complexity Analysis Of Degree Anonymization In Graphs. Information And Computation 243: 249-262 (2015). Times cited: 19
- MURAKAMI, K., UNO, T., Optimization Algorithm For K-anonymization Of Datasets With Low Information Loss. International Journal Of Information Security 17:6 631-644 (2018). Times cited: 5
- JIANG, H., ZENG, G., HU, K., A Graph-clustering Anonymity Method Implemented By Genetic Algorithm For Privacy-preserving. Jisuanji Yanjiu Yu Fazhan/computer Research And Development 53:10 2354-2364 (2016). Times cited: 4

## WEIGHTED-MAXIMUM-COMMON-SUBGRAPH-(WMCS)

- AL-KHARJI, S., TIAN, Y., AL-RODHAAN, M., A Novel (k, X)-isomorphism Method For Protecting Privacy In Weighted Social Network. 21st Saudi Computer Society National Computer Conference, Ncc 2018 : - (2018). Times cited: 5
- YUAN, M., CHEN, L., YU, P.S., MEI, H., Privacy Preserving Graph Publication In A Distributed Environment. World Wide Web 18:5 1481-1517 (2015). Times cited: 3

## CRYPTOGRAPHY

- SAMANTHULA, B.K., RAO, F.-Y., BERTINO, E., YI, X., Privacy-preserving Protocols For Shortest Path Discovery Over Outsourced Encrypted Graph Data. 16th Ieee International Conference On Information Reuse And Integration, Iri 2015 : 427-434 (2015). Times cited: 12
- ASIF, W., RAY, I.G., TAHIR, S., RAJARAJAN, M., Privacy-preserving Anonymization With Restricted Search (pars) On Social Network Data For Criminal Investigations. 19th Ieee/acis International Conference On Software Engineering, Artificial Intelligence, Networking And Parallel/distributed Computing, Snpd 2018 : 329-334 (2018). Times cited: 5

- **RANDOM-PERTURBATION**
  - YAO, L., LIU, D., WANG, X., WU, G., Preserving The Relationship Privacy Of The Published Social-network Data Based On Compressive Sensing. 25th Ieee/acm International Symposium On Quality Of Service, Iwqos 2017 : - (2017). Times cited: 3
  - LAN, L., Preserving Weighted Social Networks Privacy Using Vectors Similarity. 8th International Conference On Biomedical Engineering And Informatics, Bmei 2015 : 789-794 (2016). Times cited: 1
- **INFERENCE-ALGORITHM**
  - LIU, X., LI, M., XIA, X., LI, J., ZONG, C., ZHU, R., Spatio-temporal Features Based Sensitive Relationship Protection In Social Networks. 15th Web Information Systems And Applications Conference, Wisa 2018 11242 LNCS: 330-343 (2018). Times cited: 4
  - ZAGHIAN, A., BAGHERI, A., A Combined Model Of Clustering And Classification Methods For Preserving Privacy In Social Networks Against Inference And Neighborhood Attacks. International Journal Of Security And Its Applications 10:1 95-102 (2016). Times cited: 4
  - SRIVASTAVA, A., GEETHAKUMARI, G., Determining Privacy Utility Trade-off For Online Social Network Data Publishing. 12th Ieee International Conference Electronics, Energy, Environment, Communication, Computer, Control, Indicon 2015 : - (2016). Times cited: 0
- **DIRECTED-GRAPH**
  - MILANI FARD, A., WANG, K., Neighborhood Randomization For Link Privacy In Social Network Analysis. World Wide Web 18:1 9-32 (2015). Times cited: 30
  - HAJIAN, S., TASSA, T., BONCHI, F., Individual Privacy In Social Influence Networks. Social Network Analysis And Mining 6:1 1-14 (2016). Times cited: 6
- **LEVEL-CUT-HEURISTIC-BASED-CLUSTERING-ALGORITHM**
  - MOHAPATRA, D., PATRA, M.R., A Level-cut Heuristic-based Clustering Approach For Social Graph Anonymization. Social Network Analysis And Mining 7:1 - (2017). Times cited: 2
- **ADVERSARIAL-MACHINE-LEARNING**
  - GARCíA-RECUERO, J. Burdges, Efficient Privacy-preserving Adversarial Learning In Decentralized Online Social Networks. 9th Ieee/acm International Conference On Advances In Social Networks Analysis And Mining, Asonam 2017 : 1132-1135 (2017). Times cited: 1
- **MOBILE-SOCIAL-NETWORK**
  - CHEN, Z.-G., KANG, H.-S., YIN, S.-N., KIM, S.-R., An Efficient Privacy Protection In Mobility Social Network Services With Novel Clustering-based Anonymization. Eurasip Journal On Wireless Communications And Networking 2016:1 - (2016). Times cited: 8

- TIAN, H., LU, Y., LIU, J., YU, J., Betweenness Centrality Based K-anonymity For Privacy Preserving In Social Networks. 16th International Conference On Advances In Mobile Computing And Multimedia, Momm 2018 : 3-7 (2018). Times cited: 0

## DECISION-SUPPORT-SYSTEMS-(DSS)

- NAGARAJ, K., SRIDHAR, A., SHARVANI, G.S., Identification Of Network Communities And Assessment Of Privacy Using Hybrid Algorithm. 2nd International Conference On Computational Systems And Information Technology For Sustainable Solutions, Csitss 2017 : - (2018). Times cited: 0
- SAFIA, B., YACINE, C., Privacy Preservation In Social Networks Sequential Publishing. 32nd Ieee International Conference On Advanced Information Networking And Applications, Aina 2018 2018-May: 732-739 (2018). Times cited: 0

## GRAPH-OPERATIONS

- HARTUNG, S., TALMON, N., The Complexity Of Degree Anonymization By Graph Contractions. 12th Annual Conference On Theory And Applications Of Models Of Computation, Tamc 2015 9076: 260-271 (2015). Times cited: 7

## SYNTHETIC-GRAPH-GENERATION

- RAJAEI, M., HAGHJOO, M.S., MIYANEH, E.K., Ambiguity In Social Network Data For Presence, Sensitive-attribute, Degree And Relationship Privacy Protection. Plos One 10:6 - (2015). Times cited: 9

## MUTUAL-FRIEND-ATTACK

- MACWAN, K.R., PATEL, S.J., K -nmf Anonymization In Social Network Data Publishing. Computer Journal 61:4 601-613 (2018). Times cited: 8

## NETWORK-COHESION

- AHMED, K.W., MOURI, I.J., ZAMAN, R., YEASMIN, N., A Privacy Preserving Personalized Group Recommendation Framework. 6th International Advanced Computing Conference, Iacc 2016 : 594-598 (2016). Times cited: 4

## PAGERANK-ALGORITHM

- KUMAR, S., KUMAR, P., Upper Approximation Based Privacy Preserving In Online Social Networks. Expert Systems With Applications 88: 276-289 (2017). Times cited: 10

## UNCERTAIN-GRAPHS

- NGUYEN, H.H., IMINE, A., RUSINOWITCH, M., Anonymizing Social Graphs Via Uncertainty Semantics. 10th Acm Symposium On Information, Computer And Communications Security, Asiaccs 2015 : 495-506 (2015). Times cited: 25
- NGUYEN, H.H., IMINE, A., RUSINOWITCH, M., A Maximum Variance Approach For Graph Anonymization. Chinese Materials Congress, Cmc 2014 8930: 49-64 (2015). Times cited: 7

## GRAPH-PARTITIONING

- HUOWEN, J., HUANLIANG, X., HUIYUN, Z., A Novel Approach To Achieving κ-anonymization For Social Network Privacy Preservation Based On Vertex Connectivity.

Ieee Advanced Information Technology, Electronic And Automation Control Conference, Iaeac 2015 : 1097-1100 (2016). Times cited: 1

- **BIG-DATA**
  - GRINING, K., KLONOWSKI, M., SULKOWSKA, M., How To Cooperate Locally To Improve Global Privacy In Social Networks? On Amplification Of Privacy Preserving Data Aggregation. 16th Ieee International Conference On Trust, Security And Privacy In Computing And Communications, 11th Ieee International Conference On Big Data Science And Engineering And 14th Ieee International Conference On Embedded Software And Systems, Trustcom/bigdatase/icess 2017 : 464-471 (2017). Times cited: 1

# Subperiod 2019-2022

- **SINGULAR-VALUE-DECOMPOSITION-(SVD)**
  - BIAN, J., LI, S., Research On A Privacy Preserving Clustering Method For Social Network. 4th Ieee International Conference On Cloud Computing And Big Data Analytics, Icccbda 2019 : 29-33 (2019). Times cited: 2
  - GUO, M., CHI, C.-H., ZHENG, H., HE, J., ZHANG, X., A Subgraph Isomorphism-based Attack Towards Social Networks. 2021 Ieee/wic/acm International Conference On Web Intelligence And Intelligent Agent Technology, Wi-iat 2021 : 520-528 (2021). Times cited: 0
  - PRIYADHARSHINI, V.M., VALARMATHI, A., Adaptive Framework For Privacy Preserving In Online Social Networks. Wireless Personal Communications 121:3 2273-2290 (2021). Times cited: 0
- **PRIVACY-PRESERVATION**
  - LANGARI, R.K., SARDAR, S., AMIN MOUSAVI, S.A., RADFAR, R., Combined Fuzzy Clustering And Firefly Algorithm For Privacy Preserving In Social Networks. Expert Systems With Applications 141: - (2020). Times cited: 35
  - SIDDULA, M., LI, Y., CHENG, X., TIAN, Z., CAI, Z., Anonymization In Online Social Networks Based On Enhanced Equi-cardinal Clustering. Ieee Transactions On Computational Social Systems 6:4 809-820 (2019). Times cited: 23
  - WANG, X., CHEN, W., CHOU, J.-K., BRYAN, C., GUAN, H., PAN, R., MA, K.-L., Graphprotector: A Visual Interface For Employing And Assessing Multiple Privacy Preserving Graph Algorithms. Ieee Transactions On Visualization And Computer Graphics 25:1 193-203 (2019). Times cited: 18
  - HUANG, H., ZHANG, D., XIAO, F., WANG, K., GU, J., WANG, R., Privacy-preserving Approach Pbcn In Social Network With Differential Privacy. Ieee Transactions On Network And Service Management 17:2 931-945 (2020). Times cited: 16
  - JI, S., WANG, T., CHEN, J., LI, W., MITTAL, P., BEYAH, R., De-sag: On The De-anonymization Of Structure-attribute Graph Data. Ieee Transactions On Dependable And Secure Computing 16:4 594-607 (2019). Times cited: 13
  - KIABOD, M., DEHKORDI, M.N., BAREKATAIN, B., Tsram: A Time-saving K-degree Anonymization Method In Social Network. Expert Systems With Applications 125: 378-396 (2019). Times cited: 12
  - ROS-MARTíN, M., SALAS, J., CASAS-ROMA, J., Scalable Non-deterministic Clustering-based K-anonymization For Rich Networks. International Journal Of Information Security 18:2 219-238 (2019). Times cited: 11
  - MORTAZAVI, R., ERFANI, S.H., Gram: An Efficient (k, L) Graph Anonymization Method. Expert Systems With Applications 153: - (2020). Times cited: 10
  - GAO, T., LI, F., Sharing Social Networks Using A Novel Differentially Private Graph Model. 16th Ieee Annual Consumer Communications And Networking Conference, Ccnc 2019 : - (2019). Times cited: 10

- ZHAO, Y., LI, Z., Privacy Management In Social Network Data Publishing With Community Structure. 2nd International Conference On Healthcare Science And Engineering, Healthcare 2018 536: 141-151 (2019). Times cited: 0
- MEDKOVA, J., High-degree Noise Addition Method For The κ-degree Anonymization Algorithm. Joint 11th International Conference On Soft Computing And Intelligent Systems And 21st International Symposium On Advanced Intelligent Systems, Scis-isis 2020 : - (2020). Times cited: 0

## • ARTIFICIAL-NEURAL-NETWORK-(ANN)

- HU, X., ZHU, T., ZHAI, X., ZHOU, W., ZHAO, W., Privacy Data Propagation And Preservation In Social Media: A Real-world Case Study. Ieee Transactions On Knowledge And Data Engineering : - (2021). Times cited: 1
- HSIEH, I., LI, C., Netfense: Adversarial Defenses Against Privacy Attacks On Neural Networks For Graph Data. Ieee Transactions On Knowledge And Data Engineering : - (2021). Times cited: 0
- KAUR, I., BHARDWAJ, V., K-anonymity Enhancement For Privacy Preservation With Hybridization Of Cuckoo Search And Neural Network Using Clustering. International Journal Of Innovative Technology And Exploring Engineering 8:10 1189-1196 (2019). Times cited: 0

## • GRAPH-MATCHING-TECHNIQUE

- ZHANG, H., LI, X., XU, J., XU, L., Graph Matching Based Privacy-preserving Scheme In Social Networks. 7th International Symposium On Security And Privacy In Social Networks And Big Data, Socialsec 2021 1495 CCIS: 110-118 (2021). Times cited: 2
- HUANG, K., HU, H., ZHOU, S., GUAN, J., YE, Q., ZHOU, X., Privacy And Efficiency Guaranteed Social Subgraph Matching. Vldb Journal 31:3 581-602 (2022). Times cited: 0
- ZUO, X., LI, L., PENG, H., LUO, S., YANG, Y., Privacy-preserving Subgraph Matching Scheme With Authentication In Social Networks. Ieee Transactions On Cloud Computing : - (2020). Times cited: 0

## • MATRIX-ALGEBRA

- AHMED, F., LIU, A.X., JIN, R., Publishing Social Network Graph Eigenspectrum With Privacy Guarantees. Ieee Transactions On Network Science And Engineering 7:2 892-906 (2020). Times cited: 5
- LI, A., FANG, J., JIANG, Q., ZHOU, B., JIA, Y., A Graph Data Privacy-preserving Method Based On Generative Adversarial Networks. 21st International Conference On Web Information Systems Engineering, Wise 2020 12343 LNCS: 227-239 (2020). Times cited: 3
- GAO, T., LI, F., Differential Private Social Network Publication And Persistent Homology Preservation. Ieee Transactions On Network Science And Engineering 8:4 3152-3166 (2021). Times cited: 1

## • PERTURBATION-TECHNIQUES

- TORRA, V., SALAS, J., Graph Perturbation As Noise Graph Addition: A New Perspective For Graph Anonymization. 14th International Workshop On Data Privacy Management,

Dpm 2019, And The 3rd International Workshop On Cryptocurrencies And Blockchain Technology, Cbt 2019, Held In Conjunction With The 24th European Symposium On Research In Computer Security, Esorics 2019 11737 LNCS: 121-137 (2019). Times cited: 4

- IFTIKHAR, M., WANG, Q., LIN, Y., Dk-microaggregation: Anonymizing Graphs With Differential Privacy Guarantees. 24th Pacific-asia Conference On Knowledge Discovery And Data Mining, Pakdd 2020 12085 LNAI: 191-203 (2020). Times cited: 3
- ZHANG, C., JIANG, H., CHENG, X., ZHAO, F., CAI, Z., TIAN, Z., Utility Analysis On Privacy-preservation Algorithms For Online Social Networks: An Empirical Study. Personal And Ubiquitous Computing 25:6 1063-1079 (2021). Times cited: 1
- HORAWALAVITHANA, S., FLORES, J.G.A., SKVORETZ, J., IAMNITCHI, A., Behind The Mask: Understanding The Structural Forces That Make Social Graphs Vulnerable To Deanonymization. Ieee Transactions On Computational Social Systems 6:6 1343-1356 (2019). Times cited: 1
- SIDDARAMAPPA, H.K., MARADITHAYA, S., KUMAR, S., Secure Analysis Of Social Media Data. 2019 International Conference On Automation, Computational And Technology Management, Icactm 2019 : 315-319 (2019). Times cited: 0
- ZHANG, X., LI, J., HE, X., LIU, J., Distributed Graph Perturbation Algorithm On Social Networks With Reachability Preservation. 7th Ccf Academic Conference On Bigdata, Ccf Bigdata 2019 1120 CCIS: 194-208 (2019). Times cited: 0
- ZHAO, Y., LI, Z., Privacy Management In Social Network Data Publishing With Community Structure. 2nd International Conference On Healthcare Science And Engineering, Healthcare 2018 536: 141-151 (2019). Times cited: 0

## CLUSTER-ANALYSIS

- ZHANG, J., ZHAO, B., SONG, G., NI, L., YU, J., Maximum Delay Anonymous Clustering Feature Tree Based Privacy-preserving Data Publishing In Social Networks. 7th International Conference On Identification, Information And Knowledge In The Internet Of Things, Iiki 2018 147: 643-646 (2019). Times cited: 7
- BIAN, J., LI, S., Research On A Privacy Preserving Clustering Method For Social Network. 4th Ieee International Conference On Cloud Computing And Big Data Analytics, Icccbda 2019 : 29-33 (2019). Times cited: 2
- SU, J., CAO, Y., CHEN, Y., LIU, Y., SONG, J., Privacy Protection Of Medical Data In Social Network. Bmc Medical Informatics And Decision Making 21: - (2021). Times cited: 0
- YU, S., NIU, X., YANG, Y., SHI, X., DONG, W., DENG, Z., Personalized Privacy Protection Based On Liversity Against Connection Fingerprint Attack. 2019 Ieee International Conference On Power, Intelligent Computing And Systems, Icpics 2019 : 205-211 (2019). Times cited: 0

## SOCIAL-NETWORK-ANALYSIS-(SNA)

- ROS-MARTíN, M., SALAS, J., CASAS-ROMA, J., Scalable Non-deterministic Clustering-based K-anonymization For Rich Networks. International Journal Of Information Security 18:2 219-238 (2019). Times cited: 11

- HUMSKI, L., PINTAR, D., VRANIC, M., Analysis Of Facebook Interaction As Basis For Synthetic Expanded Social Graph Generation. Ieee Access 7: 6622-6636 (2019). Times cited: 3
- LI, P., ZHOU, F., XU, Z., LI, Y., XU, J., Privacy-preserving Graph Operations For Social Network Analysis. 6th International Symposium On Security And Privacy In Social Networks And Big Data, Socialsec 2020 1298 CCIS: 303-317 (2020). Times cited: 2

## STRUCTURED-DATA

- LI, P., ZHOU, F., XU, Z., LI, Y., XU, J., Privacy-preserving Graph Operations For Social Network Analysis. 6th International Symposium On Security And Privacy In Social Networks And Big Data, Socialsec 2020 1298 CCIS: 303-317 (2020). Times cited: 2
- TIAN, J., LIU, C., GOU, G., LI, Z., XIONG, G., GUAN, Y., Recgraph: Graph Recovery Attack Using Variational Graph Autoencoders. 2021 Ieee International Performance, Computing, And Communications Conference, Ipccc 2021 2021-October: - (2021). Times cited: 0
- MOHAPATRA, D., PATRA, M.R., Cluster-based Anonymization Of Assortative Networks. 5th International Conference On Computational Intelligence In Data Mining, Iccidm 2018 990: 709-718 (2020). Times cited: 0

## BACKGROUND-KNOWLEDGE

- CHICHA, E., BOUNA, B.A., NASSAR, M., CHBEIR, R., HARATY, R.A., OUSSALAH, M., BENSLIMANE, D., ALRAJA, M.N., A User-centric Mechanism For Sequentially Releasing Graph Datasets Under Blowfish Privacy. Acm Transactions On Internet Technology 21:1 - (2021). Times cited: 3
- TIAN, Y., YAN, J., HU, J., WU, Z., A Privacy Preserving Model In Uncertain Graph Mining. 2018 International Conference On Networking And Network Applications, Nana 2018 : 102-106 (2019). Times cited: 2
- REN, X., JIANG, D., The (P, α, K) anonymity model for privacy protection of personal information in the social networks. Wireless Communications And Mobile Computing 2022: - (2022). Times cited: 1
- KIRANMAYI, M., MAHESWARI, N., SIVAGAMI, M., Preservation Of Attribute Couplet Attack By Node Addition In Social Networks. 3rd International Conference On Computing And Communications Technologies, Iccct 2019 : 99-104 (2019). Times cited: 1
- ZHAO, Y., LI, Z., Privacy Management In Social Network Data Publishing With Community Structure. 2nd International Conference On Healthcare Science And Engineering, Healthcare 2018 536: 141-151 (2019). Times cited: 0

## REAL-WORLD-DATASETS

- GAO, T., LI, F., Sharing Social Networks Using A Novel Differentially Private Graph Model. 16th Ieee Annual Consumer Communications And Networking Conference, Ccnc 2019 : - (2019). Times cited: 10
- KUMAR, S., KUMAR, P., Privacy Preserving In Online Social Networks Using Fuzzy Rewiring. Ieee Transactions On Engineering Management : - (2021). Times cited: 3

- IFTIKHAR, M., WANG, Q., LIN, Y., Dk-microaggregation: Anonymizing Graphs With Differential Privacy Guarantees. 24th Pacific-asia Conference On Knowledge Discovery And Data Mining, Pakdd 2020 12085 LNAI: 191-203 (2020). Times cited: 3
- PRIYADHARSHINI, V.M., VALARMATHI, A., Adaptive Framework For Privacy Preserving In Online Social Networks. Wireless Personal Communications 121:3 2273-2290 (2021). Times cited: 0
- MINELLO, G., ROSSI, L., TORSELLO, A., k-Anonymity on Graphs Using the Szemerédi Regularity Lemma. Ieee Transactions On Network Science And Engineering 8:2 1283-1292 (2021). Times cited: 0
- HSIEH, I., LI, C., Netfense: Adversarial Defenses Against Privacy Attacks On Neural Networks For Graph Data. Ieee Transactions On Knowledge And Data Engineering : - (2021). Times cited: 0

## DIRECTED-GRAPH

- SIDDULA, M., LI, Y., CHENG, X., TIAN, Z., CAI, Z., Anonymization In Online Social Networks Based On Enhanced Equi-cardinal Clustering. Ieee Transactions On Computational Social Systems 6:4 809-820 (2019). Times cited: 23
- CASAS-ROMA, J., SALAS, J., MALLIAROS, F.D., VAZIRGIANNIS, M., K-degree Anonymity On Directed Networks. Knowledge And Information Systems 61:3 1743-1768 (2019). Times cited: 6
- HOANG, A.-T., CARMINATI, B., FERRARI, E., Cluster-based Anonymization Of Directed Graphs. 5th Ieee International Conference On Collaboration And Internet Computing, Cic 2019 : 91-100 (2019). Times cited: 3
- KHARE, N., KUMARAN, U., A Secure And Privacy-preserving Approach To Protect User Data Across Cloud Based Online Social Networks. International Journal Of Grid And High Performance Computing 12:2 1-24 (2020). Times cited: 2
- HORAWALAVITHANA, S., FLORES, J.G.A., SKVORETZ, J., IAMNITCHI, A., Behind The Mask: Understanding The Structural Forces That Make Social Graphs Vulnerable To Deanonymization. Ieee Transactions On Computational Social Systems 6:6 1343-1356 (2019). Times cited: 1

## GENETIC-ALGORITHM-(GA)

- YAZDANJUE, N., FATHIAN, M., AMIRI, B., Evolutionary Algorithms For K-anonymity In Social Networks Based On Clustering Approach. Computer Journal 63:7 1039-1062 (2021). Times cited: 7
- RAJABZADEH, S., SHAHSAFI, P., KHORAMNEJADI, M., A Graph Modification Approach For K-anonymity In Social Networks Using The Genetic Algorithm. Social Network Analysis And Mining 10:1 - (2020). Times cited: 5

## COLLUSION

- JAN, Y.-S., TSAI, H.-P., A Collusion Avoidance Node Selection Scheme For Social Network-based Distributed Data Storage. 20th International Conference On Mobile Data Management, Mdm 2019 2019-June: 507-512 (2019). Times cited: 0

## POWER-LAW

- SUN, Q., YU, J., JIANG, H., CHEN, Y., CHENG, X., De-anonymizing Scale-free Social Networks By Using Spectrum Partitioning Method. 7th International Conference On Identification, Information And Knowledge In The Internet Of Things, Iiki 2018 147: 441-445 (2019). Times cited: 1
- YU, D., ZHAO, H., WANG, L.-E., LIU, P., LI, X., A Hierarchical K-anonymous Technique Of Graphlet Structural Perception In Social Network Publishing. 4th International Conference On Mobile, Secure, And Programmable Networking, Mspn 2018 11005 LNCS: 224-239 (2019). Times cited: 1

## SYNTHETIC-GRAPH-GENERATION

- HUANG, H., YANG, Y., LI, Y., Psg : Local Privacy Preserving Synthetic Social Graph Generation. 17th Eai International Conference On Collaborative Computing: Networking, Applications, And Worksharing, Collaboratecom 2021 406 LNICST: 389-404 (2021). Times cited: 1
- DJOMO, R., DJOTIO NDIE, T., Towards Preventing Neighborhood Attacks: Proposal Of A New Anonymization's Approach For Social Networks Data. 10th Eai International Conference On Big Data Technologies And Applications, Bdta 2020 And 13th Eai International Conference On Wireless Internet, Wicon 2020 371 LNICST: 195-208 (2021). Times cited: 0

## FIREFLY-ALGORITHM-(FA)

- LANGARI, R.K., SARDAR, S., AMIN MOUSAVI, S.A., RADFAR, R., Combined Fuzzy Clustering And Firefly Algorithm For Privacy Preserving In Social Networks. Expert Systems With Applications 141: - (2020). Times cited: 35

## F-FACTORS

- KIABOD, M., NADERI DEHKORDI, M., BAREKATAIN, B., A Fast Graph Modification Method For Social Network Anonymization. Expert Systems With Applications 180: - (2021). Times cited: 2

## M-PRIVACY

- KADHIWALA, B., PATEL, S.J., A Novel K-anonymization Approach To Prevent Insider Attack In Collaborative Social Network Data Publishing. 15th International Conference On Information Systems Security, Iciss 2019 11952 LNCS: 255-275 (2019). Times cited: 1

## LINKAGE-ATTACK

- VINOTH, N.A.S., YOHAPRIYAA, M., JANANI, K., VIJAYPRIYA, V., Secured Health Protective Services In Social Media Network Using K-anonymity. International Journal Of Recent Technology And Engineering 8:2 Special Issue 4 244-247 (2019). Times cited: 0

## UNCERTAIN-GRAPHS

- YAN, J., ZHANG, L., TIAN, Y., WEN, G., HU, J., An Uncertain Graph Approach For Preserving Privacy In Social Networks Based On Important Nodes. 2018 International Conference On Networking And Network Applications, Nana 2018 : 107-111 (2019). Times cited: 6

- WU, Z.-Q., HU, J., TIAN, Y.-P., SHI, W.-C., YAN, J., Privacy Preserving Algorithms Of Uncertain Graphs In Social Networks . Ruan Jian Xue Bao/journal Of Software 30:4 1106-1120 (2019). Times cited: 4

- **K-DEGREE**
  - KIABOD, M., DEHKORDI, M.N., BAREKATAIN, B., Tsram: A Time-saving K-degree Anonymization Method In Social Network. Expert Systems With Applications 125: 378-396 (2019). Times cited: 12
  - MOHAPATRA, D., PATRA, M.R., Graph Anonymization Using Hierarchical Clustering. 4th International Conference On Computational Intelligence In Data Mining, Iccidm 2017 711: 145-154 (2019). Times cited: 1
- **NOISE-NODE**
  - HAMZEHZADEH, S., MAZINANI, S.M., Annm: A New Method For Adding Noise Nodes Which Are Used Recently In Anonymization Methods In Social Networks. Wireless Personal Communications 107:4 1995-2017 (2019). Times cited: 7
- **LINK-PRIVACY**
  - GUO, Y., LIU, Z., ZENG, Y., WANG, R., MA, J., Preserving Privacy For Hubs And Links In Social Networks. 2018 International Conference On Networking And Network Applications, Nana 2018 : 263-269 (2019). Times cited: 3
- **DYNAMIC-NETWORK**
  - MACWAN, K., PATEL, S., Privacy Preserving Approach In Dynamic Social Network Data Publishing. 15th International Conference On Information Security Practice And Experience, Ispec 2019 11879 LNCS: 381-398 (2019). Times cited: 3
- **SENSITIVE-WEIGHTED-EDGES**
  - GONG, W., JIN, R., LI, Y., YANG, L., MEI, J., Privacy Protection Method For Sensitive Weighted Edges In Social Networks. Ksii Transactions On Internet And Information Systems 15:2 540-557 (2021). Times cited: 0
- **ATTRIBUTE-GRAPH**
  - JI, S., WANG, T., CHEN, J., LI, W., MITTAL, P., BEYAH, R., De-sag: On The De-anonymization Of Structure-attribute Graph Data. Ieee Transactions On Dependable And Secure Computing 16:4 594-607 (2019). Times cited: 13
- **SUPERVISED-MACHINE-LEARNING-ALGORITHMS**
  - SU, J., CAO, Y., CHEN, Y., Privacy Preservation Based On Key Attribute And Structure Generalization Of Social Network For Medical Data Publication. 15th International Conference On Intelligent Computing, Icic 2019 11643 LNCS: 388-399 (2019). Times cited: 2

# Appendix B

The keywords and their PageRank scores that are included in each module of the alluvial diagram are provided in the following tables.

| Subperiod 2007-2010 | |
|---|---|
| Module 1 | PageRank Score |
| ANONYMIZATION | 0.0307 |
| PRIVACY-ATTACK-MODEL | 0.0218 |
| K-ANONYMITY | 0.0197 |
| DATA-PUBLISHING | 0.0137 |
| BACKGROUND-KNOWLEDGE | 0.0127 |
| PRIVACY-PRESERVATION | 0.0121 |
| REAL-WORLD-DATASETS | 0.0118 |
| STRUCTURAL-PROPERTIES | 0.0103 |
| SOCIAL-NETWORK-ANALYSIS-(SNA) | 0.00985 |
| COMMUNITY-DETECTION-ALGORITHMS | 0.00670 |
| SUPPRESSION | 0.00370 |
| STRUCTURED-DATA | 0.00278 |
| SOCIAL-NETWORK | 0.00206 |
| LARGE-DATASET | 0.000412 |
| Module 2 | PageRank Score |
| SPLIT-ALGORITHMS | 0.0428 |
| UNDIRECTED-GRAPH | 0.0333 |
| DATA-MINING | 0.0206 |
| RELATIONAL-DATA | 0.0200 |
| GENERALIZATION-APPROACH | 0.0166 |
| GRAPH-MATCHING-TECHNIQUE | 0.00988 |
| K-AUTOMORPHISM | 0.00677 |
| CLUSTERING-ALGORITHMS | 0.00 |
| Module 3 | PageRank Score |
| POWER-LAW | 0.0329 |
| INPUT-GRAPH | 0.0234 |
| SYNTHETIC-GRAPH-GENERATION | 0.0202 |
| PRIVACY-PRESERVING-DATA-MINING-(PPDM) | 0.0175 |
| DEGREE-SEQUENCE | 0.00686 |
| DIFFERENTIAL-PRIVACY | 0.00 |
| GRAPH-MODIFICATION-APPROACH | 0.00 |
| Module 4 | PageRank Score |
| SET-VALUED-DATA | 0.0386 |
| MATRIX-ALGEBRA | 0.0290 |
| LABEL-HIERARCHY | 0.0195 |
| GRAPH-ISOMORPHISM | 0.0136 |
| NEIGHBORHOOD-ATTACK | 0.00 |
| Module 5 | PageRank Score |
| MICRODATA | 0.0360 |
| INFORMATION-LOSS | 0.0265 |
| GRAPH-DATA | 0.0191 |
| GREEDY-ALGORITHM | 0.0152 |
| SENSITIVE-ATTRIBUTES | 0.00 |
| Module 6 | PageRank Score |
| PROBABILISTIC-GRAPH | 0.0296 |
| SHORTEST-PATH | 0.0221 |
| PAIRED-VERTICES-K-DEGREE-ANONYMITY | 0.0154 |
| RE-IDENTIFICATION-ATTACKS | 0.0108 |
| WEIGHTED-NETWORK | 0.00919 |
| LINK-PRIVACY | 0.000412 |
| Module 7 | PageRank Score |
| LINK-DISCLOSURE | 0.0230 |
| IDENTITY-DISCLOSURE | 0.0204 |
| DYNAMIC-PROGRAMMING | 0.00799 |
| RANDOMIZATION-APPROACH | 0.00410 |
| Module 8 | PageRank Score |
| SUBGRAPH | 0.0273 |
| EDGE-BETWEENNESS-CENTRALITY | 0.0191 |
| K-NEAREST-NEIGHBORS-ALGORITHM-(KNN) | 0.00489 |
| Module 9 | PageRank Score |
| GRAPH-THEORY | 0.0269 |
| GRAPH-MINING | 0.0163 |
| EDGE-WEIGHT-ANONYMIZATION | 0.00685 |
| Module 10 | PageRank Score |
| FACEBOOK | 0.0198 |
| PHISHING-ATTACK | 0.0163 |
| INTERSECTION-ATTACK | 0.00678 |
| Module 11 | PageRank Score |
| DECISION-SUPPORT-SYSTEMS-(DSS) | 0.0200 |
| IDENTITY-INFORMATION-HIDING | 0.0116 |
| EIGENVECTOR-CENTRALITY | 0.00357 |
| TOPOLOGICAL-INFORMATION | 0.00124 |
| Module 12 | PageRank Score |
| NP-HARD | 0.0227 |
| HOMOGENEITY-ATTACK | 0.00520 |
| STRUCTURAL-ATTACKS | 0.00371 |
| UTILITY | 0.00 |
| Module 13 | PageRank Score |
| NOISY-OR | 0.0170 |
| LINK-MINING | 0.00756 |
| Module 14 | PageRank Score |
| LINK-PREDICTION | 0.0132 |
| DYNAMIC-NETWORK | 0.000412 |

| Subperiod 2011-2014 | |
|---|---|
| Module 1 | PageRank Score |
| SOCIAL-NETWORK | 0.0201 |
| PRIVACY-PRESERVATION | 0.0191 |
| ANONYMIZATION | 0.0150 |
| STRUCTURAL-PROPERTIES | 0.0111 |
| GRAPH-THEORY | 0.0106 |
| K-ANONYMITY | 0.0101 |
| DATA-PUBLISHING | 0.00984 |
| DATA-MINING | 0.00958 |
| REAL-WORLD-DATASETS | 0.00841 |
| PRIVACY-ATTACK-MODEL | 0.00834 |
| PRIVACY-BREACH | 0.00643 |
| GENETIC-ALGORITHM-(GA) | 0.00382 |
| UNCERTAIN-GRAPHS | 0.00170 |
| STRUCTURAL-DISPARITY | 0.00170 |
| MUTUAL-FRIEND-ATTACK | 0.00102 |
| PERSONALIZED-ANONYMITY | 0.000848 |
| TARGET-NODES | 0.000769 |
| Module 2 | PageRank Score |
| FUZZY-SET-THEORY | 0.0130 |
| MICROAGGREGATION | 0.0111 |
| MATRIX-DECOMPOSITION | 0.0111 |
| NOISE-ADDITION | 0.0111 |
| HUBS | 0.0103 |
| BRIDGES | 0.0103 |
| CRISP | 0.0103 |
| CLUSTERING-ALGORITHMS | 0.00987 |
| GRAPH-PARTITIONING | 0.00884 |
| GENERALIZATION-APPROACH | 0.00830 |
| Module 3 | PageRank Score |
| DISTRIBUTED-ALGORITHM | 0.0113 |
| FRIENDSHIP-GRAPH | 0.00963 |
| INFORMATION-HUBS | 0.00963 |
| SOCIAL-RELATIONSHIPS | 0.00963 |
| K-MEANS | 0.00920 |
| MESSAGE-PASSING | 0.00920 |
| MODULAR-DECOMPOSITION | 0.00920 |
| Module 4 | PageRank Score |
| DECISION-SUPPORT-SYSTEMS-(DSS) | 0.0111 |
| COMBINATORIAL-GRAPH | 0.00927 |
| INAPPROXIMABILITY | 0.00927 |
| PARAMETERIZED-COMPLEXITY | 0.00927 |
| EGO-NETWORK | 0.00845 |
| TRUST | 0.00845 |
| DEGREE-SEQUENCE | 0.00643 |
| PEER-TO-PEER-NETWORKS | 0.00521 |
| Module 5 | PageRank Score |
| DIFFERENTIAL-PRIVACY | 0.00979 |
| CLUSTER-COEFFICIENT | 0.00893 |
| DIVIDE-AND-CONQUER-ALGORITHM | 0.00877 |
| GRAPH-MINING | 0.00703 |
| LAPLACE-NOISE | 0.00701 |
| GOVERNMENT-AFFAIRS-OFFICE-AUTOMATION | 0.00388 |
| WORKFLOW-GRAPH | 0.00364 |
| RANDOM-NETWORK | 0.00364 |
| Module 6 | PageRank Score |
| (K,LAMBDA)-SIMILARITY | 0.00900 |
| HELLINGER-DISTANCE | 0.00900 |
| EDGE-WEIGHT-ANONYMIZATION | 0.00900 |
| OPTIMIZATION | 0.00683 |
| WEIGHTED-NETWORK | 0.00652 |
| BACKGROUND-KNOWLEDGE | 0.00564 |
| LABEL-BAG | 0.00103 |
| KERNEL-REGRESSION | 0.000652 |
| Module 7 | PageRank Score |
| PERTURBATION-TECHNIQUES | 0.0102 |
| DESTINATION-NODES | 0.00927 |
| UNDIRECTED-GRAPH | 0.00927 |
| SOCIAL-NETWORK-ANALYSIS-(SNA) | 0.00641 |
| PARTITIONING-APPROACH | 0.00630 |
| EDGE-BETWEENNESS-CENTRALITY | 0.00219 |
| LINK-PRIVACY | 0.00207 |
| UNWEIGHTED-NETWORK | 0.00125 |
| Module 8 | PageRank Score |
| RANDOM-PERTURBATION | 0.00817 |
| INFORMATION-THEORY | 0.00778 |
| RANDOMIZATION-APPROACH | 0.00750 |
| SMALL-WORLD-NETWORK | 0.00685 |
| LARGE-DATASET | 0.00522 |
| INTEGER-PROGRAMMING | 0.00498 |
| QUANTIFICATION | 0.00360 |
| HYBRID-ALGORITHMS | 0.00112 |
| Module 9 | PageRank Score |
| WEIGHTED-MAXIMUM-COMMON-SUBGRAPH-(WMCS) | 0.0107 |
| COMPLETE-GRAPH | 0.0107 |
| DISTRIBUTED-ENVIRONMENT | 0.0107 |
| DISTRIBUTED-STORAGE | 0.0107 |

| Module 10 | PageRank Score |
| --- | --- |
| GRAPH-DATA | 0.00698 |
| GRAPH-QUERIES | 0.00597 |
| REACHABILITY | 0.00597 |
| SPARSE-GRAPH | 0.00557 |
| CHARACTERISTIC-VECTOR | 0.00557 |
| GRAPH-MODIFICATION-APPROACH | 0.00516 |
| RE-IDENTIFICATION-ATTACKS | 0.00450 |
| Module 11 | PageRank Score |
| LABELED-GRAPH | 0.00629 |
| BIPARTITE-GRAPH | 0.00578 |
| NP-HARD | 0.00511 |
| EDGE-ADDITION/DELETION | 0.00474 |
| NODE-DEGREE | 0.00442 |
| DYNAMIC-PROGRAMMING | 0.00386 |
| TABLE-GRAPHS | 0.00283 |
| RELATIONAL-CLASSIFIERS | 0.00196 |
| Module 12 | PageRank Score |
| FACEBOOK | 0.0105 |
| CONFIDENTIALITY | 0.00782 |
| CENTRALITY-MEASURES | 0.00781 |
| LINKEDIN | 0.00781 |
| Module 13 | PageRank Score |
| SUBGRAPH | 0.00905 |
| K-SUBSET-ANONYMITY | 0.00825 |
| DEGREE-CONSTRAINED-SUBGRAPHS | 0.00825 |
| DEGREE-CONSTRAINED-EDITING | 0.00825 |
| Module 14 | PageRank Score |
| SYNTHETIC-GRAPH-GENERATION | 0.00760 |
| GRAPH-MODELS | 0.00643 |
| GRAPH-RECONSTRUCTION-ATTACK | 0.00519 |
| OUTPUT-PERTURBATION | 0.00519 |
| TOPOLOGICAL-STRUCTURE-SIMILARITY | 0.00483 |
| Module 15 | PageRank Score |
| ENCRYPTION | 0.00742 |
| PRIVACY-PRESERVING-PATH-FINDING | 0.00739 |
| CONTROLLED-INFORMATION-SHARING | 0.00739 |
| PERSONALIZATION | 0.00551 |
| Module 16 | PageRank Score |
| TOPOLOGICAL-INFORMATION | 0.00741 |
| TABULAR-DATA | 0.00627 |
| METRIC-EMBEDDING | 0.00550 |
| COMPLEX-ALGORITHMS | 0.00550 |
| IDENTITY-RE-IDENTIFICATION | 0.00242 |
| Module 17 | PageRank Score |
| COMMUNITY-DETECTION-ALGORITHMS | 0.00453 |
| HYPERGRAPH-CLUSTERING | 0.00449 |
| IDENTITY-DISCLOSURE | 0.00449 |
| STRUCTURAL-ATTACKS | 0.00442 |
| MEASUREMENT | 0.00334 |
| Module 18 | PageRank Score |
| NEIGHBORHOOD-INFORMATION | 0.00709 |
| CLOUD-COMPUTING | 0.00709 |
| INDISTINGUISHABILITY-PROBABILITY | 0.00709 |
| Module 19 | PageRank Score |
| HEALTH-INFORMATION | 0.00659 |
| ROUGH-SET-THEORY | 0.00586 |
| PRIVACY-PRESERVING-DATA-MINING-(PPDM) | 0.00458 |
| CRYPTOGRAPHY | 0.00403 |
| Module 20 | PageRank Score |
| RELATIONAL-DATA | 0.00604 |
| L-DIVERSITY | 0.00526 |
| SENSITIVE-ATTRIBUTES | 0.00442 |
| GREEDY-ALGORITHM | 0.00200 |
| NEIGHBORHOOD-ATTACK | 0.00166 |
| SEMI-EDGE-ANONYMITY | 0.000999 |
| Module 21 | PageRank Score |
| UTILITY | 0.00589 |
| INFORMATION-LOSS | 0.00376 |
| EDGE-RELEVANCE | 0.00174 |
| GRAPH-ISOMORPHISM | 0.00144 |
| TIME-SERIAL-DATA | 0.00112 |
| Module 22 | PageRank Score |
| T-CLOSENESS | 0.00417 |
| TWITTER | 0.00417 |
| Module 23 | PageRank Score |
| SUPERVISED-MACHINE-LEARNING-ALGORITHMS | 0.00416 |
| UNDERLYING-GRAPH | 0.00416 |
| Module 24 | PageRank Score |
| SEMI-SUPERVISED-CLUSTERING-ALGORITHMS | 0.00403 |
| EXTREME-LEARNING-MACHINE-(ELM) | 0.00403 |
| Module 25 | PageRank Score |
| (K1,K2)-SHORTEST-PATH | 0.00392 |
| K-SHORTEST-PATH | 0.00392 |
| Module 26 | PageRank Score |
| NEIGHBORHOOD-SUBGRAPH | 0.00356 |
| DYNAMIC-NETWORK | 0.00356 |

| Module 27 | PageRank Score |
| --- | --- |
| STRUCTURED-DATA | 0.00399 |
| GRAPH-MATCHING-TECHNIQUE | 0.00296 |
| Module 28 | PageRank Score |
| HEURISTIC-ALGORITHMS | 0.00504 |
| SENSITIVE-EDGES | 0.000936 |
| Module 29 | PageRank Score |
| SHORTEST-PATH | 0.00352 |
| DIRECTED-GRAPH | 0.00234 |
| Module 30 | PageRank Score |
| ARTIFICIAL-INTELLIGENCE-(AI) | 0.00249 |
| MACHINE-LEARNING | 0.00143 |

| Subperiod 2015-2018 | |
|---|---|
| Module 1 | PageRank Score |
| SOCIAL-NETWORK | 0.0152 |
| PRIVACY-PRESERVATION | 0.0147 |
| ANONYMIZATION | 0.0112 |
| GRAPH-THEORY | 0.00865 |
| K-ANONYMITY | 0.00802 |
| STRUCTURAL-PROPERTIES | 0.00749 |
| DATA-PUBLISHING | 0.00712 |
| ARTIFICIAL-INTELLIGENCE-(AI) | 0.00691 |
| CLUSTERING-ALGORITHMS | 0.00689 |
| DATA-MINING | 0.00613 |
| INFORMATION-LOSS | 0.00577 |
| PRIVACY-BREACH | 0.00549 |
| PRIVACY-ATTACK-MODEL | 0.00524 |
| BACKGROUND-KNOWLEDGE | 0.00513 |
| RE-IDENTIFICATION-ATTACKS | 0.00366 |
| BAYES-RULE | 0.000761 |
| BIPARTITE-GRAPH | 0.000733 |
| GEO-SOCIAL-NETWORK | 0.000450 |
| SEMI-STRUCTURED-DATA | 0.000437 |
| DYNAMIC-NETWORK | 0.000363 |
| ANT COLONY OPTIMIZATION (ACO) | 0.0000912 |
| Module 2 | PageRank Score |
| CLUSTER-PURITY | 0.0118 |
| CLUSTER-ANALYSIS | 0.0118 |
| CATEGORICAL-DATA | 0.0118 |
| DBSCAN-ALGORITHM | 0.0118 |
| ROUGH-ENTROPY | 0.0118 |
| MOBILE-SOCIAL-NETWORK | 0.0107 |
| BETWEENNESS-CENTRALITY | 0.00685 |
| IMPORTANT-FEATURES | 0.00685 |
| TARGET-NODES | 0.00685 |
| Module 3 | PageRank Score |
| VECTORS-SIMILARITY | 0.00954 |
| WEIGHTED-EUCLIDEAN-DISTANCE | 0.00954 |
| VECTOR-SET-MODEL-(VSM) | 0.00954 |
| EXISTENCE-PROBABILITY | 0.00844 |
| RANDOM-PERTURBATION | 0.00718 |
| WEIGHTED-NETWORK | 0.00602 |
| RANDOMIZATION-APPROACH | 0.00569 |
| TRIADIC-CLOSURE | 0.00472 |
| INJECTING-PROBABILITY | 0.00472 |
| Module 4 | PageRank Score |
| NEIGHBORHOOD-INFORMATION | 0.00684 |
| CLOUD-COMPUTING | 0.00593 |
| UNDIRECTED-GRAPH | 0.00552 |
| K-NEIGHBORHOOD-ANONYMITY | 0.00492 |
| NEIGHBORHOOD-ATTACK | 0.00445 |
| ENCRYPTION | 0.00434 |
| GREEDY-ALGORITHM | 0.00433 |
| GRAPH-DATA | 0.00404 |
| FRIENDSHIP-ATTACK | 0.00399 |
| EDGE-CLUSTERING | 0.00356 |
| REACHABILITY-PRESERVING-ANONYMIZATION-(RPA) | 0.00245 |
| REACHABILITY | 0.00244 |
| QUANTIFICATION | 0.00229 |
| AVERAGE-SHORTEST-PATH-LENGTH-(APL) | 0.00150 |
| Module 5 | PageRank Score |
| STATISTICAL-DISCLOSURE-ATTACK-(SDA) | 0.00834 |
| POWER-LAW | 0.00834 |
| SMALL-WORLD-NETWORK | 0.00834 |
| SOCIAL-NETWORK-ANALYSIS-(SNA) | 0.00833 |
| DEGREE-SEQUENCE | 0.00644 |
| RELATIONAL-DATA | 0.00629 |
| EMAIL-NETWORK | 0.00603 |
| GRAPH-METRICS | 0.00132 |
| GRAPH-SAMPLING | 0.00105 |
| Module 6 | PageRank Score |
| HEURISTIC-ALGORITHMS | 0.00747 |
| PARAMETERIZED-COMPLEXITY | 0.00640 |
| OPTIMIZATION | 0.00602 |
| F-FACTORS | 0.00587 |
| KERNELIZATION | 0.00587 |
| COMBINATORIAL-OPTIMIZATION | 0.00485 |
| GENETIC-ALGORITHM-(GA) | 0.00485 |
| FIXED-PARAMETER-TRACTABILITY | 0.00438 |
| DEGREE-CONSTRAINED-EDITING | 0.00438 |
| Module 7 | PageRank Score |
| CLOUD-STORAGE | 0.00929 |
| GENERAL-DATA-PROTECTION-REGULATION-(GDPR) | 0.00929 |
| ELLIPTIC-CURVE | 0.00929 |
| PUBLIC-KEY-ENCRYPTION-WITH-KEYWORD-SEARCH-(PEKS) | 0.00929 |
| CRYPTOGRAPHY | 0.00812 |
| Module 8 | PageRank Score |
| PERTURBATION-TECHNIQUES | 0.00666 |
| LINK-PRIVACY | 0.00664 |
| PAIRED-VERTICES-K-DEGREE-ANONYMITY | 0.00606 |
| UNDERLYING-GRAPH | 0.00546 |
| FACEBOOK | 0.00437 |
| TWITTER | 0.00437 |
| DEGREE-OF-PAIRED-VERTICES-(DOPV)-ATTACK | 0.00406 |
| MULTI-NETWORK | 0.00406 |
| Module 9 | PageRank Score |
| BIG-DATA | 0.00533 |
| ADVERSARY | 0.00531 |
| BIG-COMPONENT | 0.00531 |
| SOCIAL-RECOMMENDATIONS | 0.00510 |
| PERSONALIZATION | 0.00510 |
| LAPLACE-NOISE | 0.00492 |
| EMBEDDED-SYSTEMS | 0.00433 |
| CORRELATION-MATRIX | 0.00158 |
| Module 10 | PageRank Score |
| UTILITY | 0.00655 |
| EDGE-ADDITION/DELETION | 0.00535 |
| GRAPH-MODIFICATION-APPROACH | 0.00532 |
| GRAPH-DEGENERACY | 0.00426 |
| CORE-NUMBER-SEQUENCE | 0.00426 |
| LARGE-DATASET | 0.00352 |
| REAL-WORLD-DATASETS | 0.00316 |
| EDGE-MEASURES | 0.00166 |
| GRAPH-MINING | 0.00132 |
| AUXILIARY-INFORMATION | 0.000768 |
| Module 11 | PageRank Score |
| WEIGHTED-MAXIMUM-COMMON-SUBGRAPH-(WMCS) | 0.00685 |
| COMPLETE-GRAPH | 0.00670 |
| DISTRIBUTED-ENVIRONMENT | 0.00670 |
| DISTRIBUTED-STORAGE | 0.00670 |
| SUBGRAPH-SIMILARITY-DETECTION | 0.00439 |
| SET-VALUED-DATA | 0.00439 |
| Module 12 | PageRank Score |
| MARKOV-CHAIN-MONTE-CARLO | 0.00855 |
| HIERARCHICAL-RANDOM-NETWORK | 0.00855 |
| MONTE-CARLO-METHODS | 0.00855 |
| FUZZY-SET-THEORY | 0.00855 |
| Module 13 | PageRank Score |
| SUPERVISED-MACHINE-LEARNING-ALGORITHMS | 0.00616 |
| ADVERSARIAL-MACHINE-LEARNING | 0.00568 |
| SOCIAL-RELATIONSHIPS | 0.00568 |
| MACHINE-LEARNING | 0.00568 |
| ADAPTIVE-RANDOM-WALK | 0.00390 |
| INTERMEDIATE-NODE | 0.00390 |
| Module 14 | PageRank Score |
| NOISE-NODE | 0.00644 |
| L-DIVERSITY | 0.00579 |
| ALPHA-ANONYMIZATION | 0.00543 |
| SENSITIVE-ATTRIBUTES | 0.00453 |
| LOSSY-JOIN | 0.00328 |
| RECURSIVE-(C,L)-DIVERSITY | 0.00268 |
| EIGENVECTOR-CENTRALITY | 0.00181 |
| Module 15 | PageRank Score |
| LEVEL-CUT-HEURISTIC-BASED-CLUSTERING-ALGORITHM | 0.00604 |
| K-DEGREE-CLOSENESS-ANONYMITY | 0.00604 |
| CENTRALITY-MEASURES | 0.00604 |
| K-DEGREE | 0.00478 |
| PASSIVE-ATTACK | 0.00166 |
| MEASUREMENT | 0.00124 |
| Module 16 | PageRank Score |
| RANDOM-NETWORK | 0.00800 |
| BOOTSTRAP-PERCOLATION | 0.00570 |
| GRAPH-MATCHING-TECHNIQUE | 0.00570 |
| DE-ANONYMIZATION | 0.00515 |
| Module 17 | PageRank Score |
| INFERENCE-ALGORITHM | 0.00556 |
| COLLECTIVE-CLASSIFICATION | 0.00370 |
| EFFICIENT-FRONTIER | 0.00370 |
| SPATIO-TEMPORAL-FEATURE | 0.00360 |
| SENSITIVE-EDGES | 0.00305 |
| Module 18 | PageRank Score |
| ANONYMIZATION-PROCESS-RESTART-ISSUE-(APRI) | 0.00572 |
| ACTIVATED-FUNCTION | 0.00572 |
| INCREMENTAL-DATA-PUBLISHING | 0.00572 |
| Module 19 | PageRank Score |
| COLLABORATIVE-NETWORK | 0.00496 |
| COMPRESSIVE-SENSING | 0.00458 |
| RELATIONSHIP-MATRIX | 0.00458 |
| CHARACTERISTIC-VECTOR | 0.00277 |
| Module 20 | PageRank Score |
| SHORTEST-PATH | 0.00542 |
| (K1,K2)-SHORTEST-PATH | 0.00382 |
| K-SHORTEST-PATH | 0.00382 |
| PRIVACY-PRESERVING-DATA-MINING-(PPDM) | 0.00255 |
| Module 21 | PageRank Score |
| GENERALIZATION-APPROACH | 0.00401 |
| UNCERTAIN-GRAPHS | 0.00361 |
| MAXIMIZING-VARIANCE | 0.00270 |
| NODE-DEGREE | 0.00236 |
| ATTRIBUTE-GRAPH | 0.000592 |
| LOCATION-PREDICTION | 0.000575 |
| Module 22 | PageRank Score |
| GRAPH-OPERATIONS | 0.00450 |
| GRAPH-CONTRACTION | 0.00450 |
| NP-HARD | 0.00396 |
| Module 23 | PageRank Score |
| DIRECTED-GRAPH | 0.00462 |
| INFLUENCE-MAXIMIZATION | 0.00387 |
| SOCIAL-INFLUENCE | 0.00387 |
| Module 24 | PageRank Score |
| DIFFERENTIAL-PRIVACY | 0.00607 |
| TOPOLOGICAL-INFORMATION | 0.00295 |
| PERSISTENT-HOMOLOGY | 0.00241 |
| Module 25 | PageRank Score |
| COMMUNITY-DETECTION-ALGORITHMS | 0.00393 |
| COMMUNITY-STRUCTURE | 0.00314 |
| IDENTITY-DISCLOSURE | 0.00131 |
| HETEROGENEOUS-NETWORK | 0.00113 |
| K-SHELLS | 0.000827 |
| EQUI-CARDINAL-CLUSTERING-(ECC) | 0.000711 |
| Module 26 | PageRank Score |
| GRAPH-PARTITIONING | 0.00372 |
| VERTEX-CONNECTIVITY | 0.00291 |
| TOPOLOGICAL-STRUCTURE-SIMILARITY | 0.00268 |
| Module 27 | PageRank Score |
| RANDOM-MATRIX | 0.00388 |
| MATRIX-ALGEBRA | 0.00388 |
| Module 28 | PageRank Score |
| LINKING-ATTRIBUTE-TABLE | 0.00375 |
| SENSITIVE-LABELS | 0.00375 |
| Module 29 | PageRank Score |
| DECISION-SUPPORT-SYSTEMS-(DSS) | 0.00325 |
| HYBRID-ALGORITHMS | 0.00230 |
| SEQUENTIAL-RELEASES | 0.00151 |

| Module 30 | PageRank Score |
|---|---|
| SUBGRAPH | 0.00334 |
| HEALTH-INFORMATION | 0.00334 |
| Module 31 | PageRank Score |
| PAGERANK-ALGORITHM | 0.00311 |
| ROUGH-SET-THEORY | 0.00311 |
| Module 32 | PageRank Score |
| IMPRECISION-BOUND | 0.00310 |
| SLICING | 0.00310 |
| Module 33 | PageRank Score |
| AUTOMATED-GROUPING | 0.00300 |
| SENSITIVE-SENTIMENTS | 0.00300 |
| Module 34 | PageRank Score |
| ACTIVE-ATTACKS | 0.00297 |
| ANTIDIMENSION | 0.00297 |
| Module 35 | PageRank Score |
| K-OUT-ANONYMITY | 0.00291 |
| IMPORTANT-NODES | 0.00291 |
| Module 36 | PageRank Score |
| SUPPRESSION | 0.00287 |
| K-MEANS | 0.00287 |
| Module 37 | PageRank Score |
| K-NMF | 0.00279 |
| MUTUAL-FRIEND-ATTACK | 0.00279 |
| Module 38 | PageRank Score |
| CONFIDENTIALITY | 0.00276 |
| SYNTHETIC-GRAPH-GENERATION | 0.00276 |
| Module 39 | PageRank Score |
| NETWORK-COHESION | 0.00264 |
| RECOMMENDER-SYSTEMS | 0.00264 |

| Subperiod 2019-2022 | |
|---|---|
| Module 1 | PageRank Score |
| PRIVACY-PRESERVATION | 0.0152 |
| SOCIAL-NETWORK | 0.0137 |
| ANONYMIZATION | 0.0109 |
| STRUCTURAL-PROPERTIES | 0.00976 |
| DATA-MINING | 0.00770 |
| DIFFERENTIAL-PRIVACY | 0.00719 |
| DATA-PUBLISHING | 0.00719 |
| GRAPH-MODIFICATION-APPROACH | 0.00714 |
| REAL-WORLD-DATASETS | 0.00686 |
| UTILITY | 0.00670 |
| GRAPH-THEORY | 0.00648 |
| INFORMATION-LOSS | 0.00637 |
| PERTURBATION-TECHNIQUES | 0.00623 |
| K-ANONYMITY | 0.00616 |
| CLUSTERING-ALGORITHMS | 0.00608 |
| PRIVACY-ATTACK-MODEL | 0.00539 |
| TOPOLOGICAL-INFORMATION | 0.00467 |
| GRAPH-MINING | 0.00458 |
| GENERALIZATION-APPROACH | 0.00436 |
| DEGREE-SEQUENCE | 0.00391 |
| MICROAGGREGATION | 0.00352 |
| RE-IDENTIFICATION-ATTACKS | 0.00302 |
| DK-GRAPH | 0.00293 |
| MEASUREMENT | 0.00135 |
| KRONECKER-GRAPHS | 0.00106 |
| PERSISTENT-HOMOLOGY | 0.000963 |
| COMMUNITY-PRIVACY | 0.000624 |
| SHORTEST-PATH | 0.000601 |
| CENTROID-SELECTION | 0.000479 |
| ALL-DISTANCE-SKETCH | 0.000309 |
| EDGE-WEIGHT-ANONYMIZATION | 0.0000556 |
| Module 2 | PageRank Score |
| STRUCTURAL-ATTACKS | 0.0108 |
| GRAPH-QUERIES | 0.0108 |
| MATRIX-DECOMPOSITION | 0.0108 |
| SET-VALUED-DATA | 0.0108 |
| ADVERSARIAL-MACHINE-LEARNING | 0.0108 |
| SINGULAR-VALUE-DECOMPOSITION-(SVD) | 0.00993 |
| DE-ANONYMIZATION | 0.00847 |
| GRAPH-ISOMORPHISM | 0.00798 |
| DISTINGUISHABILITY-PROBABILITY | 0.00478 |
| SPLIT-ALGORITHMS | 0.00478 |
| INFERENCE-ATTACK | 0.00450 |
| Module 3 | PageRank Score |
| HIGH-DEGREE-NODES | 0.00871 |
| PREDICTIVE-MODELS | 0.00871 |
| NEIGHBORHOOD-INFORMATION | 0.00871 |
| OPTIMIZATION | 0.00864 |
| ARTIFICIAL-NEURAL-NETWORK-(ANN) | 0.00708 |
| FIREFLY-ALGORITHM-(FA) | 0.00626 |
| FUZZY-CLUSTERING | 0.00626 |
| IDENTITY-DISCLOSURE | 0.00626 |
| GRAPH-NEURAL-NETWORK-(GNN) | 0.00596 |
| Module 4 | PageRank Score |
| CRYPTOGRAPHY | 0.00780 |
| HOMOMORPHIC-ENCRYPTION | 0.00730 |
| CLOUD-COMPUTING | 0.00692 |
| BIG-DATA | 0.00655 |
| MARKOV-CLUSTERING-(MCL) | 0.00615 |
| CLUSTERING-ACCURACY | 0.00615 |
| GRAPH-OPERATIONS | 0.00544 |
| SERVERS | 0.00538 |
| AUTHENTICATION | 0.00356 |
| Module 5 | PageRank Score |
| GENERATIVE-ADVERSARIAL-NETWORK-(GAN) | 0.00921 |
| FEATURE-LEARNING | 0.00921 |
| GRAPH-DATA | 0.00921 |
| ADAPTIVE-RANDOM-WALK | 0.00921 |
| MACHINE-LEARNING | 0.00797 |
| MATRIX-ALGEBRA | 0.00666 |
| Module 6 | PageRank Score |
| INTERNET-OF-THINGS-(IOT) | 0.00701 |
| AUXILIARY-INFORMATION | 0.00609 |
| PRIVACY-BREACH | 0.00560 |
| (ALPHA,-L)-DIVERSITY | 0.00539 |
| CLUSTERING-FEATURE-(CF)-TREE | 0.00539 |
| PARTITIONING-APPROACH | 0.00520 |
| POWER-LAW | 0.00487 |
| GRAPH-PARTITIONING | 0.00485 |
| FREQUENT-PATTERN-MINING | 0.00370 |
| REGULARITY-LEMMA | 0.00192 |
| GRAPHLETS | 0.00134 |
| Module 7 | PageRank Score |
| STRUCTURED-DATA | 0.00670 |
| SEQUENTIAL-CLUSTERING | 0.00577 |
| ASSORTATIVITY | 0.00577 |
| ARTIFICIAL-INTELLIGENCE-(AI) | 0.00538 |
| ANATOMY | 0.00416 |
| NODE-DEGREE | 0.00389 |
| VARIATIONAL-GRAPH-AUTOENCODERS-(VGA) | 0.00374 |
| GRAPH-RECOVERY-ATTACK | 0.00374 |
| K-DEGREE | 0.00310 |
| ONE-TIME-SCAN | 0.00233 |
| HIERARCHICAL-CLUSTERING | 0.00217 |
| Module 8 | PageRank Score |
| COLLUSION | 0.00887 |
| EMERGENCY-RESCUE | 0.00887 |
| GAME-THEORY | 0.00887 |
| FOUNTAIN-CODE-(FC) | 0.00887 |
| DISTRIBUTED-STORAGE | 0.00887 |
| Module 9 | PageRank Score |
| FACEBOOK | 0.00775 |
| DUNBAR'S-NUMBER | 0.00586 |
| GENERAL-DATA-PROTECTION-REGULATION-(GDPR) | 0.00586 |
| SOCIAL-NETWORK-ANALYSIS-(SNA) | 0.00586 |
| SYNTHETIC-GRAPH-GENERATION | 0.00586 |
| CORRELATION-MATRIX | 0.00391 |
| Module 10 | PageRank Score |
| NEURO-FUZZY | 0.00794 |
| FUZZY-SET-THEORY | 0.00794 |
| MIDDLEWARE-SERVER | 0.00733 |
| EMOTIONAL-DETECTION | 0.00733 |
| REWIRING | 0.00381 |
| Module 11 | PageRank Score |
| CLUSTER-ANALYSIS | 0.00674 |
| CONNECTION-FINGERPRINT-ATTACK | 0.00630 |
| EQUIVALENCE-CLASSES | 0.00630 |
| PERSONALIZATION | 0.00630 |
| L-DIVERSITY | 0.00503 |
| T-CLOSENESS | 0.00141 |
| Module 12 | PageRank Score |
| GRAPH-MATCHING-TECHNIQUE | 0.00589 |
| (K,T)-PRIVACY | 0.00578 |
| LABEL-GENERALIZATION | 0.00578 |
| K-AUTOMORPHISM | 0.00578 |
| 1-NEIGHBORHOOD-ATTACK | 0.00271 |
| JENSEN-SHANNON-DIVERGENCE | 0.00189 |
| Module 13 | PageRank Score |
| COMMUNITY-DETECTION-ALGORITHMS | 0.00755 |
| BACKGROUND-KNOWLEDGE | 0.00495 |
| NEIGHBORHOOD-ATTACK | 0.00367 |
| CONFIDENTIALITY | 0.00282 |
| HEALTH-INFORMATION | 0.00255 |
| COMMUNITY-STRUCTURE | 0.00235 |
| INFLUENCE-MATRIX | 0.00185 |
| Module 14 | PageRank Score |
| CORRELATION-ATTACK | 0.00638 |
| CALL-DETAIL-RECORDS | 0.00638 |
| BLOWFISH-PRIVACY | 0.00638 |
| MUTUAL-FRIEND-ATTACK | 0.00402 |
| Module 15 | PageRank Score |
| CLOUD-BASED-ONLINE-SOCIAL-NETWORKS-(COSN) | 0.00564 |
| ATTRIBUTE-BASED-ENCRYPTION-(ABE) | 0.00564 |
| PRIVACY-PRESERVING-DATA-MINING-(PPDM) | 0.00564 |
| DIRECTED-GRAPH | 0.00492 |
| Module 16 | PageRank Score |
| DIFFUSION-PROCESS | 0.00593 |
| BACKPROPAGATION-ALGORITHM | 0.00593 |
| BLOGS | 0.00593 |
| SUPERVISED-MACHINE-LEARNING-ALGORITHMS | 0.00369 |
| Module 17 | PageRank Score |
| LARGE-DATASET | 0.00584 |
| SOCIAL-RELATIONSHIPS | 0.00584 |
| HUBS | 0.00399 |
| LINK-PRIVACY | 0.00399 |
| Module 18 | PageRank Score |
| RANDOM-PROJECTION-ALGORITHM | 0.00496 |
| RANDOM-MATRIX | 0.00496 |
| RANDOMIZATION-APPROACH | 0.00449 |
| RANDOM-PERTURBATION | 0.00438 |
| Module 19 | PageRank Score |
| SENSITIVE-ATTRIBUTES | 0.00524 |
| STRUCTURE-ATTRIBUTE-GRAPH-(SAG) | 0.00437 |
| NATURAL-LANGUAGE-PROCESSING-SYSTEMS-(NLP) | 0.00437 |
| ATTRIBUTE-GRAPH | 0.00346 |
| Module 20 | PageRank Score |
| COMBINATORIAL-OPTIMIZATION | 0.00422 |
| GRAPH-GENERATION-MODEL | 0.00396 |
| AVERAGE-SHORTEST-PATH-LENGTH-(APL) | 0.00379 |
| CLUSTER-COEFFICIENT | 0.00286 |
| CUCKOO-OPTIMIZATION-ALGORITHM-(COA) | 0.00198 |
| Module 21 | PageRank Score |
| PERSONAL-GAZETTEER | 0.00531 |
| TRAJECTORY | 0.00531 |
| GLOBAL-POSITIONING-SYSTEM-(GPS) | 0.00531 |
| Module 22 | PageRank Score |
| NEIGHBORHOOD-ATTRACTION-FIREFLY-ALGORITHM-(NAFA) | 0.00479 |
| NUMBER-FACTORIZATION | 0.00479 |
| F-FACTORS | 0.00479 |
| Module 23 | PageRank Score |
| M-PRIVACY | 0.00473 |
| INSIDER-ATTACK | 0.00473 |
| COLLABORATIVE-NETWORK | 0.00473 |
| Module 24 | PageRank Score |
| LINKAGE-ATTACK | 0.00451 |
| SYBIL-ATTACK | 0.00451 |
| BLOOM-FILTER | 0.00451 |
| Module 25 | PageRank Score |
| NONDETERMINISTIC-ALGORITHMS | 0.00401 |
| RECOMMENDER-SYSTEMS | 0.00401 |
| UNDIRECTED-GRAPH | 0.00382 |
| Module 26 | PageRank Score |
| EDGE-ADDITION/DELETION | 0.00373 |
| BLOCKCHAIN | 0.00351 |
| NOISE-ADDITION | 0.00351 |
| (K,L)-ANONYMITY | 0.00109 |
| Module 27 | PageRank Score |
| PARTICLE-SWARM-OPTIMIZATION-ALGORITHM-(PSO) | 0.00387 |
| HYBRID-ALGORITHMS | 0.00387 |
| GENETIC-ALGORITHM-(GA) | 0.00307 |
| Module 28 | PageRank Score |
| UNCERTAIN-GRAPHS | 0.00325 |
| EXISTENCE-PROBABILITY | 0.00280 |
| TRIADIC-CLOSURE | 0.00116 |
| NODE-CHARACTERISTICS | 0.00106 |
| Module 29 | PageRank Score |
| GRAPHX | 0.00309 |
| REACHABILITY | 0.00309 |
| Module 30 | PageRank Score |
| K-COUPLET-ANONYMITY | 0.00297 |
| ATTRIBUTE-COUPLET-ATTACK | 0.00297 |

| Module 31 | PageRank Score |
|---|---|
| EQUI-CARDINAL-CLUSTERING-(ECC) | 0.00285 |
| TOPOLOGICAL-STRUCTURE-SIMILARITY | 0.00285 |
| Module 32 | PageRank Score |
| BETWEENNESS-CENTRALITY | 0.00265 |
| NOISE-NODE | 0.00265 |
| Module 33 | PageRank Score |
| TIME-SERIAL-DATA | 0.00257 |
| DYNAMIC-NETWORK | 0.00257 |
| Module 34 | PageRank Score |
| SENSITIVE-WEIGHTED-EDGES | 0.00234 |
| EDGE-BETWEENNESS-CENTRALITY | 0.00234 |

**Declaration of interests**

☒The authors declare that they have no known competing financial interests or personal relationships that could have appeared to influence the work reported in this paper.

☐The authors declare the following financial interests/personal relationships which may be considered as potential competing interests: